\documentclass[aps,prx,twocolumn,superscriptaddress]{revtex4-2}

\usepackage{graphicx,latexsym}
\usepackage{amsmath,amssymb}
\usepackage{amsthm}
\usepackage{bm}
\usepackage{braket}
\usepackage{longtable}
\newcommand{\up}{\uparrow}
\newcommand{\down}{\downarrow}

\usepackage{color}

\theoremstyle{remark}

\usepackage[%
 setpagesize=false,%
 bookmarks=false,%
 bookmarksdepth=tocdepth,%
 bookmarksnumbered=true,%
 colorlinks=true,%
 linkcolor=blue,%
 citecolor=blue,%
 urlcolor=magenta%
]{hyperref}

\begin{document}


\title{Topology of Discrete Quantum Feedback Control}


\author{Masaya Nakagawa}
\email{nakagawa@cat.phys.s.u-tokyo.ac.jp}
\affiliation{Department of Physics, The University of Tokyo, 7-3-1 Hongo, Bunkyo-ku, Tokyo 113-0033, Japan}
\author{Masahito Ueda}
\affiliation{Department of Physics, The University of Tokyo, 7-3-1 Hongo, Bunkyo-ku, Tokyo 113-0033, Japan}
\affiliation{Institute for Physics of Intelligence, The University of Tokyo, 7-3-1 Hongo, Bunkyo-ku, Tokyo 113-0033, Japan}
\affiliation{RIKEN Center for Emergent Matter Science (CEMS), Wako, Saitama 351-0198, Japan}


\date{\today}

\begin{abstract}
    A general framework for analyzing the topology of quantum channels of single-particle systems is developed to find a class of genuinely dynamical topological phases that can be realized by means of discrete quantum feedback control. We provide a symmetry classification of quantum channels by identifying ten symmetry classes of discrete quantum feedback control with projective measurements. We construct various types of topological feedback control by using topological Maxwell demons that achieve robust feedback-controlled chiral or helical transport against noise and decoherence. Topological feedback control thus offers a versatile tool for creating and controlling nonequilibrium topological phases in open quantum systems that are distinct from non-Hermitian and Lindbladian systems and should provide a guiding principle for topology-based design of quantum feedback control.
\end{abstract}


\maketitle

\section{Introduction}

Topology plays a key role in the classification and characterization of diverse physical phenomena that are robust against local disturbances. In particular, topological phases of matter have been discussed in a wide variety of quantum and classical systems, which are characterized by global geometric structures of eigenstates rather than local order parameters \cite{MoessnerMoore_book,Wen17,Ozawa19,Ma19,Xue22,Shankar22}. 
A quintessential example of topological quantum phenomena is the quantum Hall effect, where chiral transport immune to disorder emerges along the edges of a system \cite{Klitzing80,Laughlin81,Thouless82,Buttiker88,Hatsugai93}. Furthermore, topological phases are enriched by symmetry. For example, time-reversal symmetry permits helical edge states, where time-reversal partners flow in opposite directions with no net mass current, a phenomenon known as the quantum spin Hall effect \cite{KaneMele05_1,KaneMele05_2,Bernevig06,Wu06,Koenig07}. Classification of phases of matter through integration of topology and symmetry has been of central interest in condensed matter physics \cite{Schnyder08,Kitaev09,Chiu16}.

The general concept of topology can also be utilized to classify nonequilibrium dynamics. 
Examples include periodically driven (Floquet) systems characterized by topology of unitary time-evolution operators \cite{Thouless83,Oka09,Kitagawa10,Kitagawa10_2,Kitagawa11,Lindner11,Nathan15,RoyHarper17,Higashikawa19,Oka19,Rudner20,Harper20} and topological phases of open systems described by non-Hermitian Hamiltonians \cite{Esaki11,Gong18,Kawabata19,Zhou19,Bergholtz21,Liu22,Okuma23}. Notably, out-of-equilibrium driving not only provides a useful tool for controlling topological phases, but also induces genuinely dynamical topological phases that have no static counterpart \cite{Kitagawa10,Rudner13,Higashikawa19,Gong18}. 
Finding a new control method for topological phases thus expands the scope of research of topological phases of matter. 

Feedback control provides a versatile tool for a wide range of applications in science and technology \cite{WisemanMilburn_book}. In particular, in the research arena of high-precision measurement and control of quantum systems, feedback control has achieved suppression of quantum noise \cite{Inoue13}, stabilization of desired states \cite{Sayrin11,Vijay12}, cooling \cite{Kamba21,Tebbenjohanns21,Magrini21}, and quantum error correction \cite{NielsenChuang_book,Cramer16,Krinner22,Bluvstein23}. 
Feedback control also allows operations that would otherwise be impossible.
A prime example is Maxwell's demon, which was originally conceived as an intelligent being that can decrease entropy of a system against the second law of thermodynamics \cite{Maxwell}. Thus, Maxwell's demon can be formulated as a feedback controller \cite{Sagawa08,Parrondo15}, and various types of Maxwell demons have experimentally been realized in both classical \cite{Toyabe10,Koski14_1,Koski14_2,Saha21} and quantum \cite{Vidrighin16,Campati16,Cottet17,Nagihloo18,Masuyama18,Kumar18} systems.
In view of versatile applicability of feedback control, we are naturally led to the following question: Can we utilize feedback control to create and manipulate nonequilibrium topological phases? 
We address this problem in the following.

In this paper, we present a class of dynamical topological phases that can be realized with feedback control of quantum systems. Here, a unique role played by feedback control is that one can choose operations depending on measurement outcomes. 
This flexibility enables versatile real-time control of topological phases that would be hard to achieve by unitary dynamics in which the driving protocol is predetermined and cannot depend on the state of a system under control. 
Furthermore, feedback control allows active control of a quantum system, in contrast with passive control of open quantum systems by engineered dissipation. 

\begin{figure*}
    \includegraphics[width=17cm]{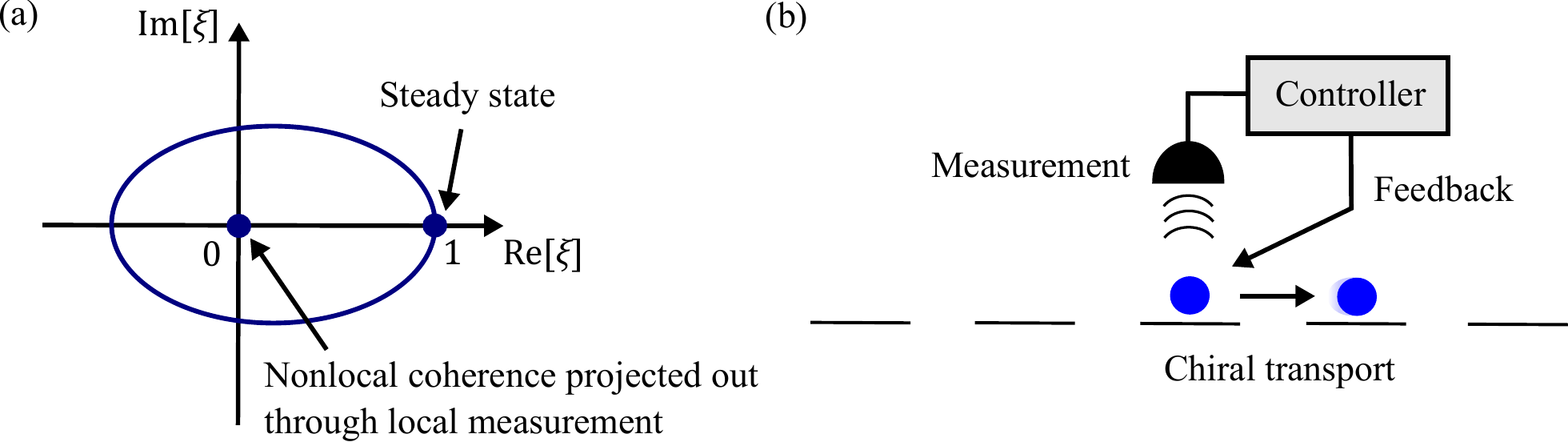}
    \caption{(a) Schematic illustration of the eigenspectrum $\{\xi\}$ (shown by a blue curve and two blue dots) of a quantum channel for topological quantum feedback control. The spectrum involves a closed loop characterized by a winding number and the highly degenerate zero eigenvalues due to local measurement through which nonlocal coherence is projected out. The unit eigenvalue corresponds to a steady state. (b) Discrete quantum feedback control comprised of a quantum measurement and a subsequent feedback operation that depends on the measurement outcome. A topologically nontrivial winding of the eigenspectrum of a quantum channel [see diagram (a)] implies chiral (unidirectional) transport of a particle on a lattice due to feedback control.}
    \label{fig_concept}
\end{figure*}

\begin{table*}
    \caption{Comparison of different classes of topological phases. The first row shows five classes of topological phases characterized by the topology of relevant operators shown in the second row. The third row indicates whether the dynamics is static, discrete, or continuous. The fourth row shows whether the system is closed or open. The fifth row shows whether the state of the system is pure or mixed. \label{table_top}}
    \begin{ruledtabular}
    \begin{tabular}{lccccc}
    Class & Ground state & Floquet & Non-Hermitian & Lindblad & Discrete quantum feedback \\ \hline
    Operator & Hamiltonian & Unitary & Non-Hermitian Hamiltonian & Lindbladian & Quantum channel \\
    Time evolution & Static & Discrete & Continuous & Continuous & Discrete \\
    Closed or open & Closed & Closed & Open & Open & Open \\
    Pure or mixed & Pure & Pure & Pure & Mixed & Mixed
    \end{tabular}
    \end{ruledtabular}
\end{table*}

We develop a general framework for analyzing the topology of quantum channels for discrete quantum feedback control. Mathematically, quantum channels are described by completely positive and trace-preserving (CPTP) maps, which provide a general form of time evolution in open quantum systems \cite{NielsenChuang_book}. Here, the time evolution of a system under feedback control is nonunitary due to measurement backaction. Consequently, the topology of CPTP maps is intrinsically non-Hermitian and beyond the theoretical framework of static (Hermitian) topological phases. We introduce topological invariants of quantum channels for single-particle systems on lattices under the general assumption of locality in measurement and feedback. For example, a nonzero winding number of the eigenspectrum of a quantum channel implies chiral transport due to feedback control, which may be regarded as a feedback-controlled counterpart of chiral edge transport in quantum Hall systems as schematically illustrated in Fig.~\ref{fig_concept}. 

Quantum feedback control creates a class of nonequilibrium open quantum systems distinct from those described by non-Hermitian Hamiltonians \cite{Gong18,Kawabata19,Zhou19} and Lindblad master equations \cite{Bardyn13,Lieu20,Kawabata22,Sa22}. 
We summarize the comparison of different classes of topological phases in Table \ref{table_top}. 
In particular, discrete quantum feedback control concerns discrete time evolution from an initial quantum state to a postfeedback quantum state. Such time evolution cannot, in general, be expressed by using a generator of continuous time evolution. 
In fact, as shown in Sec.~\ref{sec_characterization}, CPTP maps with local Kraus operators have highly degenerate zero eigenvalues, and therefore cannot be described by using any generator [see Eq.~\eqref{eq_generator}]. Physically, the existence of zero eigenvalues can be understood as a consequence of local measurement through which nonlocal coherence in the system is projected out [see Fig.~\ref{fig_concept}(a) and Sec.~\ref{sec_characterization}]. 
A non-Hermitian Hamiltonian can only describe the dynamics of pure states and requires postselection of particular measurement outcomes \cite{Ashida20}. The Lindblad quantum master equation can describe the dynamics of mixed states but does not refer to measurement outcomes \cite{WisemanMilburn_book}. In contrast, nonequilibrium topological phases created by quantum feedback control are realized by a nontrivial interplay between quantum measurement and feedback operations. We note that non-Hermitian topology has been applied to quantum channels in Ref.~\cite{Gong18}. However, its application beyond zero-dimensional systems is nontrivial and has remained elusive; in fact, the locality of Kraus operators, which is relevant in higher dimensions, is key for defining topological invariants, as discussed in Sec.~\ref{sec_characterization}.

We present prototypical models of topological feedback control by using quantum Maxwell demons, which we shall refer to as topological Maxwell demons. In one dimension, a topological Maxwell demon causes feedback-assisted chiral transport due to topology. We find that the topological Maxwell demon creates topologically distinct phases by changing the duration of feedback control. We also show that feedback control can induce the non-Hermitian skin effect, which is a hallmark of dynamical topology unique to open systems \cite{Lee16,Yao18,Gong18,Yokomizo19,Okuma20,Zhang20,Okuma23,Lin23}. The chiral transport caused by the topological Maxwell demon is protected by topology and hence robust against disorder, noise, and decoherence during the feedback process.

Symmetry plays a pivotal role in the classification of topological phases \cite{Kitaev09,Schnyder08,Chiu16,RoyHarper17,Higashikawa19,Gong18,Kawabata19,Zhou19,Lieu20,Liu22,Kawabata22,Sa22}. Here, we perform a symmetry classification of quantum feedback control. We find that only a subset of the 38 symmetry classes of non-Hermitian operators \cite{Bernard02,Kawabata19,Zhou19} is compatible with quantum channels with projective measurements. We identify ten relevant symmetry classes of quantum channels for discrete quantum feedback control (see Table \ref{table_AZdag}) and provide examples of feedback control for each class. In particular, we propose a quantum Maxwell demon with a two-point measurement as an example of symmetry-protected topological feedback control. This demon generates helical spin transport protected by symmetry, which may be regarded as a counterpart of the quantum spin Hall effect in feedback-controlled quantum systems.

\begin{table*}
    \caption{Tenfold symmetry classification of discrete quantum feedback control with projective measurements. Each symmetry class is defined according to the presence or absence of $C,K$, and $Q$ symmetries in the CPTP map. Here, $C,K$, and $Q$ symmetries of a CPTP map $\mathcal{E}$ are defined by $\mathcal{V}_C\mathcal{E}^T\mathcal{V}_C^{-1}=\mathcal{E},\mathcal{V}_K\mathcal{E}^*\mathcal{V}_K^{-1}=\mathcal{E}$, and $\mathcal{V}_Q\mathcal{E}^\dag\mathcal{V}_Q^{-1}=\mathcal{E}$, respectively, where $\mathcal{V}_X$ ($X=C, K, Q$) is a unitary superoperator. Here, $0$ indicates the absence of symmetry, and $\pm 1$ indicates the presence of symmetry with $\mathcal{V}_X\mathcal{V}_X^*=\pm\mathcal{I}$, where $\mathcal{I}$ denotes the identity superoperator. The labeling of the symmetry classes follows Ref.~\cite{Kawabata19}, where psH denotes pseudo-Hermiticity and the subscript $+$ ($-$) of psH indicates that the unitary superoperator for $C$ or $K$ symmetry commutes (anticommutes) with that of the pseudo-Hermiticity. In addition to the symmetry classes of a CPTP map $\mathcal{E}$, the corresponding Altland-Zirnbauer$^\dag$ classes of $i\mathcal{E}$ are shown in the parentheses. In the fifth column, examples of conditions for feedback (FB) control in each symmetry class are shown (see Sec.~\ref{sec_SPT} for detailed discussions). CPTP maps with $C$ or $Q$ symmetry can be realized with feedback control with a two-point measurement (2PM). Symmetry classes without $K$ symmetry with $\mathcal{V}_K\mathcal{V}_K^*=+\mathcal{I}$ require unitary symmetry (US) due to the intrinsic modular conjugation symmetry of CPTP maps. AUS$_-$ denotes antiunitary symmetry whose square gives $-1$. Some symmetry classes may be realized with an additional feedback control that enforces the symmetry condition (see examples in Sec.~\ref{sec_demons}). The rightmost column shows concrete models of topological feedback control, which are presented in Secs.~\ref{sec_model} and \ref{sec_demons}. \label{table_AZdag}}
    \begin{ruledtabular}
    \begin{tabular}{lcccll}
    Class & $C$ & $K$ & $Q$ & Example of conditions & Model \\ \hline
    A & $0$ & $0$ & $0$ & US & QND demon / Chern demon  \\ 
    psH (AIII) & $0$ & $0$ & $1$ & US, 2PM, and modified reciprocity & \\ \hline
    AI$^\dag$ & $+1$ & $0$ & $0$ & US, 2PM, and reciprocity & \\
    AI $+$ psH$_+$ (BDI$^\dag$) & $+1$ & $+1$ & $1$ & 2PM and reciprocity & \\
    AI (D$^\dag$) & $0$ & $+1$ & $0$ & No symmetry or particle-hole symmetry & Chiral Maxwell demon / SSH demon\\
    AI $+$ psH$_-$ (DIII$^\dag$) & $-1$ & $+1$ & $1$ & 2PM, reciprocity, and additional FB & Helical Maxwell demon \\
    AII$^\dag$ & $-1$ & $0$ & $0$ & US, 2PM, reciprocity, and additional FB & $\mathbb{Z}_2$ demon \\
    AII $+$ psH$_+$ (CII$^\dag)$ & $-1$ & $-1$ & $1$ & US, 2PM, AUS$_-$, reciprocity, and additional FB & \\
    AII (C$^\dag$) & $0$ & $-1$ & $0$ & US and AUS$_-$ & \\
    AII $+$ psH$_-$ (CI$^\dag$) & $+1$ & $-1$ & $1$ & US, 2PM,  AUS$_-$, and reciprocity & 
    \end{tabular}
    \end{ruledtabular}
\end{table*}

The rest of this paper is organized as follows. 
In Sec.~\ref{sec_feedback}, we introduce quantum channels that describe discrete quantum feedback control. 
In Sec.~\ref{sec_characterization}, we use the Bloch theory of translationally invariant quantum channels to show that topological invariants of quantum channels can be defined when Kraus operators are local in space. 
In Sec.~\ref{sec_model}, we construct an example of topological feedback control, which shows feedback-induced chiral transport due to its point-gap topology. 
In Sec.~\ref{sec_tenfold}, we examine the symmetry of discrete quantum feedback control and identify ten symmetry classes that are compatible with projective measurements. 
We also discuss how to construct feedback control in each symmetry class in Sec.~\ref{sec_SPT}. 
On the basis of the symmetry classification, we provide various models for symmetry-protected topological feedback control in Sec.~\ref{sec_demons}. In particular, we present topological feedback control protected by transposition symmetry, which exhibits helical spin transport due to an interplay between symmetry and topology. 
In Sec.~\ref{sec_top_CPTP}, we discuss topological phenomena arising from complete positivity and a trace-preserving property, which are unique to quantum channels and absent in general non-Hermitian systems.
In Sec.~\ref{sec_diss_fb}, we extend our theory to the case of nonunitary feedback control, thereby showing the robustness of topological feedback control against dissipation. We also propose feedback control with engineered dissipation, where dissipation is used as an essential ingredient to achieve feedback control. 
We conclude this paper in Sec.~\ref{sec_conclusion} with an outlook. 
Some technical details are relegated to the appendixes to avoid digressing from the main subject.
In Appendix \ref{sec_Bloch}, we present the details of the Bloch theory of quantum channels. 
In Appendix \ref{sec_proj}, we describe the Bloch theory of quantum feedback control with projective measurements. 
In Appendix \ref{sec_proj_channel}, we discuss a spectral property of quantum channels for projective measurements. 
In Appendix \ref{sec_sym_corresp}, we summarize the correspondence between symmetry of time-evolution operators and that of their generators. 
In Appendix \ref{sec_symOBC}, we provide a supplementary discussion of symmetry-preserving boundary conditions for a $\mathbb{Z}_2$ Maxwell demon.
In Appendix \ref{sec_spont_demon}, we present a model of topological feedback control that shows a non-Hermitian skin effect but no boundary-localized steady state.

\section{Discrete quantum feedback control\label{sec_feedback}}

We consider discrete quantum feedback control in which a feedback cycle consists of quantum measurements and subsequent unitary operations that are conditioned on measurement outcomes. Let $\hat{\rho}_{\mathrm{i}}$ be the initial density matrix of a system. We first perform a quantum measurement described by a set of measurement operators $\{ \hat{M}_m\}$ subject to the completeness relation
    \begin{equation}
        \sum_m \hat{M}_m^\dagger \hat{M}_m=\hat{I},
    \end{equation}
where the index $m$ specifies a measurement outcome and $\hat{I}$ is the identity operator \cite{WisemanMilburn_book}. The probability of an outcome $m$ being obtained is given by $p_m=\mathrm{Tr}[\hat{M}_m^\dag \hat{M}_m\hat{\rho}_{\mathrm{i}}]$, and the postmeasurement state corresponding to an outcome $m$ is given by
    \begin{equation}
        \hat{\rho}_m=\frac{\hat{M}_m \hat{\rho}_{\mathrm{i}}\hat{M}_m^\dag}{p_m},
    \end{equation}
where the nonunitary state change is caused by measurement backaction. 
Next, we let the system undergo unitary time evolution during time $\tau$, which is described by a unitary operator
    \begin{equation}
        \hat{U}_m=\mathcal{T}\exp\Bigl[-i\int_0^\tau dt\hat{H}_m(t)\Bigr],
        \label{eq_unitary}
    \end{equation}
where $\hat{H}_m(t)$ is the feedback Hamiltonian that depends on the measurement outcome $m$ and $\mathcal{T}$ is the time-ordering operator. We set $\hbar=1$ throughout this paper. 
If we take an ensemble average over measurement outcomes, then the final density matrix after one feedback loop is given by \cite{Sagawa08}
    \begin{align}
        \hat{\rho}_{\mathrm{f}}=&\sum_mp_m \hat{\rho}^\prime_m\notag\\
        =&\sum_m \hat{U}_m\hat{M}_m\hat{\rho}_{\mathrm{i}}\hat{M}_m^\dag \hat{U}_m^\dag,
        \label{eq_rho_afterfb}
    \end{align}
where $\hat{\rho}^\prime_m=\hat{U}_m\hat{\rho}_m\hat{U}_m^\dag$ is the density matrix after feedback control with measurement outcome $m$. 
Thus, a discrete quantum feedback control is described by a quantum channel
    \begin{align}
    \mathcal{E}(\hat{\rho}):=&\sum_m\hat{U}_m\hat{M}_m\hat{\rho} \hat{M}_m^\dag \hat{U}_m^\dag\notag\\
    =&\sum_m \hat{K}_m\hat{\rho} \hat{K}_m^\dag,
    \label{eq_CPTP}
    \end{align}
which is a CPTP map \cite{NielsenChuang_book}, where $\hat{K}_m:= \hat{U}_m\hat{M}_m$ is the Kraus operator satisfying $\sum_m\hat{K}_m^\dagger \hat{K}_m=\hat{I}$. 

If a feedback cycle is repeated, the final density matrix is given by applying CPTP maps multiple times. Suppose that we repeat a measurement and a feedback operation $N$ times. Let $(m_1,m_2,\cdots,m_N)$ be a sequence of measurement outcomes in this process. 
The density matrix after the $n$th measurement is given by
\begin{equation}
    \hat{\rho}_{m_1,\cdots,m_{n}}=\frac{\hat{M}_{m_{n}}\hat{\rho}_{m_1,\cdots,m_{n-1}}^\prime\hat{M}_{m_{n}}^\dag}{p(m_n|m_1,\cdots,m_{n-1})},
\end{equation}
where
\begin{equation}
    p(m_n|m_1,\cdots,m_{n-1})=\mathrm{Tr}[\hat{M}_{m_{n}}^\dag\hat{M}_{m_{n}}\hat{\rho}_{m_1,\cdots,m_{n-1}}^\prime]
\end{equation}
is the probability of an outcome $m_n$, conditioned on the past outcomes $m_1,\cdots,m_{n-1}$.
After the $n$th measurement, we perform a feedback operation and obtain the postfeedback density matrix as
\begin{equation}
    \hat{\rho}_{m_1,\cdots,m_n}^\prime=\hat{U}_{m_n}\hat{\rho}_{m_1,\cdots,m_{n}}\hat{U}_{m_n}^\dag.
\end{equation} 
Then, the final density matrix $\hat{\rho}_{\mathrm{f}}^{(N)}$, which is obtained as the averaged density matrix over all measurement outcomes, is given by
\begin{widetext}
    \begin{align}
        \hat{\rho}_{\mathrm{f}}^{(N)}=&\sum_{m_1,\cdots,m_N}p(m_N|m_1,\cdots,m_{N-1})p(m_{N-1}|m_1,\cdots,m_{N-2})\cdots p(m_2|m_1)p_{m_1}\hat{\rho}_{m_1,\cdots,m_N}^\prime\notag\\
        =&\sum_{m_1,\cdots,m_N}\hat{U}_{m_N}\hat{M}_{m_N}\cdots \hat{U}_{m_1}\hat{M}_{m_1}\hat{\rho}_{\mathrm{i}}\hat{M}_{m_1}^\dag\hat{U}_{m_1}^\dag\cdots\hat{M}_{m_N}^\dag\hat{U}_{m_N}^\dag\notag\\
        =&\mathcal{E}^N(\hat{\rho_{\mathrm{i}}}).
        \label{eq_CPTP_Ntimes}
    \end{align}
\end{widetext}
Thus, repeated feedback control is described by repeated applications of a CPTP map. 
Equation \eqref{eq_CPTP_Ntimes} suggests an analogy between periodically driven (Floquet) quantum systems \cite{Oka09,Kitagawa10,Kitagawa10_2,Kitagawa11,Lindner11,Nathan15,RoyHarper17,Higashikawa19,Oka19,Rudner20,Harper20} and feedback-controlled quantum systems: A unitary time-evolution operator for one period is repeatedly acted on a state vector in the former, while a CPTP map is repeatedly acted on a density matrix in the latter (see also Table \ref{table_top}).

The above description can be generalized to the cases in which different measurements and feedback operations are performed during a single cycle. For example, suppose that we perform a second measurement described by measurement operators $\{\hat{M}_{m_2}'\}$ after the first measurement and the subsequent feedback operation. Then, we perform the second feedback control with unitary operators $\{ \hat{U}_{m_1,m_2}'\}$, which may depend on the outcomes of the first and second measurements. The entire feedback cycle is described by a CPTP map
    \begin{align}
        \mathcal{E}'(\hat{\rho})=&\sum_{m_1,m_2}\hat{K}_{m_1,m_2}'\hat{\rho} (\hat{K}_{m_1,m_2}')^\dag,
    \end{align}
where $\hat{K}_{m_1,m_2}'=U_{m_1,m_2}'\hat{M}_{m_2}'\hat{U}_{m_1}\hat{M}_{m_1}$ is the Kraus operator. 
More general cases with nonunitary feedback operations will be discussed in Sec.~\ref{sec_diss_fb}.

Let $\mathcal{H}$ be a finite-dimensional Hilbert space of the system. Then, the density matrix $\hat{\rho}$ is an element of the space $\mathcal{L}(\mathcal{H})$ of linear operators on $\mathcal{H}$, which is a Hilbert space with dimension $(\dim\mathcal{H})^2$. The Hilbert-Schmidt inner product of two operators $\hat{A}$ and $\hat{B}$ in $\mathcal{L}(\mathcal{H})$ is denoted by $\langle \hat{A},\hat{B}\rangle:=\mathrm{Tr}[\hat{A}^\dag \hat{B}]$. A CPTP map $\mathcal{E}$ [Eq.~\eqref{eq_CPTP}] is regarded as an operator acting on the operator space $\mathcal{L}(\mathcal{H})$ (i.e., $\mathcal{E}$ is a superoperator). The adjoint superoperator $\mathcal{E}^\dag$ is defined by
    \begin{equation}
    \mathcal{E}^\dag(\hat{\sigma}):=\sum_m\hat{K}_m^\dag \hat{\sigma} \hat{K}_m,
    \end{equation}
which satisfies $\langle \hat{\sigma},\mathcal{E}(\hat{\rho})\rangle=\langle\mathcal{E}^\dag(\hat{\sigma}),\hat{\rho}\rangle$. 

We consider the eigenvalue problem of a CPTP map for feedback control:
    \begin{align}
        \mathcal{E}(\hat{\rho}_n^R)=\xi_n\hat{\rho}_n^R,\label{eq_eigenR}\\
        \mathcal{E}^\dag(\hat{\rho}_n^L)=\xi_n^*\hat{\rho}_n^L,\label{eq_eigenL}
    \end{align}
where $\xi_n\in\mathbb{C}$ denotes an eigenvalue, and $\hat{\rho}_n^R$ and $\hat{\rho}_n^L$ are the corresponding right and left eigenoperators that satisfy the biorthogonal relation $\mathrm{Tr}[(\hat{\rho}_m^L)^\dag \hat{\rho}_n^R]=0$ for $m\neq n$. In terms of the eigensystem of the CPTP map $\mathcal{E}$, the density matrix after a feedback loop in Eq.~\eqref{eq_rho_afterfb} is expanded as
    \begin{equation}
        \hat{\rho}_{\mathrm{f}}=\mathcal{E}(\hat{\rho}_{\mathrm{i}})=\sum_nc_n\xi_n\hat{\rho}_n^R,
        \label{eq_mode_expansion}
    \end{equation}
where
\begin{align}
    c_n=&\frac{\langle\hat{\rho}_n^L,\hat{\rho}_{\mathrm{i}}\rangle}{\langle\hat{\rho}_n^L,\hat{\rho}_n^R\rangle}
\end{align}
is an expansion coefficient that depends on the initial state $\hat{\rho}_{\mathrm{i}}$. Here, for simplicity, we assume that the CPTP map is diagonalizable. If the CPTP map is not diagonalizable, we can expand the density matrix by using generalized eigenmodes \cite{Ashida20}. 
Thus, the eigenvalues and the (generalized) eigenmodes of the CPTP map completely characterize the discrete quantum feedback control.

The eigensystem of a CPTP map has some general properties [see Fig.~\ref{fig_concept}(a) for a schematic illustration]. First, all the eigenvalues $\xi_n$ of a CPTP map satisfy $|\xi_n|\leq 1$ because of the contraction property of a CPTP map \cite{Rivas12}. An eigenmode $\hat{\rho}_{\mathrm{SS}}$ with eigenvalue $+1$ is a steady state of the dynamics in the sense that $\mathcal{E}(\hat{\rho}_{\mathrm{SS}})=\hat{\rho}_{\mathrm{SS}}$. Second, since a CPTP map preserves the trace of the density matrix, that is, $\mathrm{Tr}[\mathcal{E}(\hat{\rho})]=\mathrm{Tr}[\hat{\rho}]$, all the right eigenmodes except for those corresponding to steady states are traceless: $\mathrm{Tr}[\hat{\rho}_n^R]=0\ (\xi_n\neq 1)$. Third, a CPTP map preserves the Hermiticity of an operator since
\begin{equation}
[\mathcal{E}(\hat{\rho})]^\dag = \mathcal{E}(\hat{\rho}^\dag).
\label{eq_Hermiticity_preserv}
\end{equation}
It follows from Eqs.~\eqref{eq_eigenR} and \eqref{eq_Hermiticity_preserv} that if $\hat{\rho}_n^R$ is an eigenmode of $\mathcal{E}$ with eigenvalue $\xi_n$, then $(\hat{\rho}_n^R)^\dag$ is an eigenmode with eigenvalue $\xi_n^*$ since
\begin{align}
    \mathcal{E}((\hat{\rho}_n^R)^\dag)=[\mathcal{E}(\hat{\rho}_n^R)]^\dag=\xi_n^*(\hat{\rho}_n^R)^\dag.
\end{align}
Therefore, eigenvalues of a CPTP map are real or appear in complex-conjugate pairs.

\section{Topological equivalence of quantum channels\label{sec_characterization}}

In this section, we introduce the topology of quantum channels (CPTP maps) to investigate topological phases induced by feedback control.

\subsection{Setup\label{sec_top_setup}}
We consider discrete quantum feedback control of a single-particle quantum system on a $d$-dimensional lattice.
Let
\begin{equation}
    \Lambda:=\left\{\vec{R}_{\vec{i}}=\sum_{\lambda=1}^d i_\lambda\vec{a}_\lambda;\ \vec{i}=(i_1,\cdots,i_d),i_\lambda=0,\cdots,L_\lambda-1\right\}
\end{equation}
be a set of lattice sites, where $\vec{a}_\lambda\ (\lambda=1,\cdots,d)$ denote the primitive lattice vectors. Each lattice site $\vec{R}_{\vec{i}}$ is specified by a set of integers $\vec{i}=(i_1,\cdots,i_d)$. The total number of unit cells is given by $N_{\mathrm{cell}}:=L_1L_2\cdots L_d$. 
The Hilbert space of the system is spanned by an orthonormal basis set $\{\ket{\vec{i},a}\}$ ($\vec{i}=(i_1,\cdots,i_d),i_\lambda=0,\cdots,L_\lambda-1$, and $a=1,\cdots,D_{\mathrm{loc}}$), where $\ket{\vec{i},a}$ denotes a quantum state at site $\vec{i}$, with the index $a$ labeling additional degrees of freedom such as spin, orbit, sublattice, etc. The dimension of the Hilbert space in each unit cell is denoted by $D_{\mathrm{loc}}$, and thus the dimension of the entire Hilbert space $\mathcal{H}$ of the system is given by $D=N_{\mathrm{cell}}D_{\mathrm{loc}}$. (Note that we consider a one-body problem here.)
The density matrix of the system can be expanded in terms of the orthonormal basis set as
\begin{equation}
    \hat{\rho}=\sum_{\vec{i},\vec{j}}\sum_{a,b}\rho_{\vec{i},a;\vec{j},b}\ket{\vec{i},a}\bra{\vec{j},b}.
    \label{eq_rho_expansion} 
\end{equation}
We impose the periodic boundary condition (PBC)
\begin{equation}
    \ket{\vec{i}+L_\lambda\vec{e}_\lambda,a}=\ket{\vec{i},a}\ (\lambda=1,\cdots,d),
    \label{eq_PBC}   
\end{equation}
where $\vec{e}_\lambda$ is a unit vector defined by $(\vec{e}_\lambda)_{\lambda'}=\delta_{\lambda,\lambda'}$, unless otherwise specified. 

We perform feedback control of the system with a CPTP map \eqref{eq_CPTP}. We assume that the Kraus operators $\hat{K}_m$ do not change the particle number of the system. The Kraus operators are expanded by using the orthonormal basis as
\begin{equation}
    \hat{K}_m=\sum_{\vec{i},\vec{j}}\sum_{a,b}(\hat{K}_m)_{\vec{i},a;\vec{j},b}\ket{\vec{i},a}\bra{\vec{j},b}.
\end{equation}
Since a CPTP map is a superoperator acting on the operator space, it is convenient to vectorize the density matrix \eqref{eq_rho_expansion}. 
Here we invoke the Choi-Jamio\l kowski isomorphism, also known as the channel-state duality, which is a mathematical correspondence between quantum channels and bipartite states \cite{Jamiolkowski72,Choi75,Jiang13,Wilde_textbook}. In the present case, there is an isomorphic map between an operator and a vector in a doubled Hilbert space (see Appendix \ref{sec_CJ}). 
We vectorize the density matrix \eqref{eq_rho_expansion} as 
\begin{equation}
    \ket{\hat{\rho}}:=\sum_{\vec{i},\vec{j}}\sum_{a,b}\rho_{\vec{i},a;\vec{j},b}\ket{\vec{i},a}\otimes\ket{\vec{j},b}.
    \label{eq_rho_vec}
\end{equation}
Correspondingly, there is an isomorphic mapping from a CPTP map
\begin{equation}
    \mathcal{E}(\hat{\rho})=\sum_m \hat{K}_m\hat{\rho} \hat{K}_m^\dag
\end{equation}
to a non-Hermitian and nonunitary operator
\begin{equation}
    \tilde{\mathcal{E}}=\sum_m\hat{K}_m\otimes \hat{K}_m^*,
    \label{eq_Etilde}
    \end{equation}
which we shall refer to as the matrix representation of $\mathcal{E}$ (see Appendix \ref{sec_CJ}). 
Throughout this paper, we use a symbol with a tilde to indicate an operator that acts on the doubled Hilbert space.

Topological phases of matter are often studied for translationally invariant systems, where the Bloch band theory allows the momentum-space representation of a Hamiltonian \cite{MoessnerMoore_book}. Analogously, here we consider translationally invariant feedback control. We assume that a CPTP map satisfies
    \begin{equation}
    \sum_m(\hat{T}_\lambda \hat{K}_m\hat{T}_\lambda^{-1})\hat{\rho}(\hat{T}_\lambda \hat{K}_m\hat{T}_\lambda^{-1})^\dag=\sum_m\hat{K}_m\hat{\rho} \hat{K}_m^\dag
    \label{eq_trans_sym1}
    \end{equation}
    for arbitrary $\hat{\rho}$, where $\hat{T}_\lambda\ (\lambda=1,\cdots,d)$ is a translation operator:
    \begin{equation}
    \hat{T}_\lambda\ket{\vec{i},a}=\ket{\vec{i}+\vec{e}_\lambda,a}.
    \end{equation}
In the matrix representation $\tilde{\mathcal{E}}$ of the CPTP map, the translational symmetry in Eq.~\eqref{eq_trans_sym1} reads
    \begin{align}
    (\hat{T}_\lambda\otimes \hat{T}_\lambda^*)\tilde{\mathcal{E}}(\hat{T}_\lambda\otimes \hat{T}_\lambda^*)^{-1}=\tilde{\mathcal{E}}.
    \label{eq_trans_sym2}
    \end{align}
    
We use the translational symmetry to obtain the momentum-space representation of $\tilde{\mathcal{E}}$ as (see Appendix \ref{sec_mom} for derivation)
\begin{align}
\tilde{\mathcal{E}}=&\sum_{\vec{k}}\tilde{\bm{c}}_{\vec{k}}^\dag X(\vec{k})\tilde{\bm{c}}_{\vec{k}}\notag\\
=&\sum_{\vec{k}}\sum_{a,b,c,d}\sum_{\vec{\mu},\vec{\nu}}X_{a,c,\vec{\mu};b,d,\vec{\nu}}(\vec{k})\tilde{c}_{\vec{k},a,c,\vec{\mu}}^\dag \tilde{c}_{\vec{k},b,d,\vec{\nu}},
\label{eq_Etilde_k}
\end{align}
where $\tilde{\bm{c}}_{\vec{k}}^\dag=(\tilde{c}_{\vec{k},a,c,\vec{\mu}}^\dag)_{a,c,\vec{\mu}}$ with
\begin{equation}
\tilde{c}_{\vec{k},a,c,\vec{\mu}}^\dag:=\frac{1}{\sqrt{N_{\mathrm{cell}}}}\sum_{\vec{j}}\tilde{c}_{\vec{j},a,c,\vec{\mu}}^\dag e^{i\vec{k}\cdot\vec{R}_{\vec{j}}}
\label{eq_c_k}
\end{equation}
is a row vector of the Fourier transform of an auxiliary creation operator $\tilde{c}_{\vec{i},a,c,\vec{\mu}}^\dag$ defined as
\begin{equation}
\tilde{c}_{\vec{i},a,c,\vec{\mu}}^\dag(\ket{\mathrm{vac}}\otimes\ket{\mathrm{vac}})=\ket{\vec{i},a}\otimes\ket{\vec{i}+\vec{\mu},c},
\label{eq_c_i}
\end{equation}
$\ket{\mathrm{vac}}$ is the vacuum state of the system, and $X(\vec{k})$ is a momentum-dependent $N_{\mathrm{cell}}D_{\mathrm{loc}}^2\times N_{\mathrm{cell}}D_{\mathrm{loc}}^2$ matrix defined by
\begin{align}
X_{a,c,\vec{\mu};b,d,\vec{\nu}}(\vec{k}):=&
\frac{1}{N_{\mathrm{cell}}}\sum_{\vec{j},\vec{j}'}\sum_m(\hat{K}_m)_{\vec{j},a;\vec{j}',b}(\hat{K}_m)^*_{\vec{j}+\vec{\mu},c;\vec{j}'+\vec{\nu},d}\notag\\
&\times e^{-i\vec{k}\cdot(\vec{R}_{\vec{j}}-\vec{R}_{\vec{j}'})}.
\label{eq_Bloch}
\end{align}
Here, we note that the operator $\tilde{c}_{\vec{i},a,c,\vec{\mu}}^\dag$ has two additional indices $c$ and $\vec{\mu}$ compared with the original basis $\ket{\vec{i},a}$ of the Hilbert space since $\tilde{\mathcal{E}}$ acts on the doubled Hilbert space $\mathcal{H}\otimes\mathcal{H}$. Physically, this point corresponds to the fact that the density matrix has bra and ket degrees of freedom. 
The non-Hermitian matrix $X(\vec{k})$ is a counterpart of the Bloch Hamiltonian in the band theory, and therefore we call it the Bloch matrix. 
The mode expansion in Eq.~\eqref{eq_mode_expansion} of the density matrix after feedback control can be expressed in terms of the eigensystem of the Bloch matrix (see Appendix \ref{sec_mode_expansion}).

\subsection{Topology of quantum channels with local Kraus operators\label{sec_loc_Kraus}}
We are now in a position to discuss the topology of feedback control. 
We consider a homotopy between CPTP maps, that is, two CPTP maps $\mathcal{E}_0$ and $\mathcal{E}_1$ are homotopic to each other if and only if there exists a continuous one-parameter family $\Xi_s$ of CPTP maps (called a homotopy \cite{Hatcher_textbook}) that preserves the spectral gap and satisfies $\Xi_0=\mathcal{E}_0$ and $\Xi_1=\mathcal{E}_1$. Intuitively speaking, the existence of a homotopy between two CPTP maps means that they can continuously be deformed into each other.
Here, the spectral gap is either a point gap or a line gap according to the general discussion of topology of non-Hermitian operators \cite{Gong18,Kawabata19}. Specifically, a point gap at $\xi=\xi_{\mathrm{PG}}$ means that the CPTP map under consideration does not have an eigenvalue $\xi_{\mathrm{PG}}$. Similarly, a line gap for a line $l_{\mathrm{LG}}$ means that a CPTP map does not have an eigenvalue on $l_{\mathrm{LG}}$. Here, we mostly focus on the case of a point gap and discuss the case of a line gap in Sec.~\ref{sec_Chern}.

\begin{figure}
    \includegraphics[width=8.5cm]{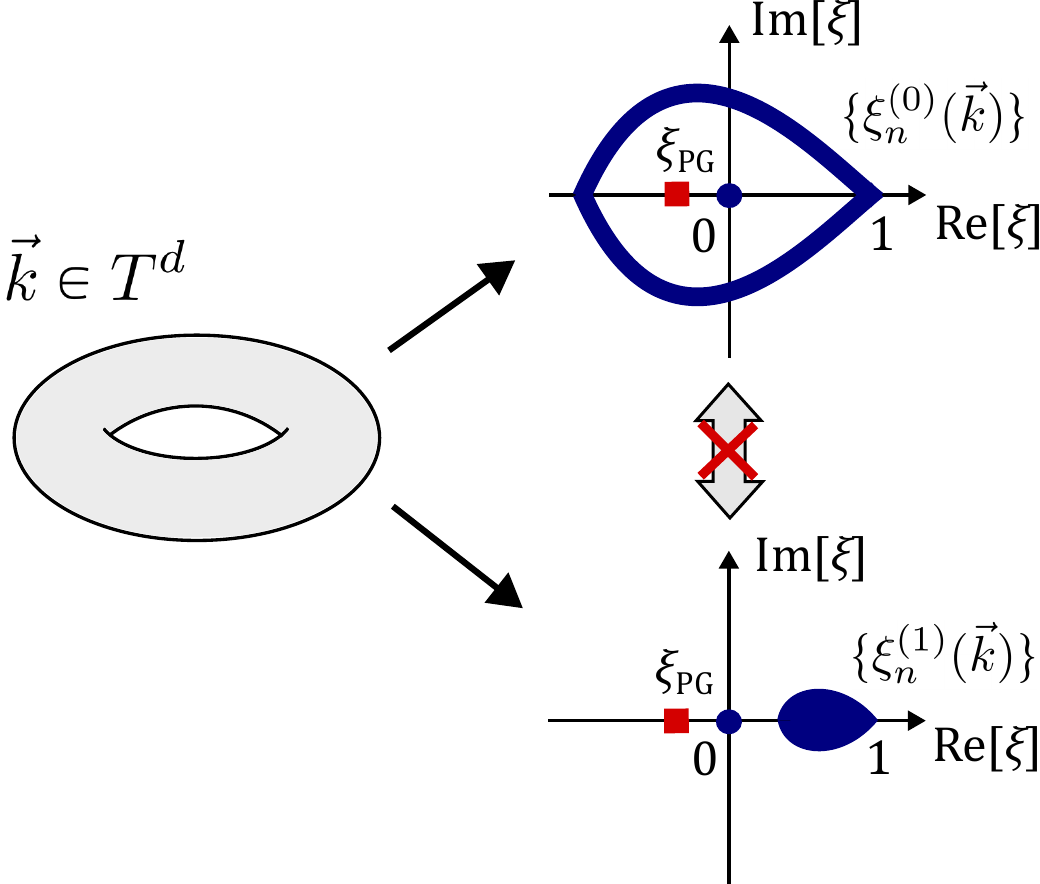}
    \caption{Eigenspectra $\{\xi_n^{(0)}(\vec{k})\}$ and $\{\xi_n^{(1)}(\vec{k})\}$ (shown by blue areas) of two translationally invariant CPTP maps $\mathcal{E}_0$ and $\mathcal{E}_1$ that can be calculated from their Bloch matrices defined on the Brillouin zone (the $d$-dimensional torus $T^d$). If the Bloch matrices have different values of a topological invariant, the two CPTP maps cannot continuously be deformed into each other without closing the spectral gap. In the figure, a point gap at $\xi_{\mathrm{PG}}$ is depicted by a red square.}
    \label{fig_homotopy}
\end{figure}

In the momentum-space representation in Eq.~\eqref{eq_Etilde_k}, the eigenspectrum of a CPTP map can be calculated from the Bloch matrix defined in Eq.~\eqref{eq_Bloch}. 
Thus, a homotopy between translationally invariant CPTP maps is obtained from that between their Bloch matrices. 
The Bloch matrix $X(\vec{k})$ defines a map from the Brillouin zone (i.e., a $d$-dimensional torus) to a space of non-Hermitian matrices. We can exploit this property to introduce a topological invariant for a CPTP map. If two CPTP maps (or their Bloch matrices) have different values of a topological invariant, they are not homotopic and cannot continuously be deformed into each other without closing the spectral gap (see Fig.~\ref{fig_homotopy}). 
Thus, topological invariants can be used to homotopically classify CPTP maps.

\begin{figure}
    \includegraphics[width=8.0cm]{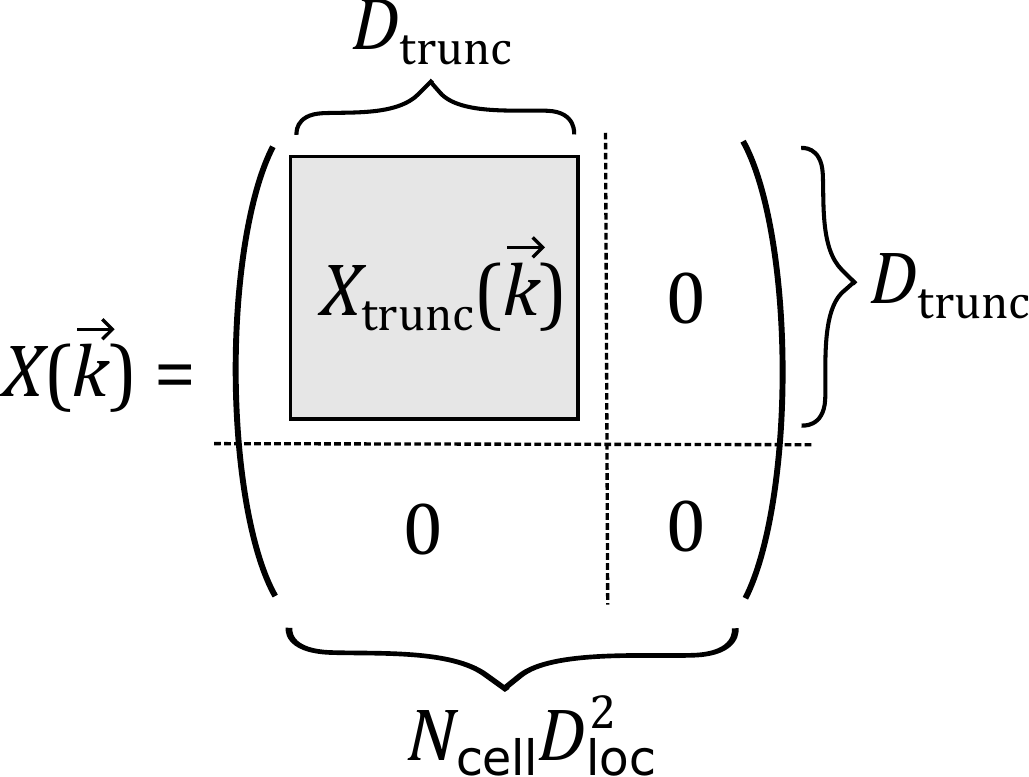}
    \caption{Schematic representation of the Bloch matrix $X(\vec{k})$ which is an $N_{\mathrm{cell}}D_{\mathrm{loc}}^2\times N_{\mathrm{cell}}D_{\mathrm{loc}}^2$ matrix. If the Kraus operators are local and have a finite support, the Bloch matrix can be truncated into a matrix $X_{\mathrm{trunc}}(\vec{k})$ with dimension $D_{\mathrm{trunc}}$ that remains finite in the limit $N_{\mathrm{cell}}\to\infty$.}
    \label{fig_trunc}
\end{figure}

Here, we note that the Bloch matrix is an $N_{\mathrm{cell}}D_{\mathrm{loc}}^2\times N_{\mathrm{cell}}D_{\mathrm{loc}}^2$ matrix. Since the size of the matrix diverges in the limit $N_{\mathrm{cell}}\to\infty$, the definition of a topological invariant of a Bloch matrix is nontrivial. We show that this problem can be circumvented by imposing locality on the Kraus operators. We assume that the Kraus operators are local in space and have a finite support. Physically, this assumption implies that the interaction between the system and the measurement apparatus (the feedback controller) is local. Let $\ell$ be the diameter of the support of the Kraus operators, and let $R$ be the range of locality of the Kraus operators. Specifically, we assume
    \begin{equation}
        (\hat{K}_m)_{\vec{i},a;\vec{j},b}(\hat{K}_m)^*_{\vec{i}',c;\vec{j}',d}=0\ \mathrm{for}\ |\vec{j}-\vec{j}'|>\ell
        \label{eq_Kraus_support}
    \end{equation}
    and
    \begin{equation}
        (\hat{K}_m)_{\vec{i},a;\vec{j},b}=0\ \mathrm{for}\ |\vec{i}-\vec{j}|>R.
        \label{eq_Kraus_range}
    \end{equation}
We also assume that $\ell$ and $R$ are independent of the system size. Then, we have 
    \begin{equation}
        (\hat{K}_m)_{\vec{i},a;\vec{j},b}(\hat{K}_m)^*_{\vec{i}',c;\vec{j}',d}=0\ \mathrm{for}\ |\vec{i}-\vec{i}'|>2R+\ell,
        \label{eq_Bloch_range1}
    \end{equation}
    since
    \begin{align}
        |\vec{i}-\vec{i}'|=&|\vec{i}-\vec{j}+\vec{j}-\vec{j}'+\vec{j}'-\vec{i}'|\notag\\
        \leq& |\vec{i}-\vec{j}|+|\vec{j}-\vec{j}'|+|\vec{i}'-\vec{j}'|.
    \end{align}
    The condition \eqref{eq_Kraus_support} leads to
    \begin{equation}
        X_{a,c,\vec{\mu};b,d,\vec{\nu}}(\vec{k})=0\ \mathrm{for}\ |\vec{\nu}|>\ell.
    \end{equation}
    Furthermore, from the relation \eqref{eq_Bloch_range1}, we obtain
    \begin{equation}
        X_{a,c,\vec{\mu};b,d,\vec{\nu}}(\vec{k})=0\ \mathrm{for}\ |\vec{\mu}|>2R+\ell,
        \label{eq_Bloch_range2}
    \end{equation}
which indicates that the Bloch matrix $X(\vec{k})$ can be truncated without changing its nonzero eigenvalues and can be replaced by a matrix $X_{\mathrm{trunc}}(\vec{k})$ with a finite dimension that does not increase when increasing the system size (see Fig.~\ref{fig_trunc}). 
Hence, the topology of a CPTP map can be diagnosed by that of its truncated Bloch matrix $X_{\mathrm{trunc}}(\vec{k})$, which remains a finite-size matrix in the $N_{\mathrm{cell}}\to\infty$ limit. Topological feedback control is thus characterized by a topological invariant for the truncated Bloch matrix.

An important consequence of the locality of Kraus operators is that the CPTP map has highly degenerate zero eigenvalues. This result can be seen by noting that
    \begin{equation}
        \hat{K}_m\ket{\vec{i},a}\bra{\vec{j},b}\hat{K}_m^\dag=0\ \mathrm{for}\ |\vec{i}-\vec{j}|>\ell,
    \end{equation}
which shows that $\ket{\vec{i},a}\bra{\vec{j},b}$ with $|\vec{i}-\vec{j}|>\ell$ is an eigenoperator of the CPTP map with eigenvalue zero. 
Thus, the truncation of a Bloch matrix and the highly degenerate zero eigenvalues can be understood as a consequence of local measurement, which projects out quantum coherence between distant sites.

The existence of zero eigenvalues of quantum channels implies a crucial difference between the dynamics described by those quantum channels and continuous time evolution described by the Lindblad quantum master equation.
In fact, a quantum channel $\mathcal{E}$ that has a zero eigenvalue cannot be expressed in terms of any generator $\mathcal{L}$, i.e.,
\begin{equation}
\mathcal{E}\neq e^{\mathcal{L}},
\label{eq_generator}
\end{equation}
since the right-hand side cannot have a zero eigenvalue. 
Thus, the topological classification of quantum channels cannot be reduced to that of the Lindbladians discussed in the literature \cite{Bardyn13,Lieu20,Kawabata22,Sa22}.

If Kraus operators are not strictly local but have an exponentially decaying tail, the topology of a quantum channel may be discussed by approximating them with local ones because topology is insensitive to small deformations. Physically, this case should hold true unless rare measurement outcomes with extremely small probabilities due to an exponentially decaying tail of wavefunctions significantly alter the physics. While it is expected to be true for most cases, it would be interesting to find the cases where rare events due to the exponentially small tails of Kraus operators play a significant role in feedback control. We do not discuss those situations further in this paper.

\subsection{Stable equivalence of quantum channels\label{sec_stable}}
\begin{figure}
    \includegraphics[width=8.5cm]{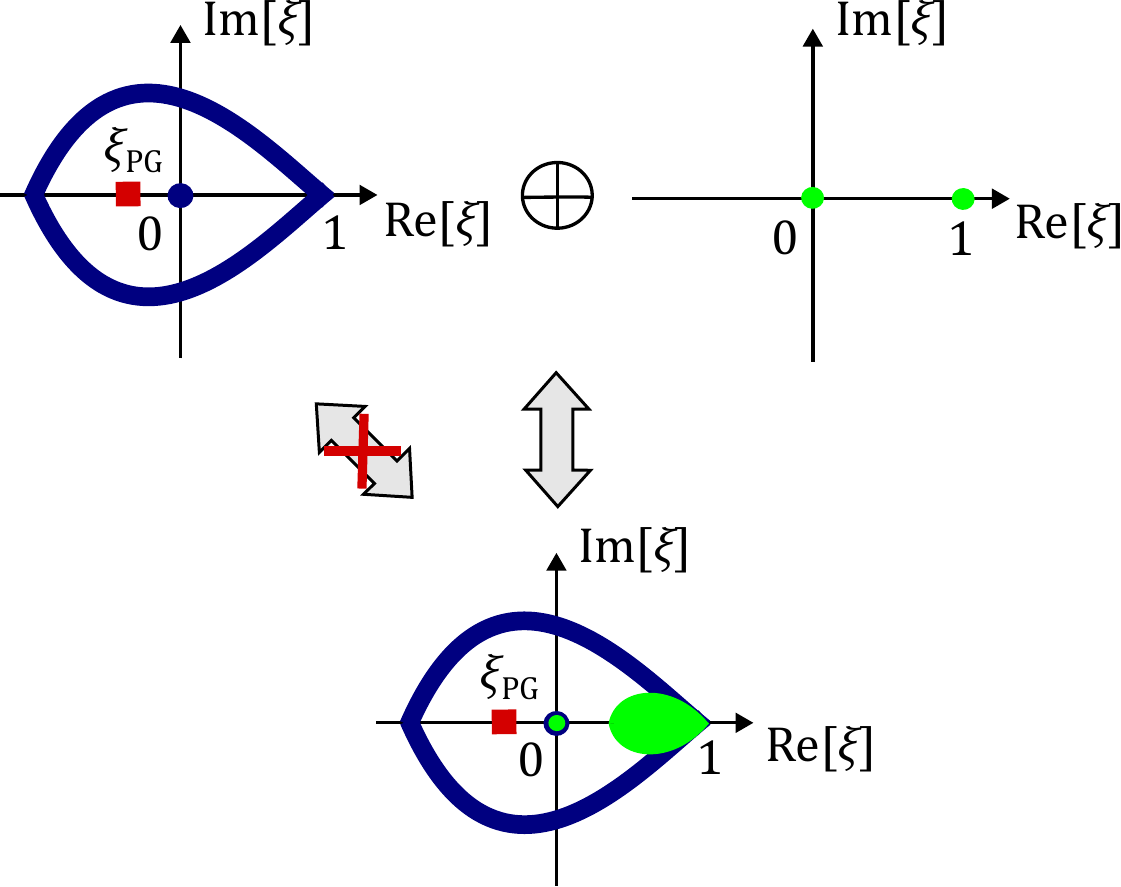}
    \caption{Stable equivalence of CPTP maps. A CPTP map with the eigenspectrum shown by blue areas in the top-left figure is not homotopic to the other CPTP map with the eigenspectrum shown by blue and green areas in the bottom figure since the numbers of bands (eigenvalues) are different. However, they can be made homotopic if we add a trivial CPTP map with the eigenspectrum shown in the top-right figure so that the numbers of eigenvalues become equal to each other.}
    \label{fig_stable_equiv}
\end{figure}

In the definition of the equivalence relation in terms of a homotopy presented in the previous subsection, two CPTP maps acting on systems with different numbers of internal degrees of freedom are not homotopic to each other. 
A more convenient classification of topological phases is obtained by introducing the notion of stable equivalence \cite{Kitaev09,RoyHarper17}. In the topological band theory, two band insulators are stably equivalent if they are continuously connected to each other after the addition of some trivial bands \cite{Kitaev09}. Analogously, we define the stable equivalence of CPTP maps as follows. First, we generalize the quantum states at each unit cell as $\ket{\vec{j},a}$ ($a=1,\cdots,D_{\mathrm{loc}},D_{\mathrm{loc}}+1,\cdots,D_{\mathrm{loc}}+n_1$) by adding $n_1$ extra local degrees of freedom, which can physically be interpreted as ancillary sublattice sites. We assume that the Kraus operators $\hat{K}_m$ act trivially on the extra degrees of freedom, i.e., $\hat{K}_m\ket{\vec{j},a}=0$ for $a=D_{\mathrm{loc}}+1,\cdots,D_{\mathrm{loc}}+n_1$. Next, we introduce a projective measurement of the ancillary degrees of freedom, which is described by
    \begin{equation}
        \mathcal{P}_{n_1}(\hat{\rho}):=\sum_{\vec{j}}\sum_{a=D_{\mathrm{loc}}+1}^{D_{\mathrm{loc}}+n_1} \hat{P}_{\vec{j},a}\hat{\rho} \hat{P}_{\vec{j},a},
        \label{eq_proj_ex}
    \end{equation}
where $\hat{P}_{\vec{j},a}:=\ket{\vec{j},a}\bra{\vec{j},a}$ is the projection operator onto the $a$th internal degree of freedom at site $\vec{j}$. The projective measurement channel \eqref{eq_proj_ex} plays the role of ``trivial bands'' in the sense that all of its nonzero eigenvalues are unity (see Appendix \ref{sec_proj_channel}). We note that the projection operators $\hat{P}_{\vec{j},a}$ are local in space in accordance with the discussion in Sec.~\ref{sec_loc_Kraus}. We define the direct sum of a quantum channel $\mathcal{E}(\hat{\rho})=\sum_m\hat{K}_m\hat{\rho} \hat{K}_m^\dag$ for the system and the projective measurement channel \eqref{eq_proj_ex} for the extra bands by
    \begin{equation}
        (\mathcal{E}\oplus\mathcal{P}_{n_1})(\hat{\rho}):=\sum_m\hat{K}_m\hat{\rho} \hat{K}_m^\dag+\sum_{\vec{j}}\sum_{a=D_{\mathrm{loc}}+1}^{D_{\mathrm{loc}}+n_1} \hat{P}_{\vec{j},a}\hat{\rho} \hat{P}_{\vec{j},a}.
    \end{equation}
Let $\mathcal{E}_1$ and $\mathcal{E}_2$ be CPTP maps. Then, the CPTP maps $\mathcal{E}_1$ and $\mathcal{E}_2$ are said to be stably equivalent if and only if $\mathcal{E}_1\oplus\mathcal{P}_{n_1}$ are homotopic to $\mathcal{E}_2\oplus\mathcal{P}_{n_2}$ for two non-negative integers $n_1$ and $n_2$ (see Fig.~\ref{fig_stable_equiv}).

\section{Topological feedback control\label{sec_model}}

\subsection{Chiral Maxwell demon\label{sec_chiral_demon}}
\begin{figure}
    \includegraphics[width=8.5cm]{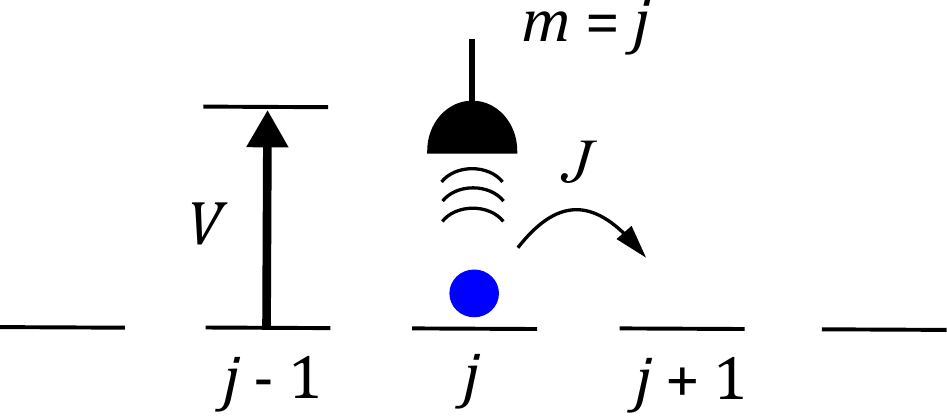}
    \caption{Schematic illustration of the chiral Maxwell demon. A quantum-mechanical particle hops on a one-dimensional lattice with hopping amplitude $J$. We perform a projective measurement of the position of the particle. If we find the particle at site $j$, we raise the potential at site $j-1$ to suppress the probability of the particle hopping to the left.}
    \label{fig_model}
\end{figure}

Having established the general framework of topology of quantum channels, here we present a prototypical example of topological quantum feedback control characterized by a topological invariant of the Bloch matrix of a quantum channel. Our model is a quantum counterpart of Maxwell's demon in Ref.~\cite{Toyabe10} (see also Ref.~\cite{LiuNakagawaUeda23}). We consider a particle hopping on a one-dimensional (1D) lattice with $L$ sites, where a quantum state at site $i$ is denoted by $\ket{i}$. We first perform a projective measurement of the position of this particle with measurement operators $\hat{M}_m=\ket{m}\bra{m}$ $(m=1,\cdots,L)$. If the measurement outcome is $j$, we raise the potential at site $j-1$ and perform feedback control with the Hamiltonian
    \begin{equation}
    \hat{H}_j=-J\sum_{i}(\ket{i}\bra{i+1}+\ket{i+1}\bra{i})+V\ket{j-1}\bra{j-1},
    \end{equation}
where $J\in\mathbb{R}$ is the hopping amplitude and $V>0$ is the height of the potential. The unitary operator for feedback control is given by $\hat{U}_j=\exp[-i\hat{H}_j\tau]$, where $\tau$ is the duration of feedback control. 
The potential barrier placed at site $j-1$ prevents the particle from going to the left, thereby achieving unidirectional (chiral) transport of the particle to the right (see Fig.~\ref{fig_model}). 
Here, we do not need any external field to cause the transport in the system; instead, the information gained from the measurement is used to drive the particle unidirectionally. 
Feedback control thus rectifies quantum fluctuations in measurement outcomes and converts them into a unidirectional transport in a manner similar to an information ratchet \cite{Sagawa10}. 
We therefore refer to this model as the chiral Maxwell demon.

The eigenvalues $\xi_n(k)\ (n=0,\cdots,L-1)$ of a quantum channel of this feedback control can be calculated by using the method described in Appendix \ref{sec_proj}. They are given by
\begin{equation}
    \xi_0(k)=\sum_{r}p(r)e^{-ikr}
    \label{eq_chiral_xi0}
\end{equation}
and
\begin{equation}
    \xi_1(k)=\xi_2(k)=\cdots=\xi_{L-1}(k)=0,
\end{equation}
where
    \begin{equation}
    p(r)=p(j+r|j):=|(\hat{U}_{j})_{j+r,j}|^2
    \end{equation}
is the probability of finding a particle at site $j+r$ after feedback control with measurement outcome $j$. This probability does not depend on $j$ because of the translational invariance. The zero eigenvalues $\xi_{n\neq 0}(k)$ are due to the projective measurement for which quantum coherence between different sites vanishes: $\hat{M}_m\ket{i}\bra{j}\hat{M}_m^\dag=0\ (i\neq j)$.

By using the eigenvalues, the topological invariant of a quantum channel is given by the winding number
    \begin{align}
        w(\xi_{\mathrm{PG}})=\int_{-\pi}^\pi\frac{dk}{2\pi i}\partial_k\ln[\xi_0(k)-\xi_{\mathrm{PG}}]\in\mathbb{Z},
    \label{eq_winding_proj}
    \end{align}
where $\xi_{\mathrm{PG}}$ denotes the location of a point gap \cite{Gong18}. Here, the eigenvalues $\xi_n(k)=0\ (n\neq 0)$ do not contribute to the winding number [since $\partial_k\xi_n(k)=0$] and thus are excluded from Eq.~\eqref{eq_winding_proj}. The winding number takes an integer value and it is invariant under continuous deformation of a feedback control protocol as long as the point gap at $\xi_{\mathrm{PG}}$ is not closed. This case is consistent with the $\mathbb{Z}$ classification of non-Hermitian matrices in 1D systems according to stable equivalence \cite{Gong18}.

\subsection{Eigenspectrum of a quantum channel under the periodic and open boundary conditions}

\begin{figure}
    \includegraphics[width=8.0cm]{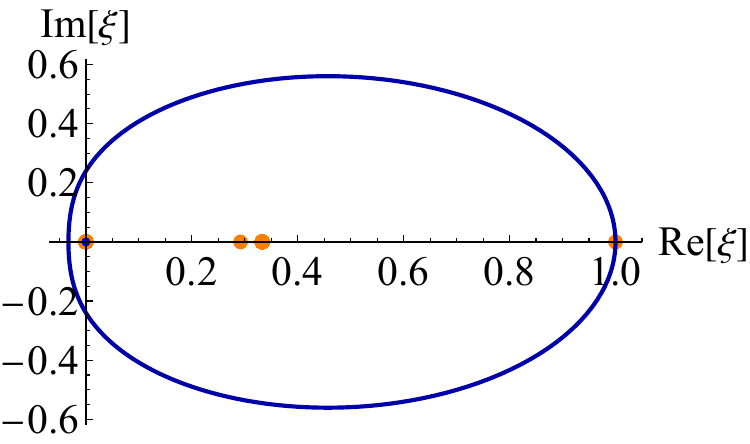}
    \caption{Eigenspectrum of a quantum channel for the chiral Maxwell demon. A blue curve and a blue dot constitute the eigenspectrum under the periodic boundary condition for infinite system size, and orange dots show the eigenspectrum under the open boundary condition for system size $L=50$. Energy is measured in units of $J=1$, and we take $V=+\infty$ for simplicity. The duration of a feedback operation is $\tau=1$.}
    \label{fig_chiral_spec}
\end{figure}

In Fig.~\ref{fig_chiral_spec}, we show the eigenspectrum of a quantum channel for the chiral Maxwell demon under the PBC. Besides the highly degenerate zero eigenvalues $\xi_{n\neq 0}(k)$, the loop structure formed by the eigenvalue $\xi_0(k)$ is evident. 
The winding number is $w(\xi_{\mathrm{PG}})=-1$ for a point gap $\xi_{\mathrm{PG}}$ inside the loop. 
The eigenvalue $\xi_0(k)$ is determined from the probability distribution of the position after feedback control as shown in Eq.~\eqref{eq_chiral_xi0}. In fact, the eigenvalue $\xi_0(k)$ is nothing but the characteristic function of the probability distribution $\{ p(r)\}$. The chiral transport caused by feedback control increases the probability of a particle being found at site $r>0$ and makes the probability distribution $\{ p(r)\}$ asymmetric with respect to $r>0$ and $r<0$. Thus, the winding number of the quantum channel is closely related to the chiral transport achieved by the feedback in analogy to the chiral edge state in the quantum Hall effect and the Hatano-Nelson model for non-Hermitian topological phases \cite{Gong18,Hatano96,Hatano97,Hatano98,Lee19}.

\begin{figure}
    \includegraphics[width=8.0cm]{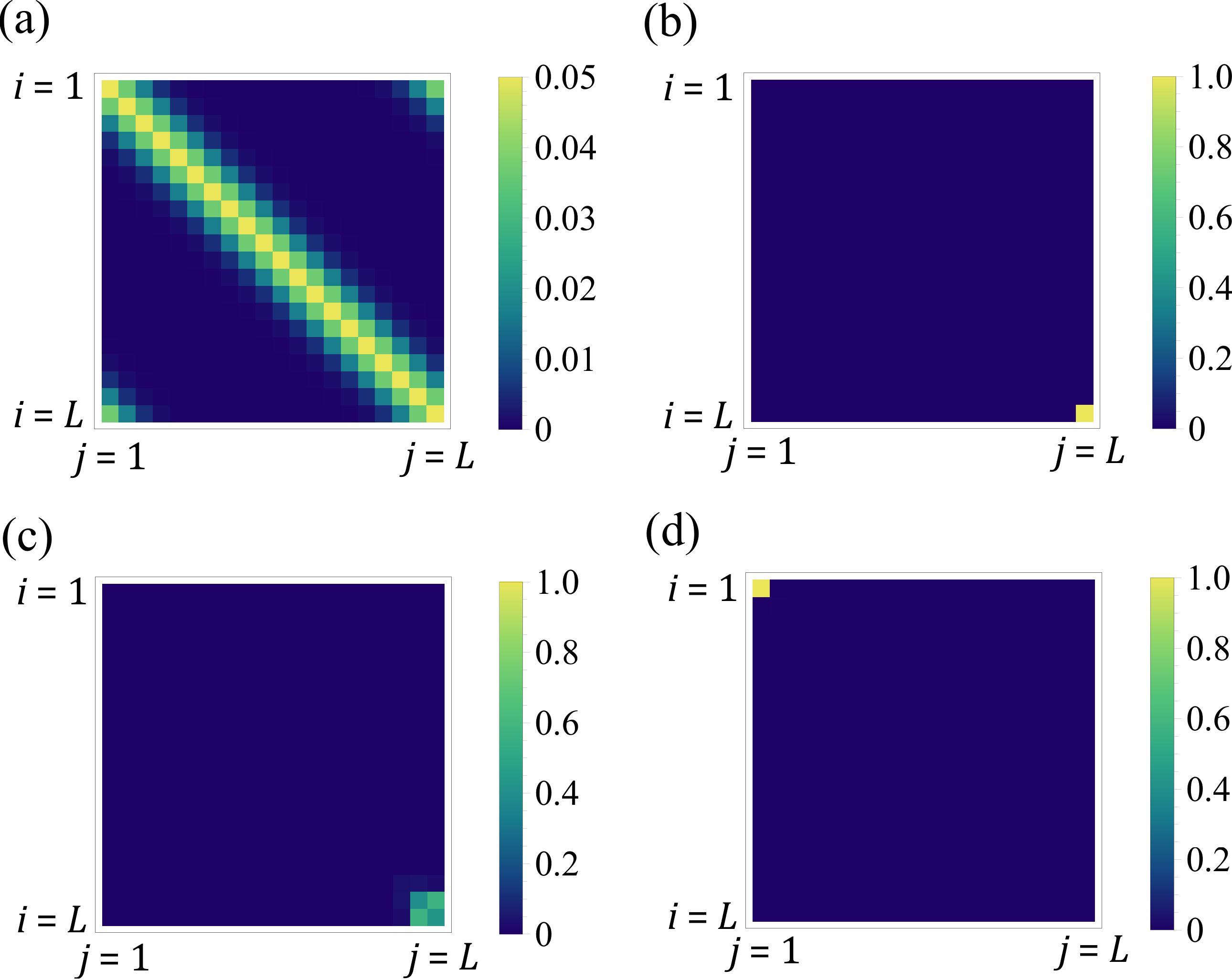}
    \caption{Magnitude $|\rho_{i,j}|$ of matrix elements of an eigenmode $\hat{\rho}_n=\sum_{i,j}\rho_{i,j}\ket{i}\bra{j}$ plotted in the position basis for a system size of $L=20$ with the hopping amplitude $J=1$, the height of the potential $V=+\infty$, and the duration $\tau=1$ of feedback control. Steady states of the chiral Maxwell demon under the PBC and the OBC are shown in panels (a) and (b), respectively. Panels (c) and (d) show typical right and left eigenmodes under the OBC with eigenvalue 0.334 (see Fig.~\ref{fig_chiral_spec}).}
    \label{fig_chiral_eigenmode}
\end{figure}

Nontrivial topology around a point gap under the PBC implies the emergence of a non-Hermitian skin effect under the open boundary condition (OBC) \cite{Gong18,Okuma20,Zhang20,Borgnia20,Okuma23,Lin23}. Orange dots in Fig.~\ref{fig_chiral_spec} show the eigenspectrum of the quantum channel under the OBC, which is drastically different from that under the PBC. Such an extraordinary sensitivity of the eigenspectrum to the boundary condition is a salient feature of the non-Hermitian skin effect. 
The steady state of the quantum channel under the PBC is uniform in space, while it is localized near the right edge of the system under the OBC [see Figs.~\ref{fig_chiral_eigenmode}(a) and \ref{fig_chiral_eigenmode}(b)].
In Figs.~\ref{fig_chiral_eigenmode}(c) and \ref{fig_chiral_eigenmode}(d), we show the typical behavior of right and left eigenmodes under the OBC. The localization of the right and left eigenmodes at the opposite boundaries of the system indeed confirms the non-Hermitian skin effect, which is a consequence of the chiral transport induced by the chiral Maxwell demon.

As seen from Fig.~\ref{fig_chiral_eigenmode}(c) and \ref{fig_chiral_eigenmode}(d), the localization length of a right eigenmode is, in general, different from that of the corresponding left eigenmode due to the difference between the dynamics near the right boundary and those near the left boundary. Specifically, the right eigenmode in Fig.~\ref{fig_chiral_eigenmode}(c) is less localized than the left eigenmode in Fig.~\ref{fig_chiral_eigenmode}(d) because of the reflection at the right boundary. This phenomenon arises from the fact that the quantum channel under the OBC is defined by the time evolution under the same boundary condition so that the complete positivity and trace preservation are ensured (see also Sec.~\ref{sec_BBC}).

We note that the localization of left and right eigenmodes $\hat{\rho}_n^L,\hat{\rho}_n^R$ at opposite edges implies that the overlap $\langle \hat{\rho}_n^L,\hat{\rho}_n^R\rangle$ decreases exponentially as a function of the system size. From the eigenmode expansion \eqref{eq_mode_expansion}, this indicates that the expansion coefficient $c_n$ becomes exponentially large, which significantly affects the relaxation dynamics of the system towards the steady state under repeated applications of a quantum channel \cite{Mori20,Haga21}. Specifically, if the non-Hermitian skin effect takes place, the relaxation time under the OBC generically diverges in the infinite-size limit \cite{Haga21}.

\subsection{Topological transition} 
\begin{figure}
    \includegraphics[width=8.5cm]{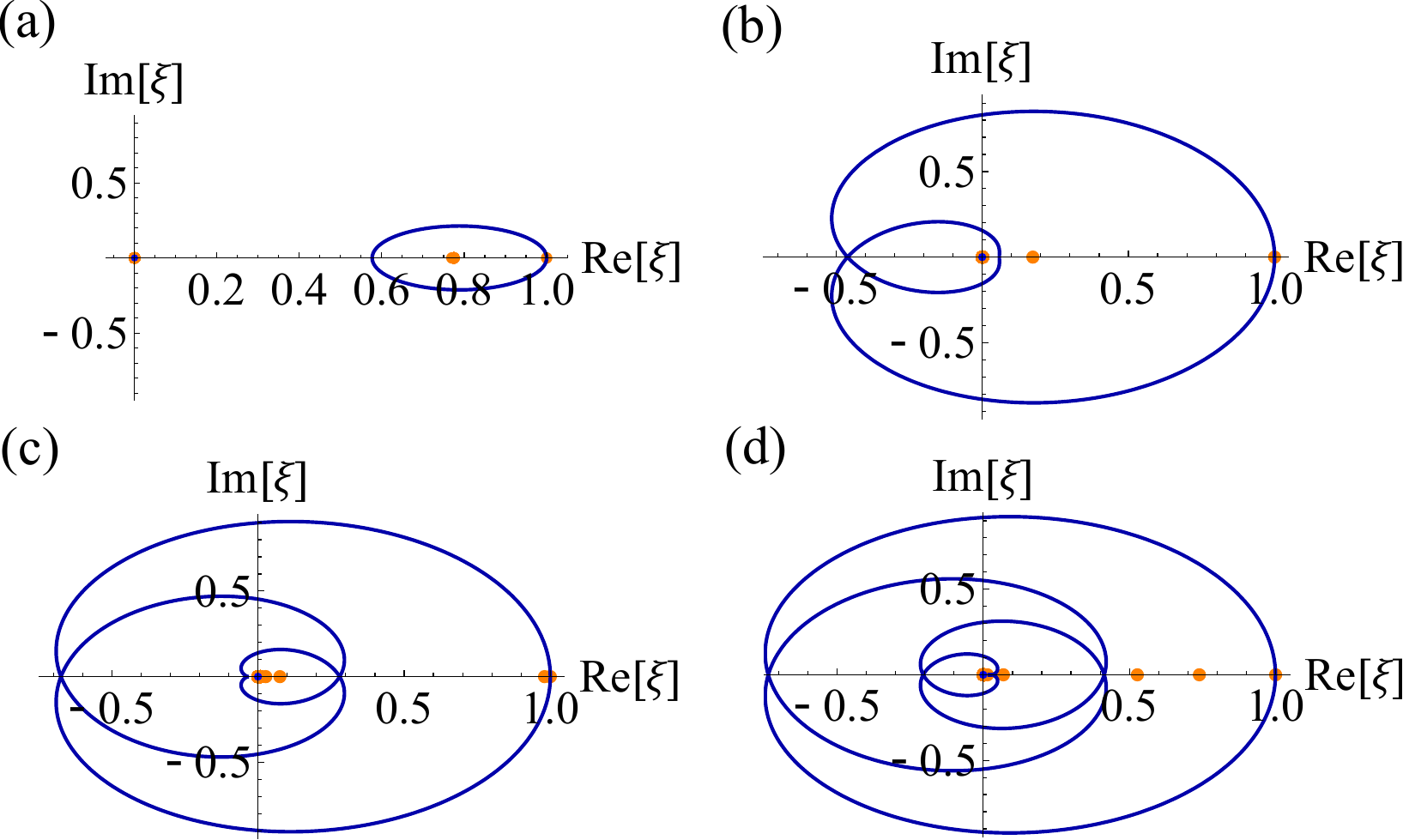}
    \caption{Eigenspectrum of the quantum channel for the chiral Maxwell demon for four different values of the duration of the feedback. A blue curve and a blue dot constitute the eigenspectrum under the PBC for infinite system size, and orange dots show the eigenspectrum under the OBC for a system size of $L=50$. Energy is measured in units of $J=1$, and the height of the potential is set to $V=+\infty$. The duration of feedback control is (a) $\tau=0.5$, (b) $\tau=2$, (c) $\tau=3$, and (d) $\tau=3.9$. The winding number $w$ around the origin is (a) $w=0$, (b) $w=-2$, (c) $w=-3$, and (d) $w=-4$.}
    \label{fig_PBCspec_duration}
\end{figure}

\begin{figure}
    \includegraphics[width=8.5cm]{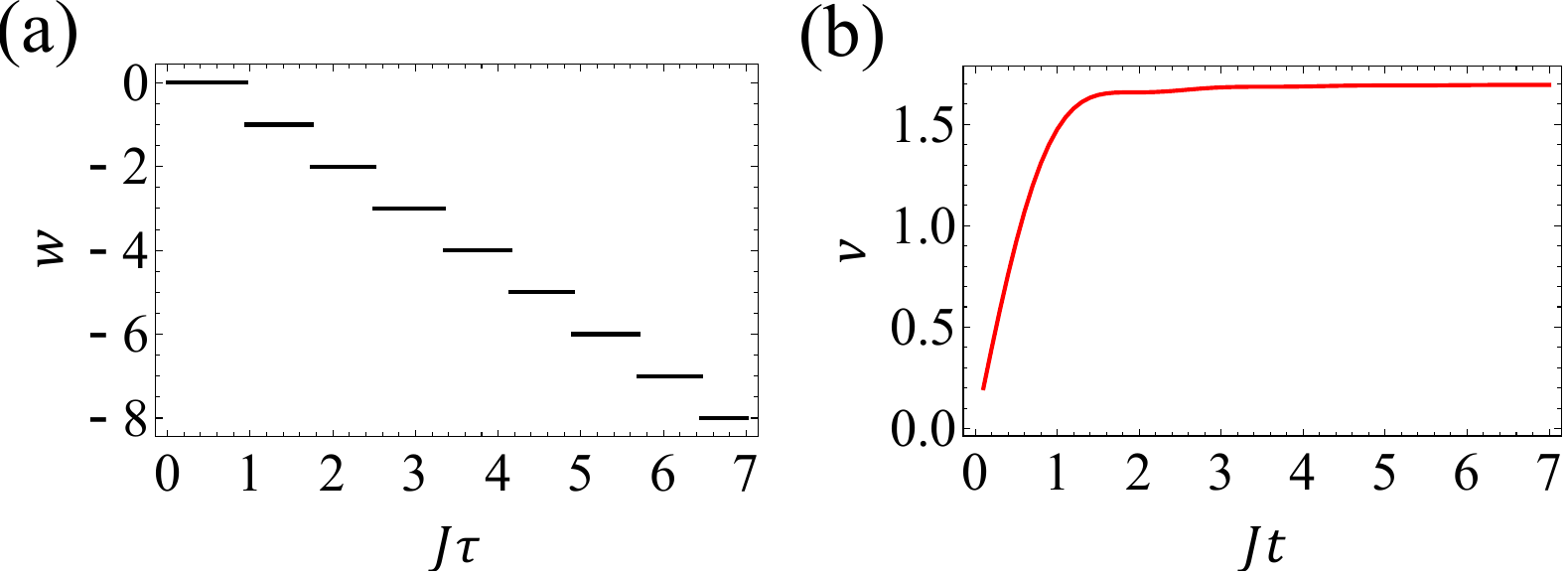}
    \caption{(a) Dependence of the winding number $w(\xi_{\mathrm{PG}}=0)$ on the duration $\tau$ of feedback control. (b) Velocity of a particle during feedback control [see Eq.~\eqref{eq_velocity}]. The height of the potential is set to $V=+\infty$.}
    \label{fig_winding_velocity}
\end{figure}

The chiral Maxwell demon described in Sec.~\ref{sec_chiral_demon} exhibits topological transitions as the duration $\tau$ of feedback control is changed. In Fig.~\ref{fig_PBCspec_duration}, we show the eigenspectrum under the PBC for four different values of the duration of feedback control. As the duration of feedback control is increased, the absolute value of the winding number also increases. In Fig.~\ref{fig_winding_velocity}(a), we show the dependence of the winding number on the duration of feedback control. The topological transitions occur at an almost constant rate as a function of the duration of feedback control. 

The topological transitions can physically be understood from the ballistic transport of a particle during feedback control. In Fig.~\ref{fig_winding_velocity}(b), we show the mean velocity
    \begin{equation}
        v(t):=\frac{d}{dt}\sum_r r|\bra{j+r}e^{-i\hat{H}_jt}\ket{j}|^2
        \label{eq_velocity}
    \end{equation}
of a particle during feedback control. As seen from the figure, the particle shows ballistic transport at a constant velocity after an initial transient dynamics. 
To understand the relation between the ballistic transport and the topological transitions, we rewrite the winding number \eqref{eq_winding_proj} by using $z=e^{-ik}$ as
\begin{align}
w(\xi_{\mathrm{PG}}=0)=&-\oint_{|z|=1}\frac{dz}{2\pi i}\frac{\xi_0'(z)}{\xi_0(z)},
\label{eq_wind_z}
\end{align}
where
\begin{equation}
\xi_0(z):=\sum_rp(r)z^r
\end{equation}
is the probability-generating function of $\{p(r)\}$, and the integration contour in Eq.~\eqref{eq_wind_z} is taken in the counterclockwise direction. From Eq.~\eqref{eq_wind_z} and by using the argument principle, we have
\begin{equation}
w(\xi_{\mathrm{PG}}=0)=N_{\mathrm{p}}-N_0,
\label{eq_wind_arg}
\end{equation}
where $N_{\mathrm{p}}$ ($N_0$) is the number of poles (zeros) of the function $\xi_0(z)$ in $|z|<1$. The expression \eqref{eq_wind_arg} gives a physical insight into topological feedback control. Let $\tilde{p}(r):=p(r-1)$ be a shifted probability distribution. Then, its probability-generating function is given by $z\xi_0(z)$. It follows from Eq.~\eqref{eq_wind_arg} that the winding number is changed by $-1$ because the number of zeros is increased by $1$. 
The constant increase of the winding number can be understood from the ballistic transport of a particle since the winding number is changed by $-1$ if the probability distribution of the displacement after feedback control is shifted by one site.

We note that the mean displacement
\begin{align}
\bar r:=&\sum_j\sum_r rp(j+r|j)p_j\notag\\
=&\sum_r rp(r)\notag\\
=&\int_0^\tau dt v(t)
\end{align}   
during the topological feedback control is not quantized in general, as inferred from Fig.~\ref{fig_winding_velocity}(b). Thus, a quantized topological invariant of a quantum channel does not necessarily guarantee the quantization of a physical observable. This situation is analogous to that of other topological nonequilibrium dynamics \cite{Oka09,Rudner20} such as nonadiabatic topological pumping \cite{Kitagawa10}.

The topological transitions are accompanied by exceptional points, at which the CPTP map cannot be diagonalized \cite{Ashida20}. In fact, when the value of the winding number around $\xi_{\mathrm{PG}}=0$ changes, $\xi_0(k)$ should vanish for some momentum $k$, and the Bloch matrix $X(k)$ cannot be diagonalized at that point (see Appendix~\ref{sec_proj}). Thus, topology causes the emergence of exceptional points of a CPTP map.

\subsection{Robustness of chiral transport against disorder}
The unidirectional transport induced by the chiral Maxwell demon is of a topological origin and therefore robust against external perturbations. Here, we demonstrate that the topological feedback control enhances the stability of particle transport. By way of illustration, we consider a task of transporting a particle from the leftmost site to the rightmost site in an $L$-site chain by using the chiral Maxwell demon. Suppose that the feedback Hamiltonian $\hat{H}_j$ is perturbed by on-site disorder potentials as
\begin{align}
\hat{H}_j =& -J\sum_{i=1}^{L-1}(\ket{i}\bra{i+1}+\ket{i+1}\bra{i})+V\ket{j-1}\bra{j-1}\notag\\
 &+\sum_{i=1}^L W_i\ket{i}\bra{i},
\end{align}
where $W_i\in\mathbb{R}$ denotes a random potential. For comparison, we consider a nontopological feedback control with a quantum channel
\begin{align}
\mathcal{E}^\prime(\hat{\rho})=\sum_{m=0,1}e^{-i\hat{H}_m^\prime\tau}\hat{M}_m^{\prime}\hat{\rho} \hat{M}_m^{\prime\dag}e^{i\hat{H}_m^\prime\tau},
\end{align}
where measurement operators are given by $\hat{M}_0^\prime=\hat{I}-\ket{L}\bra{L}$ and $\hat{M}_1^{\prime}=\ket{L}\bra{L}$ and the feedback Hamiltonians are given by
\begin{subequations}
    \begin{align}
\hat{H}_0^\prime=&-J\sum_{i=1}^{L-1}(\ket{i}\bra{i+1}+\mathrm{H.c.})+\sum_{i=1}^L W_i\ket{i}\bra{i}
    \end{align}
    and
    \begin{align}
\hat{H}_1^\prime=&0;
\end{align}
\end{subequations}
i.e., only the rightmost site is measured projectively, and the dynamics stops when a particle is found at that site. 

\begin{figure}
    \includegraphics[width=8.5cm]{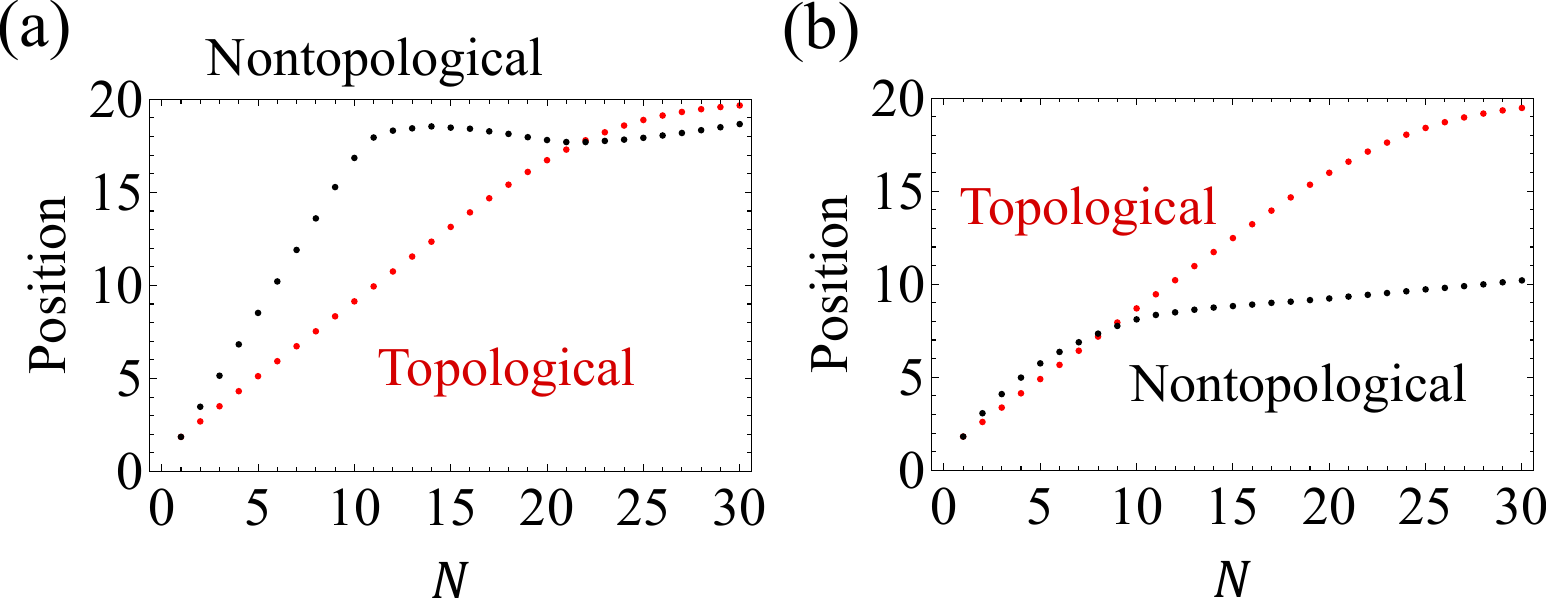}
    \caption{Expectation value of the position as a function of the number of feedback cycles. Red (black) dots show the results for topological (nontopological) feedback control. The parameters are set as $L=20$, $J=1$, $\tau=1$, and $V=10$. The disorder potential $W_i$ at each site is randomly sampled from a uniform distribution in $[-W,W]$. We take $W=0$ in panel (a) and $W=1$ in panel (b). The results are averaged over 100 samples of disorder realization.}
    \label{fig_chiral_disorder}
\end{figure}

In Fig.~\ref{fig_chiral_disorder}, we show the dynamics of the expected value of the position
\begin{align}
\langle\hat{x}\rangle_N:=\mathrm{Tr}[\hat{x}\mathcal{E}^N(\hat{\rho_{\mathrm{i}}})]
\end{align}
with $\hat{x}=\sum_j j\ket{j}\bra{j}$ and an initial state $\hat{\rho}_{\mathrm{i}}=\ket{1}\bra{1}$ as a function of the number $N$ of feedback cycles. As shown in the figure, chiral transport due to topological feedback control is robust against disorder as long as $|W_i|\ll V$, whereas particle transport in the nontopological case is quickly suppressed due to Anderson localization \cite{Anderson58} in the presence of disorder. Thus, topological feedback control can enhance the stability of particle transport against disorder.

\section{Symmetry classification of feedback control\label{sec_tenfold}}

Symmetry enriches topological phases since two topological phases may be distinguished in the presence of symmetry \cite{Kitaev09,Schnyder08,Chiu16,RoyHarper17,Higashikawa19,Gong18,Kawabata19,Zhou19,Lieu20,Liu22,Kawabata22,Sa22}.
Here, we discuss the symmetry of quantum channels for discrete quantum feedback control.
In Sec.~\ref{sec_unitary_sym}, we discuss unitary symmetry. In Sec.~\ref{sec_mod_conj_sym}, we consider antiunitary symmetry associated with every quantum channel. While the content in these two subsections can be found in the literature \cite{Buca12,Albert14,Gong18,deGroot22,Kawabata22,Sa22}, we reproduce them to make the present paper self-contained. 
In Sec.~\ref{sec_BL_sym}, we discuss the Bernard-LeClair (BL) symmetry classes of quantum channels for discrete quantum feedback control. In particular, we find that only a subset of the BL symmetry classes are compatible with projective measurements. Consequently, we obtain the tenfold symmetry classification of discrete quantum feedback control, as listed in Table \ref{table_AZdag}.

\subsection{Unitary symmetry\label{sec_unitary_sym}}

We first consider unitary symmetry of quantum channels. Similarly to the case of open quantum systems described by the Lindblad equation \cite{Buca12,Albert14}, unitary symmetry of a quantum channel can be either weak symmetry or strong symmetry \cite{deGroot22}. A CPTP map $\mathcal{E}$ has unitary symmetry if it commutes with a unitary superoperator $\mathcal{V}$, that is,
\begin{equation}
    \mathcal{V}\mathcal{E}\mathcal{V}^{-1}=\mathcal{E}.
    \label{eq_unitary_sym}
\end{equation}
Let $\tilde{\mathcal{E}}$ be the matrix representation of a CPTP map $\mathcal{E}$ [see Eq.~\eqref{eq_Etilde}]. Then, Eq.~\eqref{eq_unitary_sym} is rewritten as
\begin{equation}
    \tilde{\mathcal{V}}\tilde{\mathcal{E}}\tilde{\mathcal{V}}^{-1}=\tilde{\mathcal{E}},
    \label{eq_unitary_sym2}
\end{equation}
where $\tilde{\mathcal{V}}$ is the matrix representation of $\mathcal{V}$ which is given by a unitary operator acting on the doubled Hilbert space $\mathcal{H}\otimes\mathcal{H}$. 
We call unitary symmetry in Eqs.~\eqref{eq_unitary_sym} and \eqref{eq_unitary_sym2} weak symmetry \cite{Buca12,Albert14,deGroot22}. 
If a CPTP map has weak symmetry, it can be block diagonalized according to the symmetry eigenvalues. 
For example, translation symmetry in Eq.~\eqref{eq_trans_sym1} is weak symmetry, and the argument $\vec{k}$ of the Bloch matrix $X(\vec{k})$ corresponds to the symmetry eigenvalue (see Appendix \ref{sec_mom}).

Suppose that every Kraus operator of a CPTP map commutes with a unitary operator $\hat{V}$ as $[\hat{K}_m,\hat{V}]=0\ (\forall m)$. Then, the CPTP map satisfies Eq.~\eqref{eq_unitary_sym2} with $\tilde{\mathcal{V}}=\hat{V}\otimes \hat{I}$ or $\tilde{\mathcal{V}}=\hat{I}\otimes \hat{V}^*$. Such symmetry is called strong symmetry \cite{Buca12,Albert14,deGroot22}, which gives a more stringent condition than weak symmetry. Strong symmetry is related to a conserved quantity of the dynamics. To see this relation, we consider the case of continuous symmetry and assume that every Kraus operator $\hat{K}_m$ commutes with $\hat{V}_\theta=\exp[-i\theta \hat{A}]$ for all $\theta\in\mathbb{R}$, where $\hat{A}$ is a Hermitian operator. This assumption implies $[\hat{K}_m,\hat{A}]=0\ (\forall m)$. Then, for an arbitrary density matrix $\hat{\rho}$, we have
\begin{align}
    \mathrm{Tr}[\hat{A}\mathcal{E}(\hat{\rho})]=&\sum_m\mathrm{Tr}[\hat{K}_m^\dag \hat{A}\hat{K}_m\hat{\rho}]\notag\\
    =&\mathrm{Tr}\Bigl[\Bigl(\sum_m\hat{K}_m^\dag \hat{K}_m\Bigr)\hat{A}\hat{\rho}\Bigr]\notag\\
    =&\mathrm{Tr}[\hat{A}\hat{\rho}],
\end{align}
where we use $\sum_m\hat{K}_m^\dag \hat{K}_m=\hat{I}$ in deriving the last equality. Thus, the observable $\hat{A}$ is conserved under the CPTP map.

\subsection{Modular conjugation symmetry\label{sec_mod_conj_sym}}

A CPTP map has an antiunitary symmetry due to its Hermiticity-preserving nature \cite{Gong18}. To see this property, we define the modular conjugation \cite{Liu21, Kawabata22} by an antiunitary operator $\tilde{J}$ acting on the doubled Hilbert space $\mathcal{H}\otimes\mathcal{H}$ as
    \begin{equation}
    \tilde{J}(\rho_{\vec{i},a;\vec{j},b}\ket{\vec{i},a}\otimes\ket{\vec{j},b}):=\rho_{\vec{i},a;\vec{j},b}^*\ket{\vec{j},b}\otimes\ket{\vec{i},a}.
    \end{equation}
The modular conjugation operator $\tilde{J}$ is a combination of a swap operation on states and the complex conjugation of their coefficients. The modular conjugation satisfies $\tilde{J}^2=1$ and acts on the vectorized density matrix [see Eq.~\eqref{eq_rho_vec}] as
\begin{equation}
    \tilde{J}\ket{\hat{\rho}}=\ket{\hat{\rho}^\dag}.
    \label{eq_J_rho}
\end{equation}
Then, we have
    \begin{align}
    \tilde{J}\tilde{\mathcal{E}}\ket{\hat{\rho}}=&\tilde{J}\ket{\mathcal{E}(\hat{\rho})}\notag\\
    =&\ket{[\mathcal{E}(\hat{\rho})]^\dag}\notag\\
    =&\ket{\mathcal{E}(\hat{\rho}^\dag)}\notag\\
    =&\tilde{\mathcal{E}}\ket{\hat{\rho}^\dag}\notag\\
    =&\tilde{\mathcal{E}}\tilde{J}\ket{\hat{\rho}},
    \label{eq_J_E_rho}
    \end{align}
where we use the definition $\tilde{\mathcal{E}}\ket{\hat{\rho}}=\ket{\mathcal{E}(\hat{\rho})}$ of the matrix representation [see Eq.~\eqref{eq_mat_rep_derivation}], the property of the modular conjugation [Eq.~\eqref{eq_J_rho}], and the Hermiticity-preserving property of a CPTP map [Eq.~\eqref{eq_Hermiticity_preserv}]. 
Since Eq.~\eqref{eq_J_E_rho} holds for arbitrary $\ket{\hat{\rho}}$, we obtain
    \begin{equation}
    \tilde{J}\tilde{\mathcal{E}}\tilde{J}^{-1}=\tilde{\mathcal{E}},
    \label{eq_mod_conj_sym}
    \end{equation}
which indicates that the operator $\tilde{\mathcal{E}}$ has an antiunitary symmetry defined by the modular conjugation. Since a CPTP map must preserve the Hermiticity of the density matrix, any CPTP map has the modular conjugation symmetry given by Eq.~\eqref{eq_mod_conj_sym}.

\subsection{Tenfold symmetry classification of feedback control with projective measurement\label{sec_BL_sym}}

The topological classification of non-Hermitian matrices is performed on the basis of the BL symmetry classes \cite{Bernard02}, which are the non-Hermitian generalizations \cite{Kawabata19, Zhou19} of the Altland-Zirnbauer (AZ) symmetry classes \cite{Altland97}. Here, we discuss the BL symmetry classes of quantum channels for discrete quantum feedback control. In general, the symmetry of a CPTP map $\mathcal{E}$ that belongs to a BL symmetry class has one of the following representations \cite{Ashida20}:
\begin{subequations}
    \begin{align}
        &P\ \mathrm{symmetry}:\ \tilde{\mathcal{V}}_P\tilde{\mathcal{E}}\tilde{\mathcal{V}}_P^{-1}=-\tilde{\mathcal{E}},\ \tilde{\mathcal{V}}_P^2=\hat{I}\otimes \hat{I},\label{eq_BL_P}\\
        &C\ \mathrm{symmetry}:\ \tilde{\mathcal{V}}_C\tilde{\mathcal{E}}^T\tilde{\mathcal{V}}_C^{-1}=\epsilon_C\tilde{\mathcal{E}},\ \tilde{\mathcal{V}}_C\tilde{\mathcal{V}}_C^*=\eta_C\hat{I}\otimes \hat{I},\label{eq_BL_C}\\
        &K\ \mathrm{symmetry}:\ \tilde{\mathcal{V}}_K\tilde{\mathcal{E}}^*\tilde{\mathcal{V}}_K^{-1}=\epsilon_K\tilde{\mathcal{E}},\ \tilde{\mathcal{V}}_K\tilde{\mathcal{V}}_K^*=\eta_K\hat{I}\otimes \hat{I},\label{eq_BL_K}\\
        &Q\ \mathrm{symmetry}:\ \tilde{\mathcal{V}}_Q\tilde{\mathcal{E}}^\dag\tilde{\mathcal{V}}_Q^{-1}=\epsilon_Q\tilde{\mathcal{E}},\ \tilde{\mathcal{V}}_Q^2=\hat{I}\otimes \hat{I},\label{eq_BL_Q}
    \end{align}
\end{subequations}
where $\tilde{\mathcal{E}}$ is the matrix representation of $\mathcal{E}$ acting on the doubled Hilbert space (see Sec.~\ref{sec_top_setup}), $\tilde{\mathcal{V}}_X\ (X=P,C,K,Q)$ is a unitary operator on the doubled Hilbert space, $\epsilon_X=\pm 1\ (X=C,K,Q)$, and $\eta_X=\pm 1\ (X=C,K)$.
Here, we assume that a CPTP map does not have unitary symmetry; if a CPTP map has unitary symmetry, we first block diagonalize it according to the symmetry eigenvalues and consider the BL symmetry class of each block. 
The combinations of the four types of symmetries, Eqs.~\eqref{eq_BL_P}-\eqref{eq_BL_Q}, lead to 38 classes of non-Hermitian matrices \cite{Kawabata19}.

\begin{figure}
    \includegraphics[width=8.5cm]{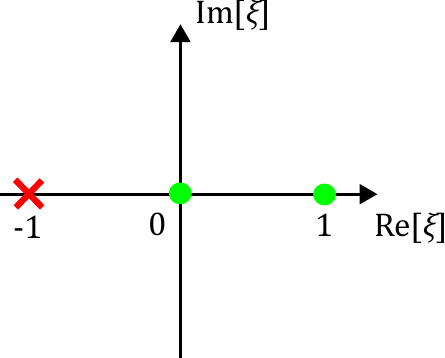}
    \caption{Eigenspectrum of a projective measurement channel (shown by green dots) consisting of eigenvalues zero and one. Since a projective measurement channel cannot have eigenvalue $-1$, its possible BL symmetry classes are restricted to the ten symmetry classes listed in Table \ref{table_AZdag}, which constitute a subset of the entire 38 symmetry classes.}
    \label{fig_proj_spec}
\end{figure}

Whereas general non-Hermitian matrices are classified in terms of the 38 BL symmetry classes \cite{Kawabata19}, some of them are not compatible with CPTP maps for projective measurements. A CPTP map for a projective measurement is given by
\begin{equation}
    \mathcal{E}_{\mathrm{proj}}(\hat{\rho})=\sum_m\hat{P}_m\hat{\rho} \hat{P}_m,
    \label{eq_CPTP_proj}
\end{equation}
where $\hat{P}_m$ is a projection operator satisfying $\hat{P}_m^\dag=\hat{P}_m$ and $\hat{P}_m\hat{P}_{m'}=\delta_{m,m'}\hat{P}_m$. A key property of the projective measurement channel \eqref{eq_CPTP_proj} is that its eigenvalues are either zero or one, which follows from the Hermiticity and idempotency of projection operators (see Appendix \ref{sec_proj_channel}). Physically, this property is related to the repeatability of projective measurements; that is, repeated projective measurements after the first projective measurement with outcome $m$ always yield the same outcome $m$ \cite{NielsenChuang_book}.
To identify the possible symmetry classes of the projective measurement channel \eqref{eq_CPTP_proj},  we note that a certain BL symmetry places a constraint on the eigenvalues of a CPTP map. Let $\xi$ be an eigenvalue of a CPTP map $\mathcal{E}$. Then, if this CPTP map has $P$ symmetry \eqref{eq_BL_P} or $C$ symmetry with $\epsilon_C=-1$, $-\xi$ is also an eigenvalue of $\mathcal{E}$ \cite{Ashida20}. Similarly, if a CPTP map $\mathcal{E}$ has $K$ ($Q$) symmetry, $\epsilon_K\xi^*$ ($\epsilon_Q\xi^*$) is also an eigenvalue of $\mathcal{E}$. 
Since a projective measurement channel \eqref{eq_CPTP_proj} cannot have an eigenvalue $-1$, neither $P$ symmetry nor $X=C,K,Q$ symmetry with $\epsilon_X=-1$ can be a symmetry of the projective measurement channel (see Fig.~\ref{fig_proj_spec}). 
Thus, the possible symmetries of a CPTP map for the projective measurement are as follows:
\begin{subequations}
    \begin{align}
        \tilde{\mathcal{V}}_C\tilde{\mathcal{E}}^T\tilde{\mathcal{V}}_C^{-1}=&\tilde{\mathcal{E}},\ \tilde{\mathcal{V}}_C\tilde{\mathcal{V}}_C^*=\eta_C\hat{I}\otimes \hat{I},\label{eq_TRSdag}\\
        \tilde{\mathcal{V}}_K\tilde{\mathcal{E}}^*\tilde{\mathcal{V}}_K^{-1}=&\tilde{\mathcal{E}},\ \tilde{\mathcal{V}}_K\tilde{\mathcal{V}}_K^*=\eta_K\hat{I}\otimes \hat{I},\label{eq_TRS}\\
        \tilde{\mathcal{V}}_Q\tilde{\mathcal{E}}^\dag \tilde{\mathcal{V}}_Q^{-1}=&\tilde{\mathcal{E}},\ \tilde{\mathcal{V}}_Q^2=\hat{I}\otimes \hat{I},\label{eq_PH}
    \end{align}
\end{subequations}
where $\eta_C,\eta_K=\pm 1$, and $\tilde{\mathcal{V}}_K,\tilde{\mathcal{V}}_K,\tilde{\mathcal{V}}_Q$ are unitary operators on the doubled Hilbert space.

Now, let us consider quantum channels for discrete quantum feedback control with projective measurements. Specifically, we consider a CPTP map of the form
\begin{equation}
    \mathcal{E}_\tau(\hat{\rho})=\sum_m \hat{U}_m(\tau)\hat{P}_m\hat{\rho} \hat{P}_m\hat{U}_m^\dag(\tau),
    \label{eq_CPTP_tau}
\end{equation}
where $\hat{U}_m(\tau)=\mathcal{T}\exp\Bigl[-i\int_0^\tau dt \hat{H}_m(t)\Bigr]$ is the feedback unitary operator generated by a Hamiltonian $\hat{H}_m(t)$. Since the relevant symmetry of feedback control described by the CPTP map \eqref{eq_CPTP_tau} should not depend on the duration $\tau$ of feedback control, it should be consistent with the symmetry in the limit of $\tau\to0$, where the CPTP map reduces to the projective measurement channel as $\lim_{\tau\to 0}\mathcal{E}_{\tau}=\mathcal{E}_{\mathrm{proj}}$. 
Thus, the BL symmetries that are compatible with the CPTP map \eqref{eq_CPTP_tau} are given by the same symmetries in Eqs.~\eqref{eq_TRSdag}-\eqref{eq_PH}. 

The three types of symmetries, Eqs.~\eqref{eq_TRSdag}-\eqref{eq_PH}, and the combinations thereof constitute ten symmetry classes of quantum channels for discrete quantum feedback control with projective measurements. Notably, they belong to a subset of the entire 38 BL symmetry classes applicable to general non-Hermitian systems \cite{Kawabata19,Zhou19}. 
Each symmetry class is specified by the presence or absence of the three types of symmetries, Eqs.~\eqref{eq_TRSdag}-\eqref{eq_PH}, and by the signs of $\eta_C$ and $\eta_K$, as shown in Table \ref{table_AZdag}. 
Here, it is sufficient to consider a single symmetry for each type [Eqs.~\eqref{eq_TRSdag}-\eqref{eq_PH}]. In fact, if a system has multiple symmetries of the same type, we can construct a unitary symmetry by combining the two BL symmetries. 
We note that the existence of two symmetries among Eqs.~\eqref{eq_TRSdag}-\eqref{eq_PH} leads to the presence of the remaining symmetry constructed from the combination of the two symmetry operations, resulting in ten classes. 

We list the tenfold symmetry classification of quantum channels in Table \ref{table_AZdag}, where each symmetry class is labeled according to the general classification of non-Hermitian symmetry classes in Ref.~\cite{Kawabata19}. We note that these ten symmetry classes are equivalent to the AZ$^\dag$ classes in Ref.~\cite{Kawabata19} if we multiply the original CPTP maps by $i$. In fact, the $C$ symmetry \eqref{eq_TRSdag} and the $K$ symmetry \eqref{eq_TRS} are equivalent to ``time-reversal symmetry$^\dag$'' and ``particle-hole symmetry$^\dag$'' of $i\tilde{\mathcal{E}}$, respectively \cite{Kawabata19NComm,Kawabata19}. In Table \ref{table_AZdag}, we also show the AZ$^\dag$ classes corresponding to the symmetry classes of CPTP maps.

Once the relevant symmetry classes are identified, we can exploit the topological classification of general non-Hermitian matrices \cite{Kawabata19,Zhou19} to examine whether there exists topological feedback control that is described by a topologically nontrivial Bloch matrix in a given symmetry class and spatial dimension. We note, however, that the Bloch matrix cannot be a general non-Hermitian matrix due to the completely positive and trace-preserving nature of quantum channels (see Appendix \ref{sec_Bloch_constraint}). In particular, the complete positivity may place nontrivial constraints on possible topological phases realized with feedback control. We will present an example of this phenomenon in Sec.~\ref{sec_sym_proj_example}.

The present symmetry classification applies to the case with multipoint projective measurements. For example, let us consider the following CPTP map for feedback control with a two-point projective measurement:
\begin{widetext}
\begin{equation}
    \mathcal{E}_{\tau_1,\tau_2}^\prime(\hat{\rho})=\sum_{m_1,m_2} \hat{U}_{m_1,m_2}'(\tau_2)\hat{P}_{m_2}\hat{U}_{m_1}(\tau_1)\hat{P}_{m_1}\hat{\rho} \hat{P}_{m_1}\hat{U}_{m_1}^\dag(\tau_1)\hat{P}_{m_2}\hat{U}_{m_1,m_2}^{\prime\dag}(\tau_2),
    \label{eq_CPTP_tau2}
\end{equation}
\end{widetext}
where $\hat{U}_{m_1,m_2}^{\prime}(\tau_2)=\mathcal{T}\exp[-i\int_0^{\tau_2}dt \hat{H}_{m_1,m_2}'(t)]$ is a unitary operator generated by a Hamiltonian $\hat{H}_{m_1,m_2}'(t)$. In the ``no-feedback'' limit, Eq.~\eqref{eq_CPTP_tau2} reduces to the projective measurement channel as $\lim_{\tau_1,\tau_2\to 0}\mathcal{E}_{\tau_1,\tau_2}^\prime=\mathcal{E}_{\mathrm{proj}}$. Thus, the allowed symmetry classes for CPTP maps \eqref{eq_CPTP_tau2} for two-point projective measurements are again given by the ten classes in Table \ref{table_AZdag}. Here, we assume that the measurement operators for the second measurement are the same as those of the first measurement given in Eq.~\eqref{eq_CPTP_tau2}. If the measurement operators are different for the two measurements, other BL symmetry classes may also be allowed. We will discuss this possibility in Sec.~\ref{sec_conclusion}.

Even if a measurement is not projective, other symmetry classes are still not allowed unless eigenvalues of a CPTP map for measurements always appear as pairs like $(\xi,-\xi)$ or $(\xi,-\xi^*)$ [see the paragraph above Eqs.~\eqref{eq_TRSdag}-\eqref{eq_PH}]. 
For example, if a measurement is not projective due to an error in the measurement process, the possible symmetry classes are not altered because we can take an error-free limit under which the symmetry is intact. 

The ten AZ$^\dag$ symmetry classes are formally equivalent to the symmetry classes for generators of Markovian quantum dynamics discussed in Ref.~\cite{Lieu20}. In fact, the symmetry classes that are given in Table \ref{table_AZdag} agree with the symmetry classes of CPTP maps that are generated by Lindbladians (see Appendix \ref{sec_sym_corresp}). However, we note that a CPTP map for feedback control cannot have a generator if it has a zero eigenvalue [see Eq.~\eqref{eq_generator}]. In general, a CPTP map generated by a Lindbladian can be classified by the 38 BL symmetry classes if one includes shifted symmetries \cite{Kawasaki22,Kawabata22,Sa22}. However, a projective measurement channel \eqref{eq_CPTP_proj} cannot have those shifted symmetries because the number of zero eigenvalues is different from that of unit eigenvalues, while shifted symmetry is accompanied by the symmetry of the eigenspectrum with respect to a certain line \cite{Kawasaki22,Kawabata22,Sa22}. Thus, we conclude that the tenfold symmetry classification exhausts all relevant symmetry classes of discrete quantum feedback control with projective measurements.

We note that a CPTP map always has modular conjugation symmetry [see Eq.~\eqref{eq_mod_conj_sym} in Sec.~\ref{sec_mod_conj_sym}]. The modular conjugation symmetry corresponds to a $K$ symmetry with $\eta_K=+1$. 
Thus, a quantum channel as a whole always has this symmetry. However, if a quantum channel possesses an additional unitary symmetry, special care must be taken because the quantum channel can then be block diagonalized according to the symmetry eigenvalues; each symmetry sector (i.e. each block) does not necessarily have modular conjugation symmetry. In this case, we must consider an appropriate BL symmetry class in each sector.
We will revisit this point when we discuss concrete examples in Sec.~\ref{sec_SPT}.

Here, we provide sufficient conditions for the BL symmetries of CPTP maps to satisfy Eqs.~\eqref{eq_TRSdag}-\eqref{eq_PH}, which are helpful to gain insight into feedback control with symmetry. The symmetries in Eqs.~\eqref{eq_TRSdag}-\eqref{eq_PH} are satisfied if Kraus operators satisfy
\begin{subequations}
    \begin{align}
        \tilde{\mathcal{V}}_C(\hat{K}_m^T\otimes \hat{K}_m^\dag)\tilde{\mathcal{V}}_C^{-1}=&\hat{K}_{f(m)}\otimes \hat{K}_{f(m)}^*,\label{eq_cond_TRSdag}\\
        \tilde{\mathcal{V}}_K(\hat{K}_m^*\otimes \hat{K}_m)\tilde{\mathcal{V}}_K^{-1}=&\hat{K}_{f(m)}\otimes \hat{K}_{f(m)}^*,\label{eq_cond_TRS}\\
        \tilde{\mathcal{V}}_Q(\hat{K}_m^\dag\otimes \hat{K}_m^T)\tilde{\mathcal{V}}_Q^{-1}=&\hat{K}_{f(m)}\otimes \hat{K}_{f(m)}^*,\label{eq_cond_PH}
    \end{align}
\end{subequations}
respectively, where $\hat{K}_m\ (m=1,2,\cdots,N_{\mathrm{K}})$ are the Kraus operators, $N_{\mathrm{K}}$ is the number of Kraus operators, and $f$ is a bijection on $\{1,2,\cdots,N_{\mathrm{K}}\}$. Suppose that the Kraus operators of feedback control are given by $\hat{K}_m=\hat{U}_m\hat{M}_m$, where $\hat{M}_m$ is a measurement operator and $\hat{U}_m=e^{-i\hat{H}_m\tau}$ is a unitary operator conditioned on a measurement outcome $m$. Then, if the two conditions
\begin{equation}
    \hat{V}_K\hat{U}_m^*\hat{V}_K^{-1}=\hat{U}_m\ \mathrm{and}\ \hat{V}_K\hat{M}_m^*\hat{V}_K^{-1}=e^{i\varphi} \hat{M}_m
    \label{eq_cond2_TRS}
\end{equation}
are satisfied for a unitary operator $\hat{V}_K$ and $\varphi\in\mathbb{R}$, Eq.~\eqref{eq_cond_TRS} is satisfied for $f(m)=m$ and $\tilde{\mathcal{V}}_K=\hat{V}_K\otimes \hat{V}_K^*$. The first condition in Eq.~\eqref{eq_cond2_TRS} is equivalent to the particle-hole symmetry of the feedback Hamiltonian (see Eq.~\eqref{eq_PHS_U} in Appendix \ref{sec_sym_corresp}). In this manner, a CPTP map with $K$ symmetry \eqref{eq_TRS} can be constructed by feedback control with symmetry \eqref{eq_cond2_TRS}. Since
\begin{align}
\tilde{\mathcal{V}}_K\tilde{\mathcal{V}}_K^*=\hat{V}_K\hat{V}_K^*\otimes \hat{V}_K^*\hat{V}_K,
\end{align}
the condition \eqref{eq_cond2_TRS} leads to a $K$ symmetry with $\eta_K=+1$ regardless of the sign of $\hat{V}_K\hat{V}_K^*$.

In contrast, the symmetries \eqref{eq_TRSdag} and \eqref{eq_PH}, respectively, include transposition and Hermitian conjugation, both of which reverse the order of operators. Because of this order-reversing nature, the symmetries \eqref{eq_TRSdag} and \eqref{eq_PH} are unlikely to be realized with the simplest protocol involving a single measurement and feedback since $(\hat{U}_m\hat{M}_m)^T=\hat{M}_m^T\hat{U}_m^T$ and $(\hat{U}_m\hat{M}_m)^\dag=\hat{M}_m^\dag \hat{U}_m^\dag$. In Sec.~\ref{sec_SPT}, we demonstrate that those symmetries can be satisfied by a protocol with a two-point measurement before and after feedback control, where the Kraus operators are given by $\hat{K}_{m}=\hat{M}_{m_2}\hat{U}_{m_1}\hat{M}_{m_1}$ with $m=(m_1,m_2)$ being a pair of measurement outcomes (see Sec.~\ref{sec_feedback}).

\section{Feedback control with symmetry\label{sec_SPT}}

In this section, we present examples of feedback control for each symmetry class shown in Table \ref{table_AZdag}.
On the basis of the intrinsic modular conjugation symmetry (see Sec.~\ref{sec_mod_conj_sym}), the ten symmetry classes of quantum channels listed in Table \ref{table_AZdag} can be categorized into three groups according the presence or absence of $K$ symmetry and the sign of $\eta_K$. The first group (AI, AI $+$ psH$_+$, and AI $+$ psH$_-$) includes $K$ symmetry with $\eta_K=+1$. Since the modular conjugation symmetry features $K$ symmetry with $\eta_K=+1$, these symmetry classes can be realized with quantum channels without additional symmetries or with additional $C$ or $Q$ symmetry. The second group (A, psH, AI$^\dag$, and AII$^\dag$) does not have $K$ symmetry. These classes can be realized in a symmetry sector of a quantum channel with unitary symmetry. The third group (AII, AII $+$ psH$_+$, and AII $+$ psH$_-$) includes $K$ symmetry with $\eta_K=-1$. Such symmetry classes can be realized by imposing $K$ symmetry with $\eta_K=-1$ on the second group.

\subsection{Class AI: Feedback without symmetry or with particle-hole symmetry\label{sec_AI}}

We begin with class AI, which includes only $K$ symmetry with $\eta_K=+1$. Since a CPTP map always has modular conjugation symmetry (see Sec.~\ref{sec_mod_conj_sym}), a quantum channel belongs to class AI if it does not have any other symmetry. For example, the chiral Maxwell demon presented in Sec.~\ref{sec_model} belongs to this class.

A CPTP map can have $K$ symmetry with $\eta_K=+1$ that is not equivalent to the modular conjugation symmetry. For example, if the measurement operators and the unitary operators for feedback control satisfy Eq.~\eqref{eq_cond2_TRS}, the corresponding CPTP map has $K$ symmetry. Since the first condition in Eq.~\eqref{eq_cond2_TRS} follows from the particle-hole symmetry of the feedback Hamiltonian, this symmetry is physically interpreted as the particle-hole symmetry in feedback control. If a CPTP map has $K$ symmetry \eqref{eq_TRS} that is not equivalent to the modular conjugation symmetry, we can construct a unitary symmetry
\begin{equation}
    \tilde{\mathcal{V}}_K \tilde{S}\tilde{\mathcal{E}}(\tilde{\mathcal{V}}_K \tilde{S})^{-1}=\tilde{\mathcal{E}},
    \label{eq_K_unitary}
\end{equation}
where $\tilde{S}$ is the swap operator defined by 
\begin{equation}
    \tilde{S}(\ket{\vec{i},a}\otimes\ket{\vec{j},b})=\ket{\vec{j},b}\otimes\ket{\vec{i},a}.
    \label{eq_swap}    
\end{equation}
The CPTP map can be block diagonalized by using the unitary symmetry \eqref{eq_K_unitary}, and the $K$ symmetry and modular conjugation symmetry are equivalent in each symmetry sector. 
Thus, if the CPTP map does not have any other symmetry, the block diagonalized CPTP map belongs to class AI.

\subsection{Classes AI $+$ psH$_+$ and AI $+$ psH$_-$: Feedback with a two-point projective measurement\label{sec_sym_proj_example}}
The symmetry classes that include $K$ symmetry with $\eta_K=+1$ are classes AI, AI $+$ psH$_+$, and AI $+$ psH$_-$, where the last two classes include $C$ and $Q$ symmetries. As mentioned in Sec.~\ref{sec_tenfold}, $C$ and $Q$ symmetries, respectively, involve transposition and Hermitian conjugation, both of which reverse the order of operators. Here we construct quantum channels with $C$ and $Q$ symmetries by using feedback control with two-point projective measurements. Specifically, we consider a scheme with two projective measurements intervened by a unitary operation. This process is described by a CPTP map
\begin{align}
    \mathcal{E}(\hat{\rho})=&\sum_{m_1,m_2}\hat{K}_{m_1,m_2}\hat{\rho} \hat{K}_{m_1,m_2}^\dag\notag\\
    =&\sum_{m_1,m_2}\hat{P}_{m_2}\hat{U}_{m_1}\hat{P}_{m_1}\hat{\rho} \hat{P}_{m_1}\hat{U}_{m_1}^\dag \hat{P}_{m_2},
    \label{eq_CPTP_2pt}
\end{align}
where $m_n=(\vec{j}_n,a_n)\ (n=1,2)$ denotes a measurement outcome for a projective measurement with the projection operator $\hat{P}_{m_n}:=\hat{P}_{\vec{j}_n,a_n}=\ket{\vec{j}_n,a_n}\bra{\vec{j}_n,a_n}$, and 
\begin{align}
\hat{K}_{m_1,m_2}:=&\hat{P}_{m_2}\hat{U}_{m_1}\hat{P}_{m_1}\notag\\
=&(\hat{U}_{\vec{j}_1,a_1})_{\vec{j}_2,a_2;\vec{j}_1,a_1}\ket{\vec{j}_2,a_2}\bra{\vec{j}_1,a_1}
\end{align}
is the Kraus operator. 
In this case, we have
\begin{align}
    \hat{K}_{m_1,m_2}\otimes\hat{K}_{m_1,m_2}^*=&p(\vec{j}_2,a_2|\vec{j}_1,a_1)\notag\\
    &\times\ket{\vec{j}_2,a_2}\bra{\vec{j}_1,a_1}\otimes\ket{\vec{j}_2,a_2}\bra{\vec{j}_1,a_1}
\end{align}
and
\begin{align}
    \hat{K}_{m_1,m_2}^T\otimes\hat{K}_{m_1,m_2}^\dag=&p(\vec{j}_2,a_2|\vec{j}_1,a_1)\notag\\
    &\times\ket{\vec{j}_1,a_1}\bra{\vec{j}_2,a_2}\otimes\ket{\vec{j}_1,a_1}\bra{\vec{j}_2,a_2},
\end{align}
where $p(\vec{j}_2,a_2|\vec{j}_1,a_1):=|(\hat{U}_{\vec{j}_1,a_1})_{\vec{j}_2,a_2;\vec{j}_1,a_1}|^2$ is the probability of a measurement outcome $(\vec{j}_2,a_2)$ being obtained in the projective measurement performed after feedback control associated with a measurement outcome $(\vec{j}_1,a_1)$. Thus, the conditions \eqref{eq_cond_TRSdag} and \eqref{eq_cond_PH} for $C$ and $Q$ symmetries follow from symmetry of the transition probability $p(\vec{j}_2,a_2|\vec{j}_1,a_1)$.

The Bloch matrix of the CPTP map \eqref{eq_CPTP_2pt} is given by
\begin{align}
    X_{a,c,\vec{\mu};b,d,\vec{\nu}}(\vec{k})=&\frac{1}{N_{\mathrm{cell}}}\sum_{\vec{j},\vec{j}'}\sum_{m_1,m_2}(\hat{K}_{m_1,m_2})_{\vec{j},a;\vec{j}',b}\notag\\
    &\times(\hat{K}_{m_1,m_2})^*_{\vec{j}+\vec{\mu},c;\vec{j}'+\vec{\nu},d}e^{-i\vec{k}\cdot(\vec{R}_{\vec{j}}-\vec{R}_{\vec{j}'})}\notag\\
    =&\delta_{\vec{\mu},\vec{0}}\delta_{\vec{\nu},\vec{0}}\delta_{a,c}\delta_{b,d}\frac{1}{N_{\mathrm{cell}}}\sum_{\vec{j}_1,\vec{j}_2}p(\vec{j}_2,a|\vec{j}_1,b)\notag\\
    &\times e^{-i\vec{k}\cdot(\vec{R}_{\vec{j}_2}-\vec{R}_{\vec{j}_1})}.
\end{align}
Thus, the Bloch matrix has the following structure:
\begin{equation}
    X(\vec{k})=
    \begin{pmatrix}
        X_{\mathrm{trunc}}(\vec{k}) & O \\
        O & O
    \end{pmatrix},
    \label{eq_Bloch_block_2pt}
\end{equation}
where
\begin{widetext}
\begin{equation}
    X_{\mathrm{trunc}}(\vec{k})=
    \begin{pmatrix}
        X_{1,1,\vec{0};1,1,\vec{0}}(\vec{k}) & X_{1,1,\vec{0};2,2,\vec{0}}(\vec{k}) & \cdots & X_{1,1,\vec{0};D_{\mathrm{loc}},D_{\mathrm{loc}},\vec{0}}(k) \\
        X_{2,2,\vec{0};1,1,\vec{0}}(\vec{k}) & X_{2,2,\vec{0};2,2,\vec{0}}(\vec{k}) & \cdots & X_{2,2,\vec{0};D_{\mathrm{loc}},D_{\mathrm{loc}},\vec{0}}(\vec{k}) \\
        \vdots & \vdots & & \vdots \\
        X_{D_{\mathrm{loc}},D_{\mathrm{loc}},\vec{0};1,1,\vec{0}}(\vec{k}) & X_{D_{\mathrm{loc}},D_{\mathrm{loc}},\vec{0};2,2,\vec{0}}(\vec{k}) & \cdots & X_{D_{\mathrm{loc}},D_{\mathrm{loc}},\vec{0};D_{\mathrm{loc}},D_{\mathrm{loc}},\vec{0}}(\vec{k}) \\
    \end{pmatrix}
\end{equation}
\end{widetext}
is the truncated Bloch matrix. Nonzero eigenvalues of the Bloch matrix are completely determined by those of the truncated Bloch matrix. The BL symmetry class of a given Bloch matrix is also determined from that of the corresponding truncated Bloch matrix.

By way of illustration, we consider a system with a spin-$1/2$ particle, which corresponds to the case with $D_{\mathrm{loc}}=2$. The truncated Bloch matrix is given by
\begin{align}
    X_{\mathrm{trunc}}(\vec{k})=
    \begin{pmatrix}
        \xi_{\up\up}(\vec{k}) & \xi_{\up\down}(\vec{k}) \\
        \xi_{\down\up}(\vec{k}) & \xi_{\down\down}(\vec{k})
    \end{pmatrix},
    \label{eq_Bloch_trunc_2pt}
\end{align}
where
\begin{align}
    \xi_{\sigma\sigma'}(\vec{k}):=&X_{\sigma,\sigma,\vec{0};\sigma',\sigma',\vec{0}}(\vec{k})\notag\\
    =&\frac{1}{N_{\mathrm{cell}}}\sum_{\vec{j}_1,\vec{j}_2}p(\vec{j}_2,\sigma|\vec{j}_1,\sigma') e^{-i\vec{k}\cdot(\vec{R}_{\vec{j}_2}-\vec{R}_{\vec{j}_1})}.
    \label{eq_2PM_xi_k}
\end{align}
Let us examine the symmetry of the truncated Bloch matrix. First, we note that the truncated Bloch matrix always has an antiunitary symmetry (the modular conjugation symmetry),
\begin{equation}
    X_{\mathrm{trunc}}^*(-\vec{k})=X_{\mathrm{trunc}}(\vec{k}),
    \label{eq_2PM_mod_conj}
\end{equation}
due to the Hermiticity-preserving nature of the CPTP map (see Sec.~\ref{sec_mod_conj_sym} and Appendix \ref{sec_Bloch_constraint}). This case corresponds to $K$ symmetry \eqref{eq_TRS} with $\eta_K=+1$. Next, we consider $C$ symmetry \eqref{eq_TRSdag}. The simplest symmetry of this type with $\eta_C=+1$ is given by
\begin{equation}
    X_{\mathrm{trunc}}^T(-\vec{k})=X_{\mathrm{trunc}}(\vec{k}),
    \label{eq_2PM_transpose}
\end{equation}
which is equivalent to the condition
\begin{equation}
    \xi_{\sigma\sigma'}(-\vec{k})=\xi_{\sigma'\sigma}(\vec{k}).
    \label{eq_cond_AIdag}
\end{equation}
The condition \eqref{eq_cond_AIdag} is satisfied if
\begin{align}
    p(\vec{j}_2,\sigma|\vec{j}_1,\sigma')=p(\vec{j}_1,\sigma'|\vec{j}_2,\sigma),
\end{align}
which can be interpreted as reciprocity in the feedback process. The truncated Bloch matrix (and hence the original Bloch matrix) with Eq.~\eqref{eq_2PM_transpose} belongs to the symmetry class AI $+$ psH$_+$ in Table \ref{table_AZdag}. In the present case with Eqs.~\eqref{eq_2PM_mod_conj} and \eqref{eq_2PM_transpose}, the truncated Bloch matrix is Hermitian [$X_{\mathrm{trunc}}^\dag(\vec{k})=X_{\mathrm{trunc}}(\vec{k})$], and therefore its eigenvalues are real. More specifically, the truncated Bloch matrix in a 1D system cannot have a nonzero winding number.

Another symmetry class AI $+$ psH$_-$ possesses $C$ symmetry with $\eta_C=-1$. An example of such symmetry is given by
\begin{equation}
    (-i\sigma_y)X_{\mathrm{trunc}}^T(-\vec{k})(-i\sigma_y)^{-1}=X_{\mathrm{trunc}}(\vec{k}),
    \label{eq_AIIdag}
\end{equation}
where $\sigma_y$ is the Pauli matrix. This type of symmetry leads to the Kramers degeneracy in non-Hermitian systems \cite{Kawabata19}. The symmetry \eqref{eq_AIIdag} is equivalent to the following three conditions:
\begin{subequations}
    \begin{align}
        \xi_{\up\up}(-\vec{k})=&\xi_{\down\down}(\vec{k}),\label{eq_cond_AIIdag1}\\
        \xi_{\up\down}(-\vec{k})=&-\xi_{\up\down}(\vec{k}),\label{eq_cond_AIIdag2}\\
        \xi_{\down\up}(-\vec{k})=&-\xi_{\down\up}(\vec{k}).\label{eq_cond_AIIdag3}
    \end{align}
\end{subequations}
It follows from Eq.~\eqref{eq_2PM_xi_k} that condition \eqref{eq_cond_AIIdag1} is satisfied if
\begin{align}
    p(\vec{j}_2,\up|\vec{j}_1,\up)=p(\vec{j}_1,\down|\vec{j}_2,\down),
\end{align}
which represents the reciprocity in which the two spin components flow in opposite directions. 
In contrast, since $p(\vec{j}_2,\sigma|\vec{j}_1,\sigma')\geq 0$, conditions \eqref{eq_cond_AIIdag2} and \eqref{eq_cond_AIIdag3} are satisfied if and only if
\begin{equation}
    p(\vec{j}_2,\up|\vec{j}_1,\down)=p(\vec{j}_2,\down|\vec{j}_1,\up)=0\ (\forall \vec{j}_1,\vec{j}_2).
\end{equation}
Thus, the truncated Bloch matrix is written as
\begin{align}
    X_{\mathrm{trunc}}(\vec{k})=
    \begin{pmatrix}
        \xi_{\up\up}(\vec{k}) & 0 \\
        0 & \xi_{\down\down}(\vec{k})
    \end{pmatrix}.
\end{align}
This matrix actually possesses a higher symmetry; the two spin sectors are completely decoupled, and the probability in each sector is separately conserved. The Bloch matrix cannot be an arbitrary non-Hermitian matrix; it must respect the complete positivity and the trace-preserving property of the original CPTP map. Thus, the CPTP properties impose nontrivial constraints on the symmetry of CPTP maps, which may also have an impact on the topological classification.

\subsection{Classes A, AI$^\dag$, AII$^\dag$, and psH: Feedback with unitary symmetry\label{sec_QND}}

\subsubsection{Class A\label{sec_A}}

Since a CPTP map always has modular conjugation symmetry, a CPTP map in a class without $K$ symmetry requires unitary symmetry by which the Bloch matrix can be block diagonalized, as explained in Sec.~\ref{sec_tenfold}. Here, we explicitly construct such examples of feedback control with unitary symmetry. 

Let $\{\hat{M}_m\}$ be measurement operators. We assume that the measurement operators commute with an observable $\hat{C}$: $[\hat{M}_m,\hat{C}]=0\ (\forall m)$. Such measurement operators describe a quantum nondemolition (QND) measurement for the observable $\hat{C}$ in the sense that all the moments of $\hat{C}$ before and after the measurement stay the same \cite{BreuerPetruccione}, that is,
\begin{align}
    \sum_m\mathrm{Tr}[\hat{C}^r\hat{M}_m\hat{\rho} \hat{M}_m^\dag]=&\sum_m\mathrm{Tr}[\hat{M}_m^\dag \hat{M}_m\hat{C}^r\hat{\rho}]\notag\\
    =&\mathrm{Tr}[\hat{C}^r\hat{\rho}]
\end{align}
for arbitrary $\hat{\rho}$ and $r\in\mathbb{N}$.
We further assume that the feedback Hamiltonian $\hat{H}_m$ commutes with $\hat{C}$, i.e., $[\hat{H}_m,\hat{C}]=0$ so that $\hat{C}$ is conserved during feedback control. Then, it follows that the Kraus operator $\hat{K}_m=\hat{U}_m\hat{M}_m$ with $\hat{U}_m=\exp[-i\hat{H}_m\tau]$ commutes with $\hat{C}$: $[\hat{K}_m,\hat{C}]=0\ (\forall m)$.
Thus, the CPTP map for this feedback control has strong symmetry (see Sec.~\ref{sec_unitary_sym}).

Because of strong symmetry, the CPTP map for feedback control can be block diagonalized according to the symmetry eigenvalues for the ket and bra spaces. Let
\begin{equation}
    \hat{C}=\sum_c c\hat{Q}_c
\end{equation}
be the spectral decomposition of $\hat{C}$, where $c$ is an eigenvalue of $\hat{C}$ and $\hat{Q}_c$ is the projection operator onto the corresponding eigenspace $\mathcal{H}_c$. 
Then, the doubled Hilbert space is decomposed as
\begin{align}
    \mathcal{H}\otimes\mathcal{H}=\bigoplus_{c,c'}\mathcal{H}_c\otimes \mathcal{H}_{c'}.
\end{align}
Accordingly, the CPTP map can be decomposed into blocks that act on subspaces with different symmetry eigenvalues. Specifically, we have
\begin{align}
    \tilde{\mathcal{E}}=\sum_m\hat{K}_m\otimes \hat{K}_m^*=\sum_{c,c'}\tilde{\mathcal{E}}_{c,c'},
\end{align}
where
\begin{align}
    \tilde{\mathcal{E}}_{c,c'}:=\sum_{m} \hat{K}_m\hat{Q}_c\otimes \hat{K}_m^*\hat{Q}_{c'}^*
\end{align}
acts nontrivially only on the subspace $\mathcal{H}_{c}\otimes\mathcal{H}_{c'}$.

The BL symmetry classes are examined for each block diagonalized matrix $\tilde{\mathcal{E}}_{c,c'}$. A crucial observation here is that every block does not necessarily have the modular conjugation symmetry since
\begin{equation}
    \tilde{J}\tilde{\mathcal{E}}_{c,c'}\tilde{J}^{-1}=\tilde{\mathcal{E}}_{c',c}.
\end{equation}
In particular, $\tilde{\mathcal{E}}_{c,c'}$ with $c\neq c'$ does not possess modular conjugation symmetry, but it is related to $\tilde{\mathcal{E}}_{c',c}$ by this symmetry. 
Thus, it belongs to class A if it does not have any other symmetry.

\subsubsection{Class AI$^\dag$}

The other classes (psH, AI$^\dag$, and AII$^\dag$) without $K$ symmetry include $C$ or $Q$ symmetry. These symmetries are realized with feedback protocols with two-point measurements similar to those in Sec.~\ref{sec_helical}. Specifically, we consider a spin-$1/2$ particle with two sublattice degrees of freedom, where a quantum state at site $\vec{j}$ with sublattice $a=A,B$ and spin $\sigma=\up,\down$ is denoted by $\ket{\vec{j},a,\sigma}$. We perform a projective position measurement on this particle and assume that this measurement does not disturb the spin state of the particle. Thus, a measurement outcome is specified only by the site index $\vec{j}$ and the sublattice index $a$, and the measurement operator is given by
\begin{equation}
    \hat{M}_{\vec{j},a}=\ket{\vec{j},a,\up}\bra{\vec{j},a,\up}+\ket{\vec{j},a,\down}\bra{\vec{j},a,\down}.
    \label{eq_QND_meas}
\end{equation}
The measurement operator satisfies $[\hat{M}_{\vec{j},a},\hat{S}^z]=0$, where
\begin{equation}
    \hat{S}^z=\frac{1}{2}\sum_{\vec{i},a}(\ket{\vec{i},a,\up}\bra{\vec{i},a,\up}-\ket{\vec{i},a,\down}\bra{\vec{i},a,\down})
\end{equation}
is the magnetization. A physical example of such a measurement is given by the position measurement of an atom with nuclear spin degrees of freedom, in which the measurement process does not alter the nuclear-spin states.

We consider a CPTP map
\begin{align}
    \mathcal{E}(\hat{\rho})=&\sum_{m_1,m_2}\hat{K}_{m_1,m_2}\hat{\rho} \hat{K}_{m_1,m_2}^\dag,
    \label{eq_2PM_addFB}
\end{align}
where $\hat{K}_{m_1,m_2}=\hat{M}_{m_2}\hat{U}_{m_1}\hat{M}_{m_1}$ is the Kraus operator, $\hat{M}_m$ is the measurement operator \eqref{eq_QND_meas} corresponding to a measurement outcome $m=(\vec{j},a)$, and $\hat{U}_{m}$ is a unitary operator for feedback operation. We assume that $[\hat{U}_m,\hat{S}^z]=0$. As in the case of a single measurement in Sec.~\ref{sec_A}, the CPTP map can be decomposed as
\begin{align}
    \tilde{\mathcal{E}}=&\sum_{m_1,m_2}\hat{K}_{m_1,m_2}\otimes \hat{K}_{m_1,m_2}^*\notag\\
    =&\tilde{\mathcal{E}}_{\up\up}+\tilde{\mathcal{E}}_{\up\down}+\tilde{\mathcal{E}}_{\down\up}+\tilde{\mathcal{E}}_{\down\down},
\end{align}
where
\begin{align}
    \tilde{\mathcal{E}}_{\sigma\sigma'}=\sum_{\vec{j}_1,a_1,\vec{j}_2,a_2}& \hat{M}_{\vec{j}_2,a_2,\sigma}\hat{U}_{\vec{j}_1,a_1}\hat{M}_{\vec{j}_1,a_1,\sigma}\notag\\
    &\otimes \hat{M}_{\vec{j}_2,a_2,\sigma'}^*\hat{U}_{\vec{j}_1,a_1}^*\hat{M}_{\vec{j}_1,a_1,\sigma'}^*,
\end{align}
and
\begin{equation}
    \hat{M}_{\vec{j},a,\sigma}:=\ket{\vec{j},a,\sigma}\bra{\vec{j},a,\sigma}.
    \label{eq_spin_resolved_meas_op}
\end{equation}

Similarly to the analysis in Sec.~\ref{sec_sym_proj_example}, the Bloch matrix for $\tilde{\mathcal{E}}_{\sigma\sigma'}$ has a block-diagonal structure \eqref{eq_Bloch_block_2pt}, and the truncated Bloch matrix is given by
\begin{equation}
    X_{\mathrm{trunc}}^{(\sigma\sigma')}(\vec{k})=
    \begin{pmatrix}
        \xi_{AA}^{(\sigma\sigma')}(\vec{k}) & \xi_{AB}^{(\sigma\sigma')}(\vec{k}) \\
        \xi_{BA}^{(\sigma\sigma')}(\vec{k}) & \xi_{BB}^{(\sigma\sigma')}(\vec{k})
    \end{pmatrix},
    \label{eq_QND_trunc}
\end{equation}
where
\begin{align}
    \xi_{ab}^{(\sigma\sigma')}(\vec{k})=\frac{1}{N_{\mathrm{cell}}}\sum_{\vec{j}_1,\vec{j}_2}q^{(\sigma\sigma')}(\vec{j}_2,a|\vec{j}_1,b) e^{-i\vec{k}\cdot(\vec{R}_{\vec{j}_1}-\vec{R}_{\vec{j}_2})}
    \label{eq_QND_xi}
\end{align}
with
\begin{align}
    q^{(\sigma\sigma')}(\vec{j}_2,a|\vec{j}_1,b)=(\hat{U}_{\vec{j}_1,b})_{\vec{j}_2,a,\sigma;\vec{j}_1,b,\sigma}(\hat{U}_{\vec{j}_1,b})_{\vec{j}_2,a,\sigma';\vec{j}_1,b,\sigma'}^*.
    \label{eq_QND_q}
\end{align}

For simplicity, here we focus on the truncated Bloch matrix \eqref{eq_QND_trunc} with $\sigma=\up$ and $\sigma'=\down$. It has $C$ symmetry with $\eta_C=+1$, i.e.,
\begin{equation}
    [X_{\mathrm{trunc}}^{(\up\down)}(-\vec{k})]^T=X_{\mathrm{trunc}}^{(\up\down)}(\vec{k})
\end{equation}
if
    \begin{gather}
        \xi_{ab}^{(\up\down)}(-\vec{k})=\xi_{ba}^{(\up\down)}(\vec{k}).
        \label{eq_QND_reciprocity_xicond}
    \end{gather}
A condition for Eq.~\eqref{eq_QND_reciprocity_xicond} is given by
    \begin{gather}
        q^{(\up\down)}(\vec{j}_2,a|\vec{j}_1,b)=q^{(\up\down)}(\vec{j}_1,b|\vec{j}_2,a).
        \label{eq_QND_reciprocity}
    \end{gather}
This condition is interpreted as reciprocity in the feedback process. Since the truncated Bloch matrix $X_{\mathrm{trunc}}^{(\up\down)}(\vec{k})$ does not have $K$ symmetry, it belongs to symmetry class AI$^\dag$.

\subsubsection{Class AII$^\dag$\label{sec_AIIdag}}

Symmetry class AII$^\dag$ involves $C$ symmetry with $\eta_C=-1$. 
Provided that
\begin{subequations}    
    \begin{align}
        \xi_{AA}^{(\up\down)}(-\vec{k})=&\xi_{BB}^{(\up\down)}(\vec{k}),\label{eq_QND_xicond1}\\
        \xi_{AB}^{(\up\down)}(-\vec{k})=&-\xi_{AB}^{(\up\down)}(\vec{k}),\label{eq_QND_xicond2}\\
        \xi_{BA}^{(\up\down)}(-\vec{k})=&-\xi_{BA}^{(\up\down)}(\vec{k}),\label{eq_QND_xicond3}
    \end{align}
    \end{subequations}
the truncated Bloch matrix $X_{\mathrm{trunc}}^{(\up\down)}(\vec{k})$ has $C$ symmetry with $\eta_C=-1$:
\begin{equation}
    (-i\sigma_y)[X_{\mathrm{trunc}}^{(\up\down)}(-\vec{k})]^T(-i\sigma_y)^{-1}=X_{\mathrm{trunc}}^{(\up\down)}(\vec{k}).
\end{equation}
The conditions for this symmetry are given by
\begin{subequations}
    \begin{align}
        q^{(\up\down)}(\vec{j}_2,A|\vec{j}_1,A)=&q^{(\up\down)}(\vec{j}_1,B|\vec{j}_2,B),\label{eq_QND_qcond1}\\
        q^{(\up\down)}(\vec{j}_2,A|\vec{j}_1,B)=&-q^{(\up\down)}(\vec{j}_1,A|\vec{j}_2,B),\label{eq_QND_qcond2}\\
        q^{(\up\down)}(\vec{j}_2,B|\vec{j}_1,A)=&-q^{(\up\down)}(\vec{j}_1,B|\vec{j}_2,A).\label{eq_QND_qcond3}
    \end{align}
\end{subequations}
Here, in sharp contrast to Eqs.~\eqref{eq_cond_AIIdag2} and \eqref{eq_cond_AIIdag3} in Sec.~\ref{sec_sym_proj_example}, the conditions \eqref{eq_QND_xicond2} and \eqref{eq_QND_xicond3} do not lead to $q^{(\up\down)}(\vec{j}_2,A|\vec{j}_1,B)=q^{(\up\down)}(\vec{j}_2,B|\vec{j}_1,A)=0$ since $q^{(\up\down)}(\vec{j}_2,a|\vec{j}_1,b)$ is not positive semidefinite.

\subsubsection{Class psH}

Finally, we consider the condition for $Q$ symmetry, which is also referred to as pseudo-Hermiticity (psH) \cite{Kawabata19}. An example of psH of the truncated Bloch matrix is given by
\begin{equation}
    \sigma_x[X_{\mathrm{trunc}}^{(\up\down)}(\vec{k})]^\dag\sigma_x^{-1}=X_{\mathrm{trunc}}^{(\up\down)}(\vec{k}),
    \label{eq_QND_pH}
\end{equation}
which is equivalent to
\begin{subequations}    
\begin{align}
    [\xi_{AA}^{(\up\down)}(\vec{k})]^*=&\xi_{BB}^{(\up\down)}(\vec{k}),\label{eq_QND_pH_xicond1}\\
    [\xi_{AB}^{(\up\down)}(\vec{k})]^*=&\xi_{AB}^{(\up\down)}(\vec{k}),\label{eq_QND_pH_xicond2}\\
    [\xi_{BA}^{(\up\down)}(\vec{k})]^*=&\xi_{BA}^{(\up\down)}(\vec{k}).\label{eq_QND_pH_xicond3}
\end{align}
\end{subequations}
Conditions \eqref{eq_QND_pH_xicond1}-\eqref{eq_QND_pH_xicond3} are satisfied if
\begin{subequations}
    \begin{align}
        [q^{(\up\down)}(\vec{j}_2,A|\vec{j}_1,A)]^*=&q^{(\up\down)}(\vec{j}_1,B|\vec{j}_2,B),\label{eq_QND_pH_qcond1}\\
        [q^{(\up\down)}(\vec{j}_2,A|\vec{j}_1,B)]^*=&q^{(\up\down)}(\vec{j}_1,A|\vec{j}_2,B),\label{eq_QND_pH_qcond2}\\
        [q^{(\up\down)}(\vec{j}_2,B|\vec{j}_1,A)]^*=&q^{(\up\down)}(\vec{j}_1,B|\vec{j}_2,A).\label{eq_QND_pH_qcond3}
    \end{align}
\end{subequations}
In terms of the feedback unitary operators, sufficient conditions for Eqs.~\eqref{eq_QND_pH_qcond1}-\eqref{eq_QND_pH_qcond3} are given by
\begin{subequations}
    \begin{align}
        (\hat{U}_{\vec{j}_1,A})_{\vec{j}_2,A,\sigma;\vec{j}_1,A,\sigma}^*=&(\hat{U}_{\vec{j}_2,B})_{\vec{j}_1,B,\sigma;\vec{j}_2,B,\sigma},\label{eq_QND_pH_Ucond1}\\
        (\hat{U}_{\vec{j}_1,B})_{\vec{j}_2,A,\sigma;\vec{j}_1,B,\sigma}^*=&(\hat{U}_{\vec{j}_2,B})_{\vec{j}_1,A,\sigma;\vec{j}_2,B,\sigma},\label{eq_QND_pH_Ucond2}\\
        (\hat{U}_{\vec{j}_1,A})_{\vec{j}_2,B,\sigma;\vec{j}_1,A,\sigma}^*=&(\hat{U}_{\vec{j}_2,A})_{\vec{j}_1,B,\sigma;\vec{j}_2,A,\sigma}.\label{eq_QND_pH_Ucond3}
    \end{align}    
\end{subequations}
The psH in Eq.~\eqref{eq_QND_pH} is understood as a consequence of the modified reciprocity expressed by Eqs.~\eqref{eq_QND_pH_Ucond1}-\eqref{eq_QND_pH_Ucond3}.

\subsection{Classes AII, AII $+$ psH$_+$, and AII $+$ psH$_-$: Feedback with $K$ symmetry and $\eta_K=-1$}

The remaining symmetry classes include $K$ symmetry with $\eta_K=-1$. To implement this symmetry, we consider a system with spin ($\sigma=\up,\down$), orbit ($\ell=s,p$), and sublattice ($a=A,B$) degrees of freedom. We perform a two-point measurement described by measurement operators
\begin{equation}
    \hat{M}_{\vec{j},a,\ell}=\ket{\vec{j},a,\ell,\up}\bra{\vec{j},a,\ell,\up}+\ket{\vec{j},a,\ell,\down}\bra{\vec{j},a,\ell,\down}.
\end{equation}
We note that these measurement operators do not alter the spin degrees of freedom. We assume that feedback Hamiltonians conserve the magnetization. Then, a CPTP map for feedback control with a two-point measurement is given by the form \eqref{eq_2PM_addFB}, where the measurement outcomes are labeled by $m_n=(\vec{j}_n,a_n,\ell_n)\ (n=1,2)$. The truncated Bloch matrix is obtained in a manner similar to that in Sec.~\ref{sec_QND}.

We impose $K$ symmetry with $\eta_K=-1$ on the truncated Bloch matrix as
\begin{equation}
    (-i\tau_y)[X_{\mathrm{trunc}}^{(\up\down)}(-\vec{k})]^*(-i\tau_y)^{-1}=X_{\mathrm{trunc}}^{(\up\down)}(\vec{k}),
    \label{eq_AII}
\end{equation}
where $\tau_y$ is the Pauli matrix for the orbital degrees of freedom. This symmetry is equivalent to the conditions
\begin{subequations}    
    \begin{align}
        [\xi_{a,p;b,p}^{(\up\down)}(-\vec{k})]^*=&\xi_{a,s;b,s}^{(\up\down)}(\vec{k}),\label{eq_AII_xicond1}\\
        [\xi_{a,s;b,p}^{(\up\down)}(-\vec{k})]^*=&-\xi_{a,p;b,s}^{(\up\down)}(\vec{k}),\label{eq_AII_xicond2}
    \end{align}
\end{subequations}
where
\begin{align}
    \xi_{a,\ell;b,\ell'}^{(\sigma\sigma')}(\vec{k})=\frac{1}{N_{\mathrm{cell}}}\sum_{\vec{j}_1,\vec{j}_2}q^{(\sigma\sigma')}(\vec{j}_2,a,\ell|\vec{j}_1,b,\ell') e^{-i\vec{k}\cdot(\vec{R}_{\vec{j}_1}-\vec{R}_{\vec{j}_2})}
    \label{eq_AII_xi}
\end{align}
with
\begin{align}
    q^{(\sigma\sigma')}(\vec{j}_2,a,\ell|\vec{j}_1,b,\ell') =&(\hat{U}_{\vec{j}_1,b,\ell'})_{\vec{j}_2,a,\ell,\sigma;\vec{j}_1,b,\ell',\sigma}\notag\\
    &\times(\hat{U}_{\vec{j}_1,b,\ell'})_{\vec{j}_2,a,\ell,\sigma';\vec{j}_1,b,\ell',\sigma'}^*.
    \label{eq_AII_q}
\end{align}
The conditions \eqref{eq_AII_xicond1} and \eqref{eq_AII_xicond2} are satisfied if
\begin{subequations}
    \begin{align}
        [q^{(\up\down)}(\vec{j}_1,a,p|\vec{j}_2,b,p)]^*=&q^{(\up\down)}(\vec{j}_1,a,s|\vec{j}_2,b,s),\label{eq_AII_qcond1}\\
        [q^{(\up\down)}(\vec{j}_1,a,s|\vec{j}_2,b,p)]^*=&-q^{(\up\down)}(\vec{j}_1,a,p|\vec{j}_2,b,s),\label{eq_AII_qcond2}
    \end{align}
\end{subequations}
which represent a type of symmetry in the orbital space. 

If we only impose the symmetry \eqref{eq_AII}, the truncated Bloch matrix belongs to class AII. If we further impose $C$ symmetry by using the sublattice degrees of freedom as in Sec.~\ref{sec_QND}, we can also realize the truncated Bloch matrix in class AII $+$ psH$_+$ and class AII $+$ psH$_-$.

\section{topological Maxwell demons\label{sec_demons}}

In this section, building on the symmetry classification of quantum channels developed in Sec.~\ref{sec_tenfold}, we present various models of topological feedback control. In Secs.~\ref{sec_SSH}-\ref{sec_Z2}, we focus on quantum feedback control of a particle in a 1D lattice and construct topological quantum channels protected by their point-gap topology and symmetry. According to the general topological classification of non-Hermitian operators in Ref.~\cite{Kawabata19}, a non-Hermitian operator in a 1D system can host nontrivial point-gap topology in classes A, AI, AI $+$ psH$_-$, AII$^\dag$, and AII. 
In Sec.~\ref{sec_SSH}, we consider topological feedback control in class AI, where we can combine particle-hole symmetry with modular conjugation symmetry to construct a unitary symmetry. In Sec.~\ref{sec_helical}, we construct a model of topological feedback control in class AI $+$ psH$_-$ and demonstrate that it shows helical spin transport protected by $C$ symmetry. In Sec.~\ref{sec_QND_demon}, we consider topological feedback control in class A, where strong unitary symmetry enables us to block diagonalize a CPTP map. In Sec.~\ref{sec_Z2}, we impose strong unitary symmetry and $C$ symmetry on a CPTP map to construct topological feedback control in class AII$^\dag$, which exhibits the $\mathbb{Z}_2$ skin effect.
Finally, in Sec.~\ref{sec_Chern}, we show an example of quantum feedback control with nontrivial line-gap topology in two dimensions.

\subsection{Su-Schrieffer-Heeger demon\label{sec_SSH}}

\begin{figure}[t]
    \includegraphics[width=8.5cm]{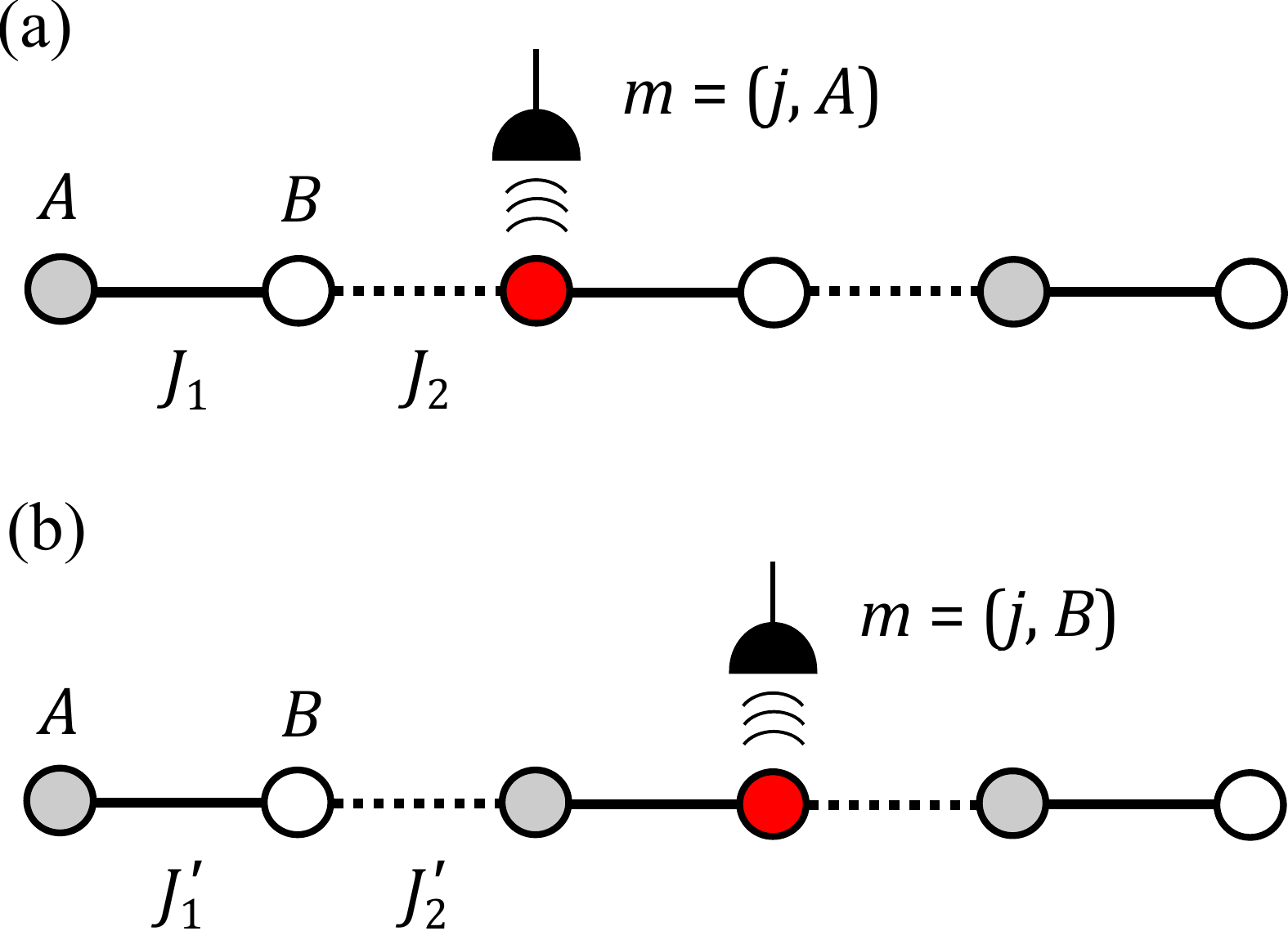}
    \caption{Schematic illustration of SSH demon. We first perform a projective measurement of the position and the sublattice. In panel (a), we show that if a particle is detected in sublattice $A$, we perform feedback control governed by the Hamiltonian $H_A$. Panel (b) shows that if a particle is detected in sublattice $B$, we perform feedback control governed by the Hamiltonian $H_B$.}
    \label{fig_SSH}
\end{figure}

First, we present a model of topological feedback control with $K$ symmetry with $\eta_K=+1$ that is distinct from the modular conjugation symmetry. This model provides a concrete example for the discussion presented in Sec.~\ref{sec_AI}. We consider a 1D lattice with two sublattices. Each unit cell is labeled by an index $j$ and each sublattice is labeled by $a=A,B$. We first perform a projective position measurement, which completely resolves the sublattice structure. A measurement outcome is specified by $m=(j,a)$, and the corresponding measurement operator is given by $\hat{M}_m=\ket{j,a}\bra{j,a}$. Next, we perform feedback control with Hamiltonian $\hat{H}_a\ (a=A,B)$, which depends only on the measurement outcome of the sublattice index. The feedback Hamiltonians are given by the Su-Schrieffer-Heeger (SSH) model \cite{Su79}:
\begin{align}
    \hat{H}_{A}=&-J_1\sum_j(\ket{j,A}\bra{j,B}+\mathrm{H.c.})\notag\\
    &-J_2\sum_j(\ket{j+1,A}\bra{j,B}+\mathrm{H.c.}),\\
    \hat{H}_{B}=&-J_1^\prime\sum_j(\ket{j,A}\bra{j,B}+\mathrm{H.c.})\notag\\
    &-J_2^\prime\sum_j(\ket{j+1,A}\bra{j,B}+\mathrm{H.c.}),
\end{align}
where $J_1,J_2,J_1^\prime,J_2^\prime\in\mathbb{R}$ are the hopping amplitudes. See Fig.~\ref{fig_SSH} for a schematic illustration of the feedback protocol. We note that the SSH model has time-reversal symmetry
\begin{subequations}
\begin{equation}
    \hat{H}_a^*=\hat{H}_a,
    \label{eq_SSH_TRS}
\end{equation}
particle-hole symmetry
\begin{equation}
    \hat{\Sigma}_z\hat{H}_a^*\hat{\Sigma}_z=-\hat{H}_a,
    \label{eq_SSH_PHS}
\end{equation}
and chiral symmetry
\begin{equation}
    \hat{\Sigma}_z\hat{H}_a\hat{\Sigma}_z=-\hat{H}_a,
    \label{eq_SSH_CS}
\end{equation}
\end{subequations}
where
\begin{equation}
    \hat{\Sigma}_z=\sum_j(\ket{j,A}\bra{j,A}-\ket{j,B}\bra{j,B}).
\end{equation}
Note that $\hat{\Sigma}_z^2=\hat{I}$.
Consequently, the unitary time-evolution operator $\hat{U}_a=\exp[-i\hat{H}_a\tau]$ has the following symmetries (see Appendix \ref{sec_sym_corresp}):
\begin{subequations}
    \begin{gather}
        \hat{U}_a^T=\hat{U}_a,\label{eq_SSH_TRS_U}\\
        \hat{\Sigma}_z\hat{U}_a^*\hat{\Sigma}_z=\hat{U}_a,\label{eq_SSH_PHS_U}
    \end{gather}
    and
    \begin{gather}
        \hat{\Sigma}_z\hat{U}_a^\dag\hat{\Sigma}_z=\hat{U}_a,\label{eq_SSH_CS_U}
    \end{gather}
\end{subequations}
which follow from Eqs.~\eqref{eq_SSH_TRS}-\eqref{eq_SSH_CS}, respectively. 
Among the symmetries \eqref{eq_SSH_TRS_U}-\eqref{eq_SSH_CS_U}, only particle-hole symmetry \eqref{eq_SSH_PHS_U} leads to the symmetry of a quantum channel for this feedback control. 
Since the measurement operator satisfies
\begin{equation}
    \hat{\Sigma}_z\hat{M}_m\hat{\Sigma}_z=\hat{M}_m,
\end{equation}
particle-hole symmetry \eqref{eq_SSH_PHS_U} leads to $K$ symmetry
\begin{equation}
    (\hat{\Sigma}_z\otimes\hat{\Sigma}_z^*)\sum_m(\hat{K}_m^*\otimes \hat{K}_m)(\hat{\Sigma}_z\otimes\hat{\Sigma}_z^*)=\sum_m(\hat{K}_m\otimes \hat{K}_m^*)
    \label{eq_SSH_Ksym}
\end{equation}
of the CPTP map, where $\hat{K}_m=\hat{K}_{(j,a)}=\hat{U}_a\hat{M}_m$ is the Kraus operator.

As mentioned in Sec.~\ref{sec_tenfold}, the CPTP map has modular conjugation symmetry \eqref{eq_mod_conj_sym}, which gives another $K$ symmetry different from \eqref{eq_SSH_Ksym}. Combining the two $K$ symmetries, we obtain the unitary symmetry of the CPTP map as
\begin{equation}
    (\hat{\Sigma}_z\otimes\hat{\Sigma}_z^*)\tilde{S}\sum_m(\hat{K}_m\otimes \hat{K}_m^*)\tilde{S}^{-1}(\hat{\Sigma}_z\otimes\hat{\Sigma}_z^*)=\sum_m(\hat{K}_m\otimes \hat{K}_m^*),
    \label{eq_SSH_weak_sym}
\end{equation}
where $\tilde{S}$ is the swap operator defined by Eq.~\eqref{eq_swap}. Hence, the CPTP map can be block diagonalized according to the symmetry eigenvalues of the unitary symmetry \eqref{eq_SSH_weak_sym}. The eigenvalues of $(\hat{\Sigma}_z\otimes\hat{\Sigma}_z^*)\tilde{S}$ are given by $\pm 1$. The eigenvectors for the eigenvalue $+1$ are given by
\begin{subequations}
    \begin{gather}
        \ket{j,A}\otimes\ket{j,A},\\
        \ket{j,B}\otimes\ket{j,B},\\
        \frac{1}{\sqrt{2}}(\ket{j,A}\otimes\ket{j',B}+\ket{j',B}\otimes\ket{j,A}),
    \end{gather}
\end{subequations}
and those for the eigenvalue $-1$ are given by
\begin{align}
    \frac{1}{\sqrt{2}}(\ket{j,A}\otimes\ket{j',B}-\ket{j',B}\otimes\ket{j,A}).
\end{align}
In each eigenspace of the operator $(\hat{\Sigma}_z\otimes\hat{\Sigma}_z^*)\tilde{S}$, modular conjugation symmetry \eqref{eq_mod_conj_sym} and particle-hole symmetry \eqref{eq_SSH_Ksym} are equivalent to each other since $(\hat{\Sigma}_z\otimes\hat{\Sigma}_z^*)=\pm \tilde{S}$. Thus, the topological characterization of a CPTP map in those eigenspaces reduces to that of a CPTP map with a single $K$ symmetry. Thus, each symmetry sector of the CPTP map belongs to class AI.

\begin{figure}[t]
    \includegraphics[width=8.5cm]{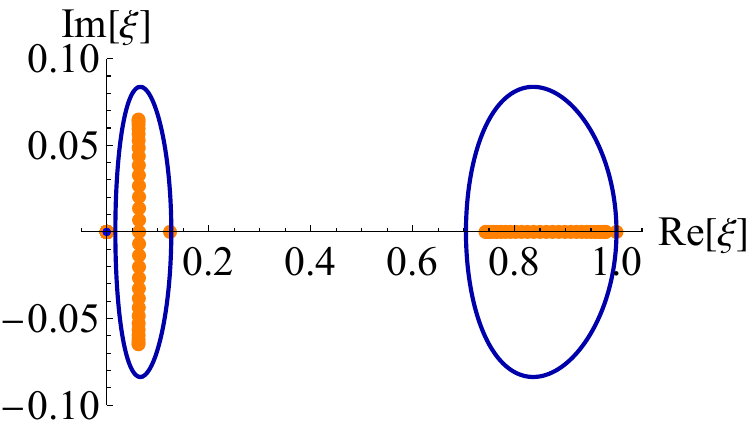}
    \caption{Eigenspectrum of a quantum channel for the SSH demon. Blue curves and the blue dot at the origin constitute the eigenspectrum under the PBC in the infinite-size limit, and orange dots constitute the eigenspectrum under the OBC with the number of unit cells $L=30$. The hopping amplitudes are set to $J_1=1,J_2=0.5,J_1'=0.4$, and $J_2'=0.4$. The duration of feedback is $\tau=1$.}
    \label{fig_SSH_spec}
\end{figure}

Figure \ref{fig_SSH_spec} shows the eigenspectrum of a CPTP map for feedback control with the SSH model under the PBC as well as that under the OBC. The eigenspectrum under the PBC forms two loops in the complex plane, each of which has winding number $-1$ for a point gap inside the loop. This nontrivial point-gap topology makes the eigenspectrum under the OBC drastically different from that in the PBC, leading to the non-Hermitian skin effect induced by the SSH demon.

\subsection{Helical Maxwell demon\label{sec_helical}}
\begin{figure*}
    \includegraphics[width=17cm]{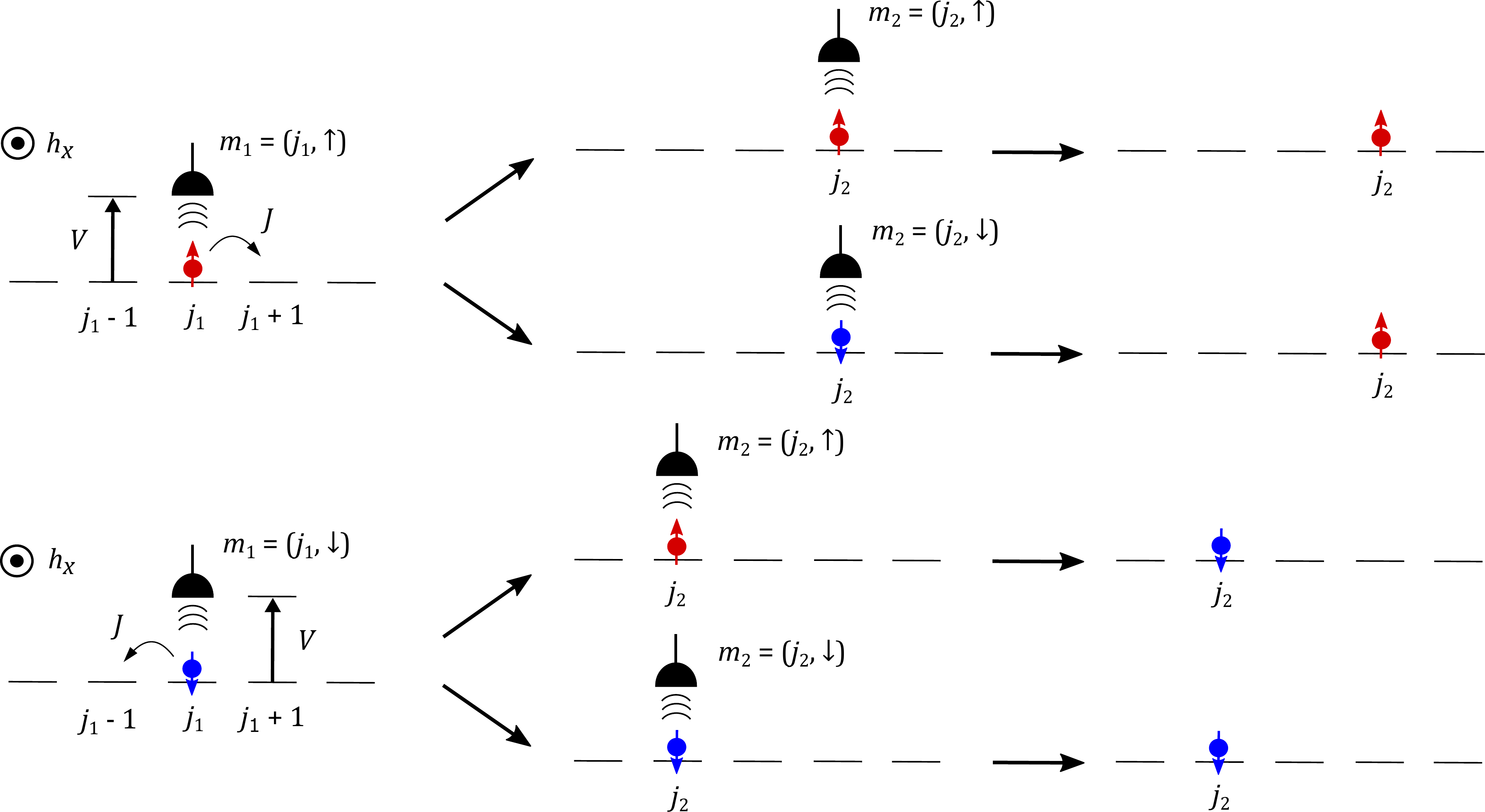}
    \caption{Schematic illustration of the helical Maxwell demon for a spin-$1/2$ particle on a 1D lattice. We first perform a projective measurement of the position and spin of the particle. If we obtain an outcome $m_1=(j_1,\up)$ ($m_1=(j_1,\down)$), we raise the potential at site $j_1-1$ ($j+1$) and let the particle undergo unitary time evolution under the feedback Hamiltonian given in Eq.~\eqref{eq_helical_hamilup} [Eq.~\eqref{eq_helical_hamildown}]. Next, we perform the second projective measurement of the position and spin. If the measurement outcome of spin is different from that of the first measurement, we apply a $\pi$ pulse of a magnetic field to flip the spin with unit probability. Thus, helical spin transport is achieved.}
    \label{fig_model_helical}
\end{figure*}

Next, we construct a model of symmetry-protected topological feedback control that belongs to class AI $+$ psH$_-$. We consider a spin-$1/2$ particle on a 1D lattice with $L$ sites. The Hilbert space of this system is spanned by a basis $\{\ket{j,\sigma}\}$ ($j=1,\cdots,L$, $\sigma=\up,\down$). We consider feedback control of this particle with a two-point measurement scheme (see Fig.~\ref{fig_model_helical} for a schematic illustration). We first perform a projective measurement of the position and spin with the projection operators $\hat{P}_{j,\sigma}=\ket{j,\sigma}\bra{j,\sigma}$. Let $m_1=(j_1,\sigma_1)$ be an outcome of this measurement. If we obtain a measurement outcome $(j_1,\up)$, we raise a spin-independent potential at site $j_1-1$ to prevent the particle from going to the left. If we obtain a measurement outcome $(j_1,\down)$, we raise a spin-independent potential at site $j_1+1$ to prevent the particle from going to the right. The feedback Hamiltonians are given by
\begin{align}
    \hat{H}_{j_1,\up}=&-J\sum_{j}\sum_\sigma(\ket{j,\sigma}\bra{j+1,\sigma}+\ket{j+1,\sigma}\bra{j,\sigma})\notag\\
    &+V\sum_\sigma\ket{j_1-1,\sigma}\bra{j_1-1,\sigma}\notag\\
    &+h_x\sum_j(\ket{j,\up}\bra{j,\down}+\ket{j,\down}\bra{j,\up})
    \label{eq_helical_hamilup}
\end{align}
and
\begin{align}
    \hat{H}_{j_1,\down}=&-J\sum_{j}\sum_\sigma(\ket{j,\sigma}\bra{j+1,\sigma}+\ket{j+1,\sigma}\bra{j,\sigma})\notag\\
    &+V\sum_\sigma\ket{j_1+1,\sigma}\bra{j_1+1,\sigma}\notag\\
    &+h_x\sum_j(\ket{j,\up}\bra{j,\down}+\ket{j,\down}\bra{j,\up}),
    \label{eq_helical_hamildown}
\end{align}
where $J\in\mathbb{R}$ is the hopping amplitude, $V>0$ is the height of the potential, and $h_x$ denotes the strength of a transverse magnetic field that mixes the spin components. Feedback control is described by a unitary operator $\hat{U}_{j_1,\sigma_1}=\exp[-i\hat{H}_{j_1,\sigma_1}\tau]$. After feedback control, we perform the second projective measurement with projection operators $\hat{P}_{j,\sigma}$. Let $m_2=(j_2,\sigma_2)$ be an outcome of this second measurement. If $\sigma_2=\sigma_1$, we do nothing. If $\sigma_2\neq\sigma_1$, we flip the spin of the particle by applying a $\pi$ pulse of the magnetic field. Through this feedback control, a spin-up particle is transported to the right while a spin-down particle is transported to the left, which is analogous to helical edge states in the quantum spin Hall effect \cite{KaneMele05_1,KaneMele05_2,Bernevig06,Wu06,Koenig07}.

The quantum channel for the entire process of this feedback control is given by
\begin{equation}
    \mathcal{E}(\hat{\rho})=\sum_{m_1,m_2}\hat{K}_{m_1,m_2}\hat{\rho} \hat{K}_{m_1,m_2}^\dag,
    \label{eq_CPTP_helical}
\end{equation}
where the Kraus operators are given by
\begin{subequations}
    \begin{align}
        \hat{K}_{(j_1,\up),(j_2,\up)}=&\ket{j_2,\up}\bra{j_2,\up}\hat{U}_{j_1,\up}\ket{j_1,\up}\bra{j_1,\up},\\
        \hat{K}_{(j_1,\down),(j_2,\down)}=&\ket{j_2,\down}\bra{j_2,\down}\hat{U}_{j_1,\down}\ket{j_1,\down}\bra{j_1,\down},\\
        \hat{K}_{(j_1,\up),(j_2,\down)}=&\ket{j_2,\up}\bra{j_2,\down}\hat{U}_{j_1,\up}\ket{j_1,\up}\bra{j_1,\up},\\
        \hat{K}_{(j_1,\down),(j_2,\up)}=&\ket{j_2,\down}\bra{j_2,\up}\hat{U}_{j_1,\down}\ket{j_1,\down}\bra{j_1,\down}.
    \end{align}
\end{subequations}
The Bloch matrix of this quantum channel has the structure in Eq.~\eqref{eq_Bloch_block_2pt} and is given by
\begin{equation}
    X(k)=
    \begin{pmatrix}
        X_{\mathrm{trunc}}(k) & O \\
        O & O
    \end{pmatrix},
\end{equation}
where
\begin{equation}
    X_{\mathrm{trunc}}(k)=
    \begin{pmatrix}
        \xi_{\up\up}(k)+\xi_{\down\up}(k) & 0 \\
        0 & \xi_{\down\down}(k)+\xi_{\up\down}(k)
    \end{pmatrix}
    \label{eq_helical_trunc}
\end{equation}
and
\begin{align}
    \xi_{\sigma\sigma'}(k)=&\frac{1}{L}\sum_{j_1,j_2}|\bra{j_2,\sigma}\hat{U}_{j_1,\sigma'}\ket{j_1,\sigma'}|^2 e^{-ik(j_2-j_1)}.
\end{align}
Note that we can transfer the off-diagonal components in Eq.~\eqref{eq_Bloch_trunc_2pt} to the diagonal components by an additional feedback control after the second measurement. The truncated Bloch matrix has symmetry
\begin{equation}
    (-i\sigma_y)X_{\mathrm{trunc}}^T(-k)(-i\sigma_y)^{-1}=X_{\mathrm{trunc}}(k),
\end{equation}
which corresponds to $C$ symmetry \eqref{eq_TRSdag} of the CPTP map with $\eta_C=-1$, since the relations
\begin{subequations}
    \begin{equation}
        \xi_{\up\up}(-k)=\xi_{\down\down}(k)
        \label{eq_cond_sym_helical1}
    \end{equation}
    and
    \begin{equation}
        \xi_{\up\down}(-k)=\xi_{\down\up}(k)
        \label{eq_cond_sym_helical2}
    \end{equation}
\end{subequations}
hold due to the symmetry of the feedback protocol. 
Thus, the Bloch matrix belongs to class AI $+$ psH$_-$.

\begin{figure}[t]
    \includegraphics[width=8.5cm]{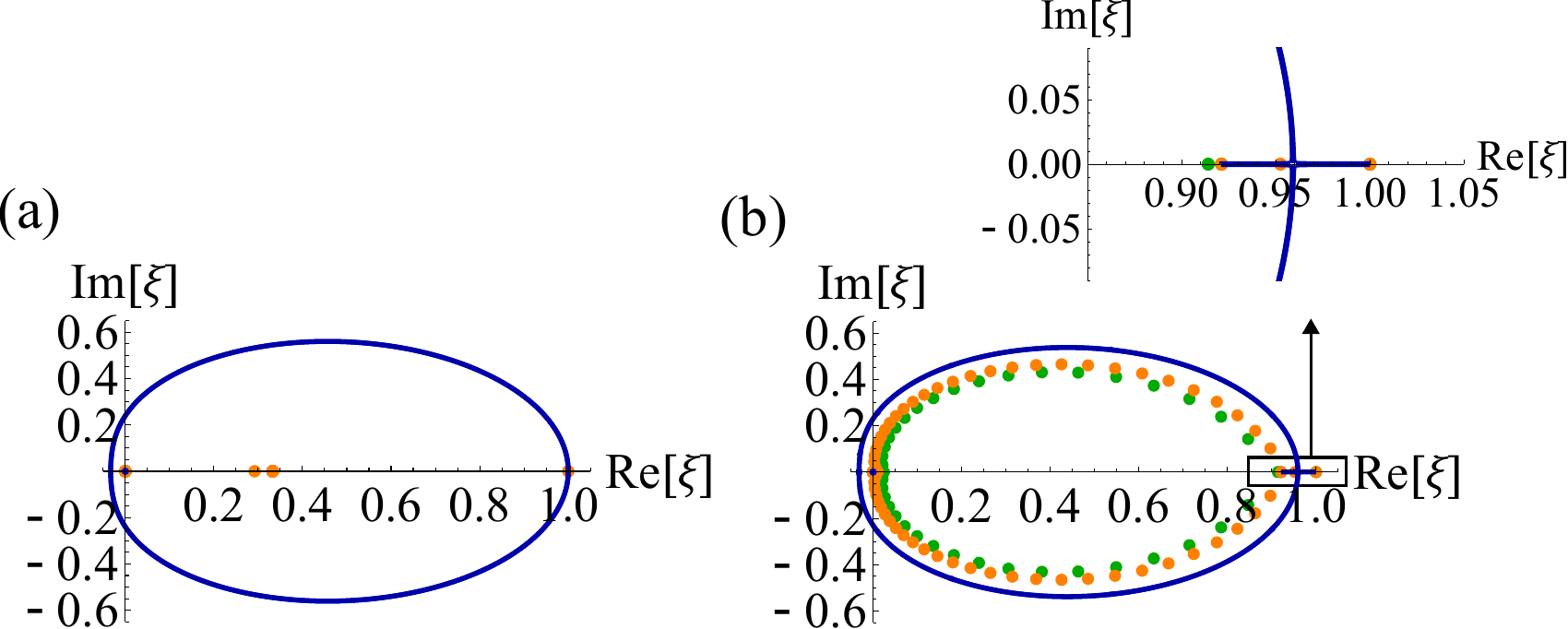}
    \caption{(a) Eigenspectrum of a quantum channel for the helical Maxwell demon with additional feedback performed after the second measurement. (b) Eigenspectrum of a quantum channel for the helical Maxwell demon with no further feedback performed after the second measurement. An enlarged eigenspectrum near the steady state is also shown. Blue curves and the blue dot at the origin constitute the eigenspectrum under the PBC, and orange (green) dots form the eigenspectrum under the OBC with $L=30$ ($L=20$). The hopping amplitude is set to $J=1$, and the transverse magnetic field is set to $h_x=0.2$. The duration of feedback is $\tau=1$.}
    \label{fig_helical_spec}
\end{figure}

From the block-diagonal structure \eqref{eq_helical_trunc}, a topological invariant of the quantum channel for this system is given by the spin winding number
\begin{align}
    w_{\mathrm{spin}}(\xi_{\mathrm{PG}}):=&\frac{1}{2}\Biggl[\int_{-\pi}^\pi\frac{dk}{2\pi i}\partial_k\ln[\xi_{\up\up}(k)+\xi_{\up\down}(k)-\xi_{\mathrm{PG}}]\notag\\
    &-\int_{-\pi}^\pi\frac{dk}{2\pi i}\partial_k\ln[\xi_{\down\down}(k)+\xi_{\down\up}(k)-\xi_{\mathrm{PG}}]\Biggr]\notag\\
    =&\int_{-\pi}^\pi\frac{dk}{2\pi i}\partial_k\ln[\xi_{\up\up}(k)+\xi_{\up\down}(k)-\xi_{\mathrm{PG}}],
\label{eq_winding_helical}
\end{align}
where $\xi_{\mathrm{PG}}$ denotes the location of a point gap and we use Eqs.~\eqref{eq_cond_sym_helical1} and \eqref{eq_cond_sym_helical2} in deriving the second equality. In Fig.~\ref{fig_helical_spec}(a), we show the eigenspectra of the CPTP map under the PBC and the OBC. Each eigenvalue is doubly degenerate due to the Kramers degeneracy ensured by $C$ symmetry with $\eta_C=-1$ \cite{Kawabata19}. The loop structure of the eigenspectrum under the PBC leads to the winding number $w_{\mathrm{spin}}(\xi_{\mathrm{PG}})=-1$ for a point gap $\xi_{\mathrm{PG}}$ inside the loop. Reflecting the nontrivial point-gap topology, the eigenspectrum under the OBC is drastically changed from that under the PBC, and the right eigenmodes of the spin-up (spin-down) component are localized near the right (left) edge of the system. This phenomenon is a manifestation of the symmetry-protected non-Hermitian skin effect \cite{Okuma20}, where the Kramers doublets are localized near the opposite edges. Physically, the symmetry-protected non-Hermitian skin effect is a consequence of the helical spin transport due to feedback control. Thus, the helical Maxwell demon considered here achieves robust feedback-assisted spin transport protected by nontrivial topology of a quantum channel.

The point-gap topology of the helical Maxwell demon is protected by $C$ symmetry. To see this property, we show the eigenspectrum of another CPTP map,
\begin{equation}
    \mathcal{E}'(\hat{\rho})=\sum_{(j_1,\sigma_1),(j_2,\sigma_2)}\hat{P}_{j_2,\sigma_2}\hat{U}_{j_1,\sigma_1}\hat{P}_{j_1,\sigma_1}\hat{\rho} \hat{P}_{j_1,\sigma_1}\hat{U}_{j_1,\sigma_1}^\dag \hat{P}_{j_2,\sigma_2},
\end{equation}
where the difference from Eq.~\eqref{eq_CPTP_helical} is the absence of feedback after the second measurement. The truncated Bloch matrix of this CPTP map is given by
\begin{equation}
    X_{\mathrm{trunc}}'(k)=
    \begin{pmatrix}
        \xi_{\up\up}(k) & \xi_{\up\down}(k) \\
        \xi_{\down\up}(k) & \xi_{\down\down}(k)
    \end{pmatrix},
    \label{eq_helical_trunc2}
\end{equation}
where the hybridization between spin components implies the absence of $C$ symmetry with $\eta_C=-1$. The eigenspectra of the CPTP map $\mathcal{E}'$ are shown in Fig.~\ref{fig_helical_spec}(b). Because of the lack of $C$ symmetry, the Kramers degeneracy of the eigenvalues no longer holds. In fact, the two degenerate loops of eigenvalues for $h_x=0$ are hybridized due to the real-complex transition near the steady state. As a consequence of spin mixing, the non-Hermitian skin effect disappears under the OBC and hence the topology becomes trivial. Thus, the symmetry protection of topological feedback control is demonstrated.

\subsection{QND demon\label{sec_QND_demon}}

Here, we present a model of topological feedback control with strong unitary symmetry by following the construction in Sec.~\ref{sec_QND}. We consider a spin-$1/2$ particle on a 1D lattice with two sublattices. A quantum state at site $j$ with sublattice $a=A,B$ and spin $\sigma=\up,\down$ is denoted by $\ket{j,a,\sigma}$. We perform a projective position measurement on this particle and assume that this measurement does not disturb the spin state of the particle. Thus, a measurement outcome is specified only by the site index $j$ and the sublattice index $a$; therefore, the measurement operator is given by
\begin{equation}
    \hat{M}_{j,a}=\ket{j,a,\up}\bra{j,a,\up}+\ket{j,a,\down}\bra{j,a,\down}.
    \label{eq_QND_demon_meas}
\end{equation}
The measurement operator satisfies $[\hat{M}_{j,a},\hat{S}^z]=0$, where
\begin{equation}
    \hat{S}^z=\frac{1}{2}\sum_{i,a}(\ket{i,a,\up}\bra{i,a,\up}-\ket{i,a,\down}\bra{i,a,\down})
    \label{eq_Sz}
\end{equation}
is the magnetization. 
Thus, the position measurement does not influence the spin degrees of freedom, and it serves as a QND measurement for $\hat{S}^z$ (see Sec.~\ref{sec_A}).

If the measurement outcome is $m=(j,a)$, we perform feedback control with Hamiltonian $\hat{H}_m$ given by
\begin{align}
    \hat{H}_{j,A}=&-J_1\sum_{i,\sigma}(\ket{i,A,\sigma}\bra{i,B,\sigma}+\mathrm{H.c.})\notag\\
    &-J_2\sum_{i,\sigma}(\ket{i+1,A,\sigma}\bra{i,B,\sigma}+\mathrm{H.c.})\notag\\
    &+V\sum_\sigma\ket{j-1,B,\sigma}\bra{j-1,B,\sigma}\notag\\
    &-\sum_i(h_z+h_s)(\ket{i,A,\up}\bra{i,A,\up}-\ket{i,A,\down}\bra{i,A,\down})\notag\\
    &-\sum_i(h_z-h_s)(\ket{i,B,\up}\bra{i,B,\up}-\ket{i,B,\down}\bra{i,B,\down}),\label{eq_QND_HamilA}
\end{align}
and
\begin{align}
    \hat{H}_{j,B}=&-J_1\sum_{i,\sigma}(\ket{i,A,\sigma}\bra{i,B,\sigma}+\mathrm{H.c.})\notag\\
    &-J_2\sum_{i,\sigma}(\ket{i+1,A,\sigma}\bra{i,B,\sigma}+\mathrm{H.c.})\notag\\
    &+V\sum_\sigma\ket{j,A,\sigma}\bra{j,A,\sigma}\notag\\
    &-\sum_i(h_z+h_s)(\ket{i,A,\up}\bra{i,A,\up}-\ket{i,A,\down}\bra{i,A,\down})\notag\\
    &-\sum_i(h_z-h_s)(\ket{i,B,\up}\bra{i,B,\up}-\ket{i,B,\down}\bra{i,B,\down}),\label{eq_QND_HamilB}
\end{align}
where $J_1,J_2$ are the hopping amplitudes, $V$ is the height of the potential, and $h_z$ ($h_s$) denotes the strength of a uniform (staggered) magnetic field. 
The feedback Hamiltonian conserves the magnetization, i.e., $[\hat{H}_m,\hat{S}^z]=0$. 
The Hamiltonians \eqref{eq_QND_HamilA} and \eqref{eq_QND_HamilB} are generalized versions of the chiral Maxwell demon discussed in Sec.~\ref{sec_model}. The CPTP map for this feedback control is given by Eq.~\eqref{eq_CPTP} with $\hat{K}_m=\hat{U}_m\hat{M}_m$ and $\hat{U}_m=\exp[-i\hat{H}_m\tau]$.

Since $[\hat{M}_m,\hat{S}^z]=[\hat{H}_m,\hat{S}^z]=0$, the CPTP map for this feedback control has strong symmetry (see Sec.~\ref{sec_QND}). 
Hence, the CPTP map can be decomposed into four parts that act on subspaces with different symmetry eigenvalues. Specifically, we have
\begin{align}
    \tilde{\mathcal{E}}=&\sum_m\hat{K}_m\otimes \hat{K}_m^*\notag\\
    =&\tilde{\mathcal{E}}_{\up\up}+\tilde{\mathcal{E}}_{\up\down}+\tilde{\mathcal{E}}_{\down\up}+\tilde{\mathcal{E}}_{\down\down},
\end{align}
where
\begin{align}
    \tilde{\mathcal{E}}_{\sigma\sigma'}:=\sum_{j,a} \hat{U}_{j,a}\hat{M}_{j,a,\sigma}\otimes \hat{U}_{j,a}^*\hat{M}_{j,a,\sigma'}^*,
\end{align}
and
\begin{equation}
    \hat{M}_{j,a,\sigma}:=\ket{j,a,\sigma}\bra{j,a,\sigma}.
    \label{eq_spin_resolved_meas_op_QNDdemon}
\end{equation}

\begin{figure}[t]
    \includegraphics[width=8.5cm]{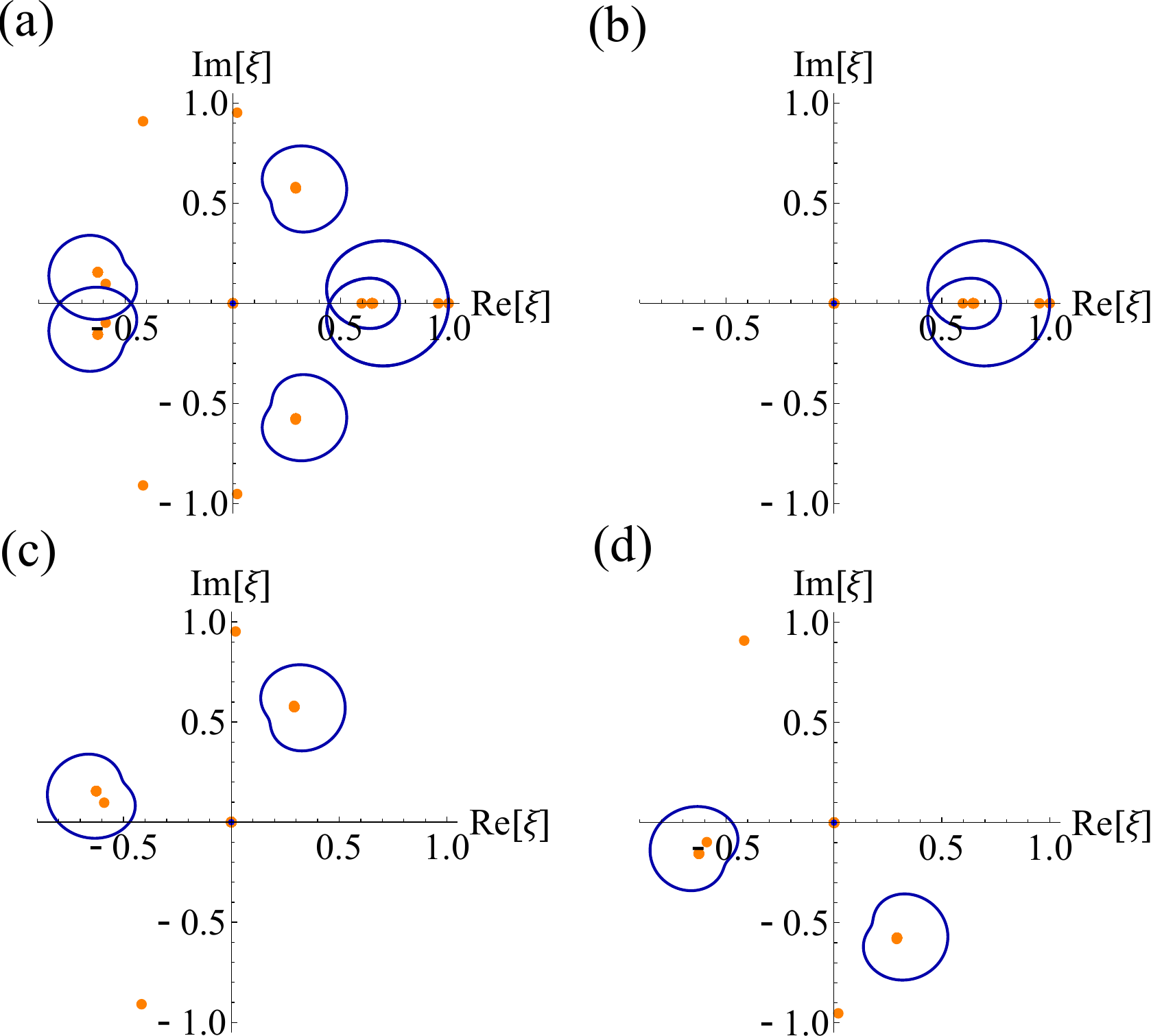}
    \caption{Eigenspectrum of a quantum channel for feedback control with strong unitary symmetry. In each figure, blue curves and the blue dot at the origin constitute the eigenspectrum under the PBC in the infinite-size limit, and orange dots form the eigenspectrum under the OBC with the number of unit cells $L=40$. (a) Full eigenspectrum of the CPTP map $\tilde{\mathcal{E}}$. (b) Eigenspectrum of $\tilde{\mathcal{E}}_{\up\up}$, which is identical to that of $\tilde{\mathcal{E}}_{\down\down}$. (c) Eigenspectrum of $\tilde{\mathcal{E}}_{\up\down}$. (d) Eigenspectrum of $\tilde{\mathcal{E}}_{\down\up}$. The parameters of the model are set to $J_1=J_2=1$, $V=+\infty$, $h_z=0.5$, and $h_s=1$. The duration of feedback is $\tau=2$.}
    \label{fig_QND_spec}
\end{figure}

In Fig.~\ref{fig_QND_spec}, we plot the eigenspectra of the quantum channel $\tilde{\mathcal{E}}$ under the PBC and the OBC together with those of $\tilde{\mathcal{E}}_{\sigma\sigma'}$. The topological characterization of the quantum channel can be applied to each symmetry sector. The PBC eigenspectrum of $\tilde{\mathcal{E}}_{\sigma\sigma'}$ with $\sigma=\sigma'$ contains a single loop with winding number $-2$, while that of $\tilde{\mathcal{E}}_{\sigma\sigma'}$ with $\sigma\neq\sigma'$ contains two loops, each of which has winding number $-1$. 
Here, the uniform and staggered magnetic fields lift the degeneracy of the eigenspectra of $\tilde{\mathcal{E}}_{\sigma\sigma'}$'s. 
Importantly, the eigenspectrum of $\tilde{\mathcal{E}}_{\sigma\sigma'}$ with $\sigma\neq\sigma'$ is no longer symmetric with respect to the real axis, implying that the modular conjugation symmetry is broken in these symmetry sectors. Thus, $\tilde{\mathcal{E}}_{\sigma\sigma'}$ with $\sigma\neq\sigma'$ offers a topologically nontrivial superoperator in class A.

\subsection{$\mathbb{Z}_2$ demon\label{sec_Z2}}

So far, we have discussed topological feedback controls in classes AI, AI $+$ psH$_-$, and A, all of which are characterized by an integer-valued winding number in 1D. In contrast, the topological classification for class AII$^\dag$ implies the existence of a $\mathbb{Z}_2$ topological invariant in 1D \cite{Kawabata19}. Here we present topological feedback control that is characterized by a $\mathbb{Z}_2$ topological invariant.

\begin{figure}[t]
    \includegraphics[width=8.5cm]{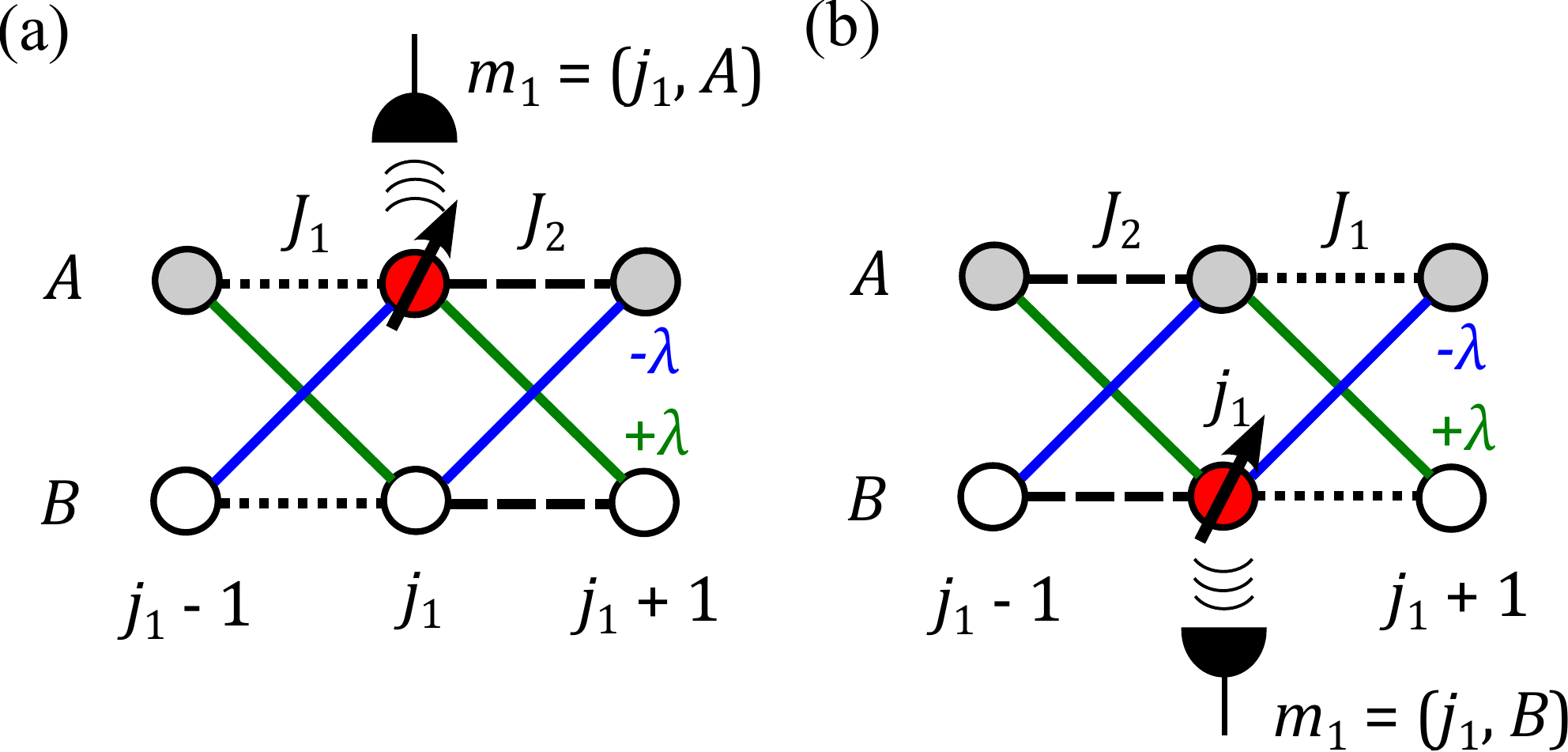}
    \caption{Schematic illustration of the $\mathbb{Z}_2$ demon. We first perform a projective joint measurement of the position and the sublattice without disturbing the spin degrees of freedom. In panel (a), if a particle is detected at site $j_1$ and sublattice $A$, we perform feedback control by using the Hamiltonian $H_{j_1,A}$. In panel (b), if a particle is detected at site $j_1$ and sublattice $B$, we perform feedback control by using the Hamiltonian $H_{j_1,B}$. After unitary evolution under the feedback Hamiltonian $H_{j_1,a_1}$ $(a_1=A,B)$, we perform the second projective measurement and additional feedback with a magnetic field [see Eq.~\eqref{eq_Z2_additionalFB}].}
    \label{fig_Z2_model}
\end{figure}

As shown in Sec.~\ref{sec_AIIdag}, a quantum channel in class AII$^\dag$ requires unitary symmetry and $C$ symmetry with $\eta_C=-1$ of the truncated Bloch matrix. To construct feedback control that fulfills these requirements, we consider a spin-$1/2$ particle on a 1D lattice with two sublattices as in Sec.~\ref{sec_QND_demon}. We first perform a spin-nondisturbing projective joint measurement of the position and the sublattice, where the measurement operator is given by Eq.~\eqref{eq_QND_demon_meas}. For a measurement outcome $m_1=(j_1,a_1)$, we perform feedback control with Hamiltonian $\hat{H}_{m_1}$, which has symmetry
\begin{equation}
    (-i\hat{\Sigma}_y)\hat{H}_{m_1}^*(-i\hat{\Sigma}_y)=\hat{H}_{m_1},
    \label{eq_Z2_TRS}
\end{equation}
where
\begin{equation}
    \hat{\Sigma}_y=i\sum_{j,\sigma}(\ket{j,B,\sigma}\bra{j,A,\sigma}-\ket{j,A,\sigma}\bra{j,B,\sigma}).
\end{equation}
Example of such feedback Hamiltonians are given by
\begin{widetext}
    \begin{align}
        \hat{H}_{j_1,A}=&-J_1\sum_{a=A,B}\sum_{\sigma=\up,\down}(\ket{j_1,a,\sigma}\bra{j_1-1,a,\sigma}+\mathrm{H.c.})
        -J_2\sum_{a}\sum_{\sigma}(\ket{j_1+1,a,\sigma}\bra{j_1,a,\sigma}+\mathrm{H.c.})\notag\\
        &+\lambda\sum_\sigma(\ket{j_1,B,\sigma}\bra{j_1-1,A,\sigma}+\ket{j_1+1,B,\sigma}\bra{j_1,A,\sigma}-\ket{j_1,A,\sigma}\bra{j_1-1,B,\sigma}-\ket{j_1+1,A,\sigma}\bra{j_1,B,\sigma}+\mathrm{H.c.}),
    \end{align}
    and
    \begin{align}
        \hat{H}_{j_1,B}=&-J_2\sum_{a=A,B}\sum_{\sigma=\up,\down}(\ket{j_1,a,\sigma}\bra{j_1-1,a,\sigma}+\mathrm{H.c.})
        -J_1\sum_{a}\sum_{\sigma}(\ket{j_1+1,a,\sigma}\bra{j_1,a,\sigma}+\mathrm{H.c.})\notag\\
        &+\lambda\sum_\sigma(\ket{j_1,B,\sigma}\bra{j_1-1,A,\sigma}+\ket{j_1+1,B,\sigma}\bra{j_1,A,\sigma}-\ket{j_1,A,\sigma}\bra{j_1-1,B,\sigma}-\ket{j_1+1,A,\sigma}\bra{j_1,B,\sigma}+\mathrm{H.c.}),
    \end{align}
\end{widetext}
as schematically illustrated in Fig.~\ref{fig_Z2_model}. For computational simplicity, here we assume that $\hat{H}_{j_1,a}$ acts only on three unit cells labeled by $j_1-1,j_1,j+1$; however, this assumption is not essential. After the unitary time evolution described by $\hat{U}_{j_1,a_1}=\exp[-i\hat{H}_{j_1,a_1}\tau]$, we perform the second projective joint measurement of the position and the sublattice with the same measurement operators \eqref{eq_QND_demon_meas} and suppose that we obtain an outcome $m_2=(j_2,a_2)$. Then, using the set of measurement outcomes $m_1$ and $m_2$, we perform additional feedback control with the following unitary operator:
\begin{align}
    \hat{U}_{m_1,m_2}^\prime=
    \begin{cases}
        \exp[-i\pi \hat{S}^z] & \mathrm{if}\ j_1>j_2\ \mathrm{and}\ a_1\neq a_2 \\
        \hat{I} & \mathrm{otherwise}.
    \end{cases}
    \label{eq_Z2_additionalFB}
\end{align}
This additional feedback control can be realized by a magnetic field, and it imprints a phase depending on the spin degrees of freedom. This feedback control is essential for realizing $C$ symmetry of a quantum channel (see below).

The quantum channel for the entire process is given by
\begin{align}
    \tilde{\mathcal{E}}=&\sum_{m_1,m_2}\hat{K}_{m_1,m_2}\otimes \hat{K}_{m_1,m_2}^*\notag\\
    =&\tilde{\mathcal{E}}_{\up\up}+\tilde{\mathcal{E}}_{\up\down}+\tilde{\mathcal{E}}_{\down\up}+\tilde{\mathcal{E}}_{\down\down},
    \label{eq_Z2_decomp}
\end{align}
where $\hat{K}_{m_1,m_2}=\hat{U}_{m_1,m_2}^\prime \hat{M}_{m_2}\hat{U}_{m_1}\hat{M}_{m_1}$ is the Kraus operator and
\begin{align}
    \tilde{\mathcal{E}}_{\sigma\sigma'}=&\hat{U}_{m_1,m_2}^\prime \hat{M}_{j_2,a_2,\sigma}\hat{U}_{m_1}\hat{M}_{j_1,a_1,\sigma}\notag\\
    &\otimes(\hat{U}_{m_1,m_2}^\prime \hat{M}_{j_2,a_2,\sigma'}\hat{U}_{m_1}\hat{M}_{j_1,a_1,\sigma'})^*
\end{align}
with Eq.~\eqref{eq_spin_resolved_meas_op_QNDdemon}. The decomposition of $\tilde{\mathcal{E}}$ into $\tilde{\mathcal{E}}_{\sigma\sigma'}$'s is due to the conservation of $\hat{S}^z$ during the measurement and feedback processes (i.e., strong symmetry).

From the general discussion presented in Sec.~\ref{sec_QND}, the truncated Bloch matrix $X_{\mathrm{trunc}}^{(\sigma\sigma')}(k)$ is given by Eq.~\eqref{eq_QND_trunc} with Eq.~\eqref{eq_QND_xi}; however, because of the additional feedback, Eq.~\eqref{eq_QND_q} is changed to
\begin{align}
    q^{(\sigma\sigma')}(j_2,a_2|j_1,a_1)=&\mathrm{sgn}[\sigma\sigma']^{\theta(j_1-j_2-1/2)}\notag\\
    \times(\hat{U}_{j_1,a_1})&_{j_2,a_2,\sigma;j_1,a_1,\sigma}(\hat{U}_{j_1,a_1})_{j_2,a_2,\sigma';j_1,a_1,\sigma'}^*,
    \label{eq_QND_q2}
\end{align}
for $a_1\neq a_2$, where $\theta(x)$ is the Heaviside unit-step function and
\begin{equation}
    \mathrm{sgn}[\sigma\sigma']:=
    \begin{cases}
        1 & (\sigma=\sigma');\\
        -1 & (\sigma\neq\sigma').
    \end{cases}
\end{equation}
The unitary operator with the feedback Hamiltonian $\hat{H}_{m_1}$ satisfies the following reciprocity relations because of the symmetry between $\hat{H}_{j_1,A}$ and $\hat{H}_{j_1,B}$:
\begin{subequations}
    \begin{align}
        &(\hat{U}_{j_1,A})_{j_2,A,\sigma;j_1,A,\sigma}(\hat{U}_{j_1,A})_{j_2,A,\sigma';j_1,A,\sigma'}^*\notag\\
        &=(\hat{U}_{j_2,B})_{j_1,B,\sigma;j_2,B,\sigma}(\hat{U}_{j_2,B})_{j_1,B,\sigma';j_2,B,\sigma'}^*,\label{eq_Z2_Ucond1}\\
        &(\hat{U}_{j_1,B})_{j_2,A,\sigma;j_1,B,\sigma}(\hat{U}_{j_1,B})_{j_2,A,\sigma';j_1,B,\sigma'}^*\notag\\
        &=(\hat{U}_{j_2,B})_{j_1,A,\sigma;j_2,B,\sigma}(\hat{U}_{j_2,B})_{j_1,A,\sigma';j_2,B,\sigma'}^*,\label{eq_Z2_Ucond2}\\
        &(\hat{U}_{j_1,A})_{j_2,B,\sigma;j_1,A,\sigma}(\hat{U}_{j_1,A})_{j_2,B,\sigma';j_1,A,\sigma'}^*\notag\\
        &=(\hat{U}_{j_2,A})_{j_1,B,\sigma;j_2,A,\sigma}(\hat{U}_{j_2,A})_{j_1,B,\sigma';j_2,A,\sigma'}^*.\label{eq_Z2_Ucond3}
    \end{align}    
\end{subequations}
In addition, we have
\begin{equation}
    q^{(\sigma\sigma')}(j,A|j,B)=q^{(\sigma\sigma')}(j,B|j,A)=0
\end{equation}
from the symmetry of the unitary operator
\begin{align}
    (-i\hat{\Sigma}_y)\hat{U}_{m_1}^T(-i\hat{\Sigma}_y)^{-1}=\hat{U}_{m_1},
\end{align}
which follows from the symmetry \eqref{eq_Z2_TRS} of the feedback Hamiltonian. 
As a result, conditions \eqref{eq_QND_qcond1}-\eqref{eq_QND_qcond3} are satisfied, and thus $\tilde{\mathcal{E}}_{\sigma\sigma'}$ belongs to class AII$^\dag$ for $\sigma\neq\sigma'$ and class AI $+$ psH$_+$ for $\sigma=\sigma'$. 

\begin{figure}[t]
    \includegraphics[width=8.5cm]{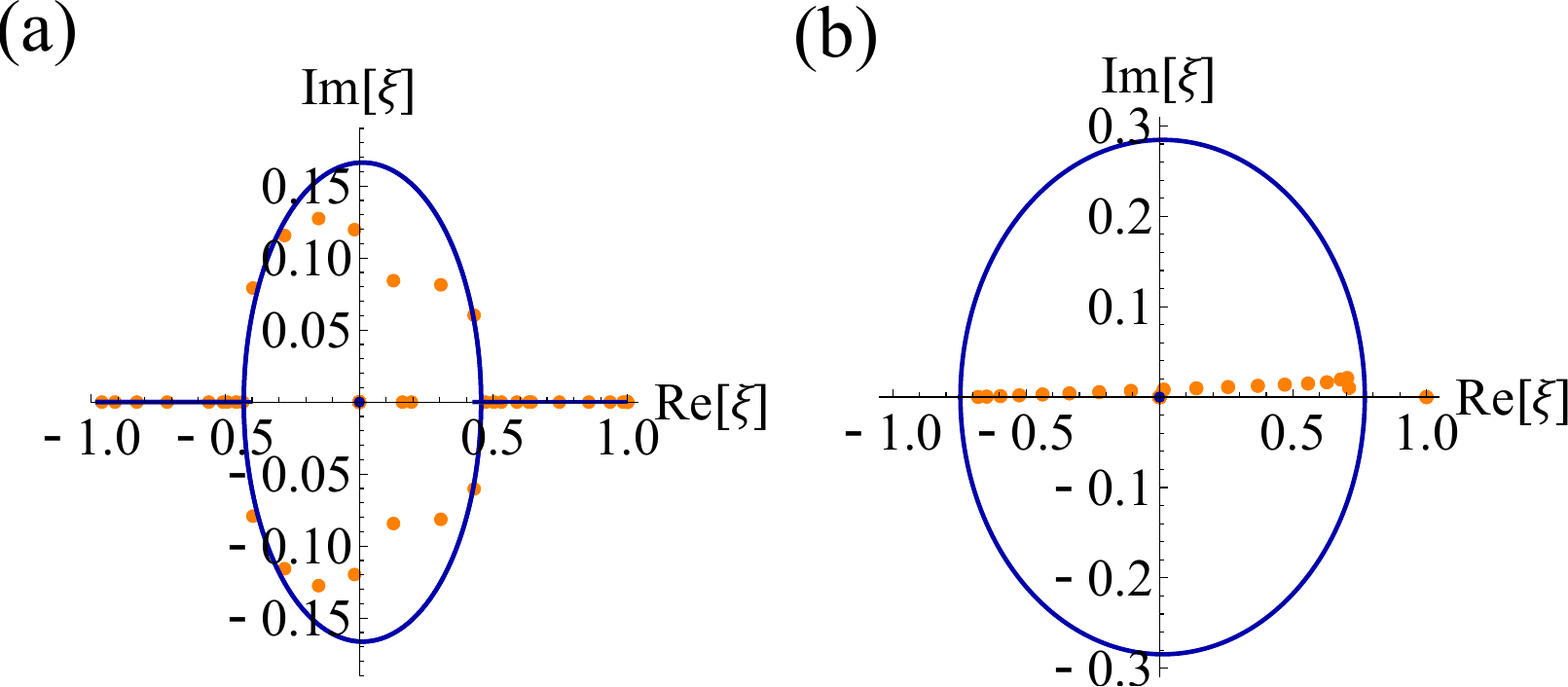}
    \caption{Eigenspectrum of a quantum channel for the $\mathbb{Z}_2$ demon. In each figure, blue curves and the blue dot at the origin constitute the eigenspectrum under the PBC in the infinite-size limit, and orange dots form the eigenspectrum under the OBC with the number of unit cells $L=20$. (a) Eigenspectrum of $\tilde{\mathcal{E}}_{\up\up}$. (b) Eigenspectrum of $\tilde{\mathcal{E}}_{\up\down}$. The parameters of the model are set to $J_1=0.8,J_2=1$, and $\lambda=0.5$. The duration of feedback is $\tau=1$.}
    \label{fig_Z2_spec}
\end{figure}

The topological feedback control in class AII$^\dag$ exhibits the $\mathbb{Z}_2$ skin effect \cite{Okuma20}. To see this case, we show the eigenspectrum of the quantum channel for the $\mathbb{Z}_2$ demon under the PBC and the OBC in Fig.~\ref{fig_Z2_spec}. The block-diagonalized CPTP map $\tilde{\mathcal{E}}_{\up\up}$ does not possess $C$ symmetry with $\eta_C=-1$, and therefore, the Kramers degeneracy of eigenvalues is not ensured. Consequently, the loop structure in the eigenspectrum of $\tilde{\mathcal{E}}_{\up\up}$ shown in Fig.~\ref{fig_Z2_spec}(a) collapses near the real axis due to real-complex transitions, and it does not support nontrivial topology. In contrast, the block $\tilde{\mathcal{E}}_{\up\down}$ has $C$ symmetry with $\eta_C=-1$, leading to the Kramers-degenerated eigenspectrum shown in Fig.~\ref{fig_Z2_spec}(b). The point-gap topology of this spectrum is characterized by a $\mathbb{Z}_2$ topological invariant \cite{Okuma20}
\begin{align}
    (-1)^{\nu(\xi_{\mathrm{PG}})}=&\mathrm{sgn}\Biggl\{\frac{\mathrm{Pf}[(X_{\mathrm{trunc}}^{(\up\down)}(\pi)-\xi_{\mathrm{PG}})(-i\sigma_y)]}{\mathrm{Pf}[(X_{\mathrm{trunc}}^{(\up\down)}(0)-\xi_{\mathrm{PG}})(-i\sigma_y)]}\notag\\
    \times\exp\Bigl[-\frac{1}{2}&\int_{k=0}^{k=\pi}d\log\det\{(X_{\mathrm{trunc}}^{(\up\down)}(k)-\xi_{\mathrm{PG}})(-i\sigma_y)\}\Bigr]\Biggr\}.
\end{align}
Indeed, the eigenspectrum of $\tilde{\mathcal{E}}_{\up\down}$ under the OBC forms a line rather than a closed loop, which is a hallmark of the $\mathbb{Z}_2$ skin effect. Interestingly, the $\mathbb{Z}_2$ skin effect can occur only in $\tilde{\mathcal{E}}_{\sigma\sigma'}$ with $\sigma\neq\sigma'$ since $\tilde{\mathcal{E}}_{\sigma\sigma}$ has modular conjugation symmetry and is therefore characterized by the winding number rather than the $\mathbb{Z}_2$ invariant (see Sec.~\ref{sec_helical}). Physically, the eigenspectra in Fig.~\ref{fig_Z2_spec} indicate that off-diagonal elements of the density matrix in terms of the $\hat{S}^z$ basis are accumulated on the edges of the system, whereas diagonal elements are delocalized over the system. Such distinct behavior is induced by the additional feedback in Eq.~\eqref{eq_Z2_additionalFB}, which nontrivially acts only on the off-diagonal elements.

Here, we remark on the boundary condition of the $\mathbb{Z}_2$ demon. To see the non-Hermitian skin effect in symmetry-protected topological feedback control, a quantum channel under the OBC must respect the protecting symmetry.  We give details on how to implement a symmetry-preserving OBC in Appendix \ref{sec_symOBC}.

The distinct topology of $\mathcal{E}_{\sigma\sigma}$ and $\mathcal{E}_{\sigma\sigma'}$ $(\sigma'\neq\sigma)$ can also be detected through dynamics. Suppose that the initial density matrix of the system is given by $\hat{\rho}_{\mathrm{i}}=\ket{j_0,a_0,+}\bra{j_0,a_0,+}$, where $\ket{j_0,a_0,+}=(\ket{j_0,a_0,\up}+\ket{j_0,a_0,\down})/\sqrt{2}$ is a site-localized state with its spin polarized along the $x$ direction. Then, the particle-number density and the spin density of the final state after $N$ feedback cycles are given by
\begin{subequations}
    \begin{align}
    \mathrm{Tr}[\hat{n}_{j,a}\mathcal{E}^N(\hat{\rho}_{\mathrm{i}})]=&\frac{1}{2}(\bra{j,a,\uparrow}\mathcal{E}_{\uparrow\uparrow}^N(\hat{\rho}_{\mathrm{i}})\ket{j,a,\uparrow}\notag\\
    &+\bra{j,a,\downarrow}\mathcal{E}_{\downarrow\downarrow}^N(\hat{\rho}_{\mathrm{i}})\ket{j,a,\downarrow}),\\
    \mathrm{Tr}[\hat{S}_{j,a}^x\mathcal{E}^N(\hat{\rho}_{\mathrm{i}})]=&\mathrm{Re}[\bra{j,a,\uparrow}\mathcal{E}_{\uparrow\downarrow}^N(\hat{\rho}_{\mathrm{i}})\ket{j,a,\downarrow}],\\
    \mathrm{Tr}[\hat{S}_{j,a}^y\mathcal{E}^N(\hat{\rho}_{\mathrm{i}})]=&\mathrm{Im}[\bra{j,a,\uparrow}\mathcal{E}_{\uparrow\downarrow}^N(\hat{\rho}_{\mathrm{i}})\ket{j,a,\downarrow}],\\
    \mathrm{Tr}[\hat{S}_{j,a}^z\mathcal{E}^N(\hat{\rho}_{\mathrm{i}})]=&\frac{1}{4}(\bra{j,a,\uparrow}\mathcal{E}_{\uparrow\uparrow}^N(\hat{\rho}_{\mathrm{i}})\ket{j,a,\uparrow}\notag\\
    &-\bra{j,a,\downarrow}\mathcal{E}_{\downarrow\downarrow}^N(\hat{\rho}_{\mathrm{i}})\ket{j,a,\downarrow}),
\end{align}
\end{subequations}
where
\begin{subequations}
    \begin{align}
\hat{n}_{j,a}=&\sum_{\sigma}\ket{j,a,\sigma}\bra{j,a,\sigma},\\
\hat{S}_{j,a}^x=&\frac{\ket{j,a,\up}\bra{j,a,\down}+\ket{j,a,\down}\bra{j,a,\up}}{2},\\
\hat{S}_{j,a}^y=&\frac{\ket{j,a,\up}\bra{j,a,\down}-\ket{j,a,\down}\bra{j,a,\up}}{2i},\\
\hat{S}_{j,a}^z=&\frac{\ket{j,a,\up}\bra{j,a,\up}-\ket{j,a,\down}\bra{j,a,\down}}{2}.
    \end{align}
\end{subequations}
Thus, topology of the off-diagonal sectors $\mathcal{E}_{\sigma\sigma'}$ $(\sigma'\neq\sigma)$ is reflected in the dynamics of the transverse components of spin, whereas the dynamics of the number density and the longitudinal component of spin are governed by the diagonal sectors $\mathcal{E}_{\sigma\sigma}$.

\begin{figure*}[t]
    \includegraphics[width=18cm]{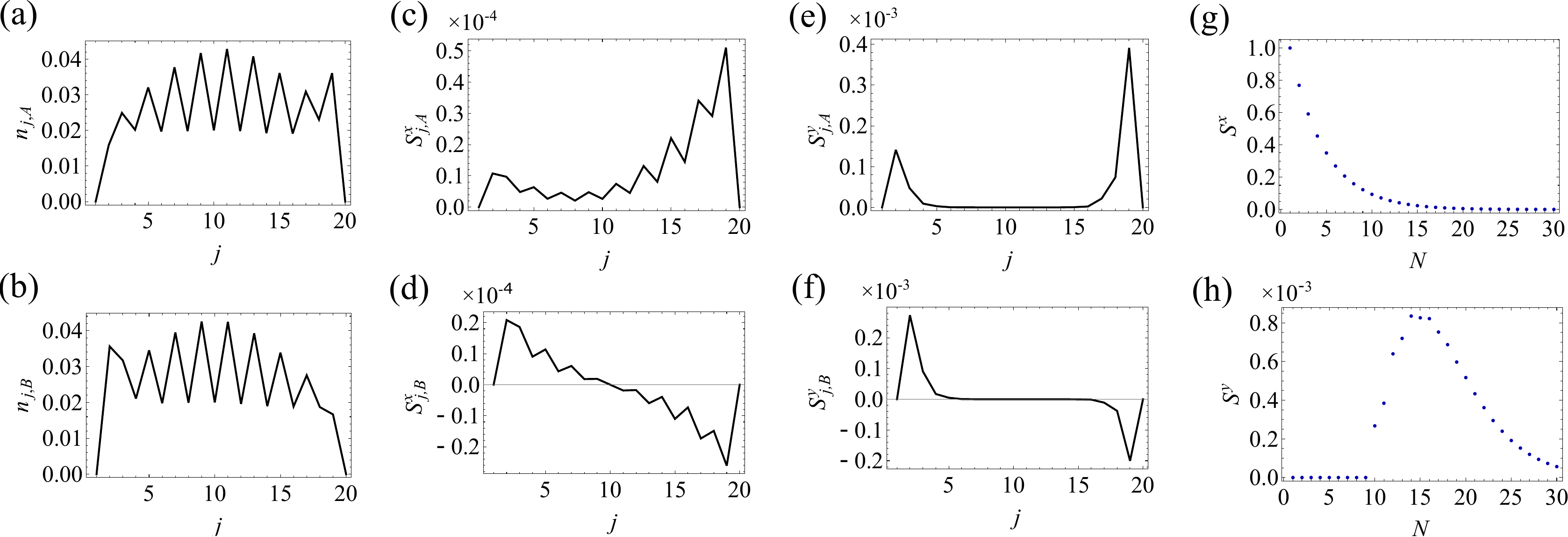}
    \caption{Dynamical signatures of nontrivial topology of a $\mathbb{Z}_2$ demon. We show physical quantities after $N$ feedback cycles. (a) [(b)] Number density in sublattice $A$ ($B$) for $N=40$. (c) [(d)] Density of the $x$ component of spin in sublattice $A$ ($B$) for $N=30$. (e) [(f)] Density of the $y$ component of spin in sublattice $A$ ($B$) for $N=15$. (g) [(h)] Dynamics of the total $x$ ($y$) component of spin. The initial state is given by $\hat{\rho}_{\mathrm{i}}=\ket{j_0,A,+}\bra{j_0,A,+}$ with $j_0=L/2$ (see text). The parameters of the model are the same as in Fig.~\ref{fig_Z2_spec}.}
    \label{fig_Z2_dynamics}
\end{figure*}

In Fig.~\ref{fig_Z2_dynamics}, we plot the number density and the transverse spin components of a particle after $N$ repeated applications of feedback control with the $\mathbb{Z}_2$ demon. The numerical results clearly show that the transverse spin components are localized at the edges due to the skin effect, whereas the number density is delocalized over the system. 
Thus, the $\mathbb{Z}_2$ skin effect of the topological feedback control manifests itself only in the off-diagonal components of the density matrix.
Note that the transverse spin components are not conserved due to the additional feedback in Eq.~\eqref{eq_Z2_additionalFB}, and they vanish in the steady state of the quantum channel [see Figs.~\ref{fig_Z2_dynamics}(g) and \ref{fig_Z2_dynamics}(h)]. Therefore, the nontrivial topology of the off-diagonal sectors appears only in the transient dynamics under repeated applications of feedback control.

\subsection{Chern demon\label{sec_Chern}}
In the previous subsections, we have focused on the models of quantum feedback control that are characterized by point-gap topology. In this subsection, we present an example of a topological Maxwell demon that exhibits nontrivial line-gap topology. Our main idea is to simulate the Haldane model \cite{Haldane88} of a Chern insulator by using feedback control. We consider a spin-$1/2$ particle on a honeycomb lattice and perform a projective measurement of the position of the particle so that the measurement does not disturb the spin state as in Secs.~\ref{sec_QND_demon} and \ref{sec_Z2}. The measurement operator is thus given by 
\begin{equation}
    \hat{M}_{\vec{j},a}=\ket{\vec{j},a,\up}\bra{\vec{j},a,\up}+\ket{\vec{j},a,\down}\bra{\vec{j},a,\down},
    \label{eq_Chern_demon_meas}
\end{equation}
where $a=A,B$ denotes a sublattice of the honeycomb lattice, and we record a measurement outcome $m_1=(\vec{j}_1,a_1)$. After the measurement, we let the system undergo unitary time evolution during time $\tau$ according to the tight-binding Hamiltonian
\begin{align}
\hat{H}=&J_1\sum_{\sigma=\up,\down}\sum_{\langle \vec{i},\vec{j}\rangle}(\ket{\vec{i},A,\sigma}\bra{\vec{j},B,\sigma}+\mathrm{H.c.})\notag\\
&+J_2\sum_{\sigma=\up,\down}\sum_{\langle\langle \vec{i},\vec{j}\rangle\rangle}\sum_{a=A,B}(\ket{\vec{i},a,\sigma}\bra{\vec{j},a,\sigma}+\mathrm{H.c.}),
\label{eq_H_honeycomb}
\end{align}
where $J_1\in\mathbb{R}$ and $J_2\in\mathbb{R}$ are the hopping amplitudes for nearest-neighbor and next-nearest-neighbor sites, respectively. Note that Hamiltonian $\hat{H}$ does not depend on a measurement outcome. Next, we perform the second projective measurement described by the measurement operator \eqref{eq_Chern_demon_meas} to obtain an outcome $m_2=(\vec{j}_2,a_2)$. The set of measurement outcomes $\{m_1,m_2\}$ informs us of the initial and final positions of the particle traveling under Hamiltonian $\hat{H}$. Using this information of measurement outcomes, we apply feedback control with the unitary operator $\hat{U}_{m_1,m_2}^\prime=\exp[-i\phi_{m_1,m_2} \hat{S}^z]$, where $\phi_{m_1,m_2}\in\mathbb{R}$ and $\hat{S}^z$ is the magnetization obtained by replacing $i$ with $\vec{i}$ in Eq.~\eqref{eq_Sz}.

The quantum channel $\mathcal{E}$ that describes this quantum feedback control has the decomposition
\begin{equation}
\mathcal{E}=\mathcal{E}_{\up\up}+\mathcal{E}_{\up\down}+\mathcal{E}_{\down\up}+\mathcal{E}_{\down\down}
\end{equation}
similar to Eq.~\eqref{eq_Z2_decomp}, and each block is given by
\begin{widetext}
    \begin{align}
\mathcal{E}_{\sigma\sigma}(\hat{\rho})=&\sum_{m_1,m_2}\ket{\vec{j}_2,a_2,\sigma}\bra{\vec{j}_2,a_2,\sigma}e^{-i\hat{H}\tau}\ket{\vec{j_1},a_1,\sigma}\bra{\vec{j}_1,a_1,\sigma}\hat{\rho} \ket{\vec{j}_1,a_1,\sigma}\bra{\vec{j}_1,a_1,\sigma}e^{i\hat{H}\tau}\ket{\vec{j}_2,a_2,\sigma}\bra{\vec{j}_2,a_2,\sigma}\notag\\
=&\sum_{m_1,m_2}p(\vec{j}_2,a_2|\vec{j}_1,a_1)\ket{\vec{j}_2,a_2,\sigma}\bra{\vec{j}_1,a_1,\sigma}\hat{\rho} \ket{\vec{j}_1,a_1,\sigma}\bra{\vec{j}_2,a_2,\sigma}
    \end{align}
and
    \begin{align}
\mathcal{E}_{\sigma\sigma'}(\hat{\rho})=&\sum_{m_1,m_2}e^{-i\phi_{m_1,m_2}\mathrm{sgn}[\sigma]}p(\vec{j}_2,a_2|\vec{j}_1,a_1)\ket{\vec{j}_2,a_2,\sigma}\bra{\vec{j}_1,a_1,\sigma}\hat{\rho} \ket{\vec{j}_1,a_1,\sigma'}\bra{\vec{j}_2,a_2,\sigma'},\label{eq_Chern_Eupdown}
    \end{align}
\end{widetext}
where $\sigma'\neq\sigma$, $p(\vec{j}_2,a_2|\vec{j}_1,a_1):=|\bra{\vec{j}_2,a_2,\sigma}e^{-i\hat{H}\tau}\ket{\vec{j}_1,a_1,\sigma}|^2$, and $\mathrm{sgn}[\sigma]$ is defined by $\mathrm{sgn}[\up]=-\mathrm{sgn}[\down]=+1$.
Equation \eqref{eq_Chern_Eupdown} indicates that one can imprint an arbitrary phase factor $e^{-i\phi_{m_1,m_2}}$ on the transition amplitude $p(\vec{j}_2,a_2|\vec{j}_1,a_1)$ by using feedback control. 
From a general perspective, this approach gives a significant advantage to realize topological phases with feedback control.
In the present case, we set $\phi_{m_1,m_2}$ to nonzero if and only if $m_1$ and $m_2$ are nearest-neighbor sites, and we choose $\phi_{m_1,m_2}=\pm\phi\in\mathbb{R}$, where the plus (minus) sign is taken if the hopping from $m_1$ to $m_2$ is in the counterclockwise (clockwise) direction. This choice of the phase factor resembles that in the Haldane model \cite{Haldane88}, thereby making the topologically trivial model \eqref{eq_H_honeycomb} topological. Because of the phase factor, block $\mathcal{E}_{\sigma\sigma'}$ belongs to class A while block $\mathcal{E}_{\sigma\sigma}$ belongs to class AI due to modular conjugation symmetry.

\begin{figure}[t]
    \includegraphics[width=8.5cm]{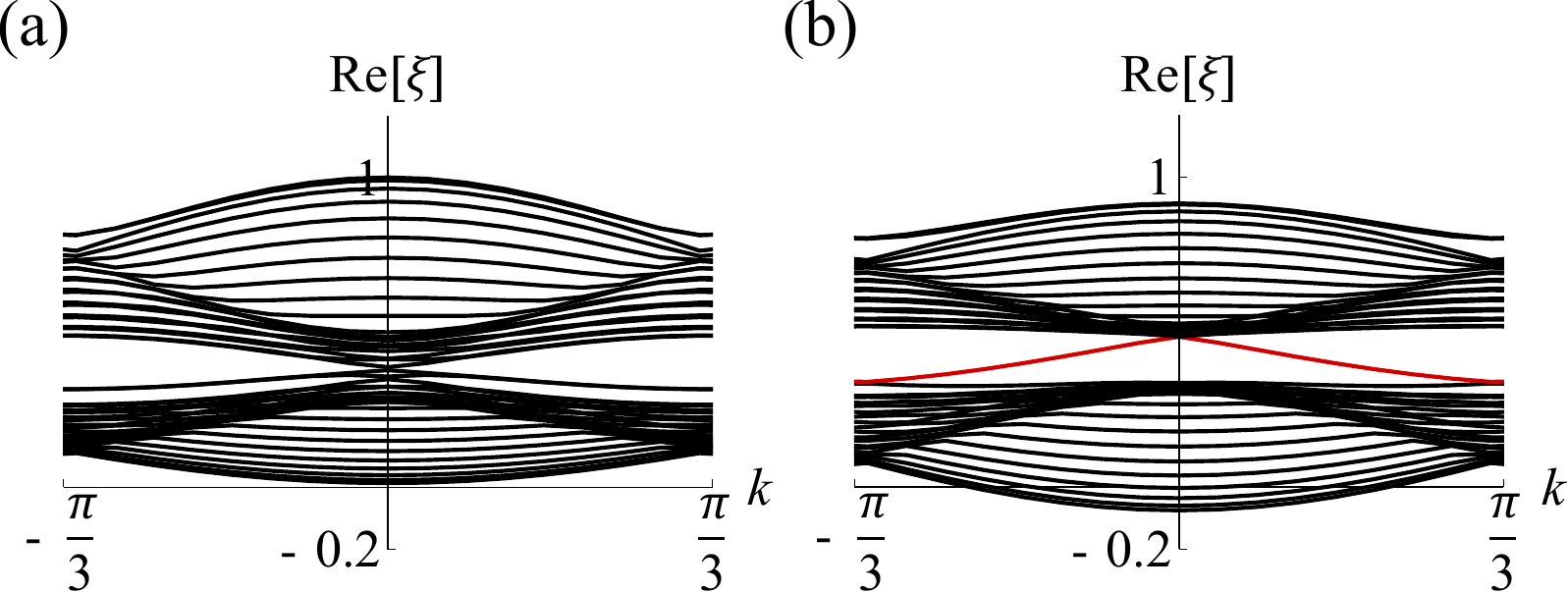}
    \caption{Dispersion relations of the real part of eigenvalues of (a) block $\mathcal{E}_{\up\up}$ and (b) block $\mathcal{E}_{\up\down}$ of a quantum channel for the Chern demon. The parameters of the model are set to $J_1=1,J_2=0.3,\tau=0.5$, and $\phi=\pi/2$. The unit of length is chosen to be the distance between nearest-neighbor sites. We employ the armchair edge in which one direction is periodic and the other direction is open with ten unit cells. The horizontal axis is the momentum along the edge. The eigenvalues of edge-localized states are colored in red in panel (b). The imaginary part of the eigenvalues is negligible.}
    \label{fig_Chern_disp}
\end{figure}

In Fig.~\ref{fig_Chern_disp}, we plot the dispersion relation of blocks $\mathcal{E}_{\up\up}$ and $\mathcal{E}_{\up\down}$ of the Chern demon constructed above. As seen from the figure, the phase $\phi$ imprinted by feedback control opens a gap in the eigenspectrum of $\mathcal{E}_{\up\down}$, and gapless edge modes that connect the bulk bands emerge under the OBC. This feature is analogous to the physics of the Haldane model. Thus, the quantum channel is characterized by nontrivial line-gap topology of $\mathcal{E}_{\up\down}$, which results in the emergence of boundary-localized edge modes.

Here, special attention should be given to the physical difference between a Chern insulator and the Chern demon presented in this section. A Chern insulator is characterized by line-gap topology of a Hamiltonian, which is a generator of the dynamics of a system. Since the derivative of the dispersion relation with respect to momentum gives the group velocity, the gapless spectrum of edge states leads to chiral modes, which propagate unidirectionally along the edges. In contrast, the Chern demon is defined by line-gap topology of a quantum channel, which is a time-evolution operator itself rather than its generator. Thus, the physical meaning of the gapless edge-mode spectrum of the Chern demon is different from that of a Chern insulator. In particular, the boundary-localized modes of the Chern demon do not necessarily propagate along the edges of the system. This physical implication merits further investigation.

\section{Topological phenomena arising from complete positivity and trace preservation\label{sec_top_CPTP}}
Although topological feedback control exploits non-Hermitian topology of quantum channels, it is crucially different from quantum systems with non-Hermitian Hamiltonians due to the fact that a quantum channel must satisfy complete positivity and trace preservation, i.e., it is a CPTP map \cite{NielsenChuang_book}. Here, we discuss how these properties lead to topological phenomena that are absent in conventional non-Hermitian systems.

\subsection{Boundary-localized steady state\label{sec_boundary_SS}}
Complete positivity and trace preservation ensure that a quantum channel gives a map between density matrices, which leads to the existence of a steady state with the unit eigenvalue \cite{Wolf12}. Consequently, the topological Maxwell demons discussed in Secs.~\ref{sec_model} and \ref{sec_demons} (except for those in Secs.~\ref{sec_Z2} and \ref{sec_Chern}) exhibit boundary-localized steady states under the OBC. This feature is in stark contrast to topological phases described by non-Hermitian Hamiltonians, where the conservation of the norm of a state is not guaranteed and therefore a steady state does not exist, in general. Although those boundary-localized steady states in topological feedback control appear to be closely related to the non-Hermitian skin effect, we can construct a model that shows the non-Hermitian skin effect with no localized steady state (see Appendix \ref{sec_spont_demon}). In this sense, the emergence of boundary-localized steady states is a special feature of quantum channels and should be distinguished from the non-Hermitian skin effect. We note that a similar phenomenon is observed in classical stochastic processes where the probability conservation is satisfied, unlike general non-Hermitian systems \cite{Sawada24}.

\subsection{Breakdown of conventional bulk-boundary correspondence in non-Hermitian skin effects\label{sec_BBC}}
The bulk-boundary correspondence in non-Hermitian systems with nontrivial point-gap topology states that a nonzero winding number of the eigenspectrum under the PBC leads to boundary-localized skin modes under the OBC whose eigenvalues generically reside inside the region delimited by the PBC spectrum \cite{Okuma20}. However, eigenvalues of the boundary-localized steady states discussed in Sec.~\ref{sec_boundary_SS} must be unity by definition, and they do not belong to the internal region enclosed by the PBC spectrum (see, e.g., Figs.~\ref{fig_chiral_spec}, \ref{fig_PBCspec_duration}, and \ref{fig_SSH_spec}). More strikingly, it is possible that some eigenvalues under the OBC lie outside the region enclosed by the PBC spectrum (see the case of the QND demon shown in Fig.~\ref{fig_QND_spec}). These facts indicate that the conventional bulk-boundary correspondence in non-Hermitian skin effects does not hold in topological feedback control.

The breakdown of the conventional bulk-boundary correspondence is attributed to the requirement of complete positivity and trace preservation for a quantum channel. The Hamiltonian under the OBC in a conventional non-Hermitian system is obtained by eliminating the matrix elements of the PBC Hamiltonian across the boundary of the system. However, such a straightforward implementation of the OBC for a quantum channel violates complete positivity and trace preservation. Instead, a quantum channel under the OBC should be defined as follows. Suppose that a quantum channel for feedback control under the PBC is given by 
\begin{align}
\mathcal{E}_{\mathrm{PBC}}(\hat{\rho})=\sum_m e^{-i\hat{H}_{m,\mathrm{PBC}}\tau}\hat{M}_m\hat{\rho}\hat{M}_m^\dag e^{i\hat{H}_{m,\mathrm{PBC}}\tau},
\label{eq_E_PBC}
\end{align}
where $\hat{H}_{m,\mathrm{PBC}}$ is a feedback Hamiltonian under the PBC. Then, the corresponding quantum channel under the OBC is defined as
\begin{align}
    \mathcal{E}_{\mathrm{OBC}}(\hat{\rho})=\sum_m e^{-i\hat{H}_{m,\mathrm{OBC}}\tau}\hat{M}_m\hat{\rho}\hat{M}_m^\dag e^{i\hat{H}_{m,\mathrm{OBC}}\tau},
    \label{eq_E_OBC}
\end{align}
by using time evolution governed by the Hamiltonian $\hat{H}_{m,\mathrm{OBC}}$ under the OBC so that $\mathcal{E}_{\mathrm{OBC}}$ gives a CPTP map. By comparing Eqs.~\eqref{eq_E_PBC} and \eqref{eq_E_OBC}, one can see that the matrix elements of the quantum channel near the boundary are reconstructed to ensure complete positivity and trace preservation.

Since the OBC for a quantum channel is defined in a manner different from that for a Hamiltonian, the existing proofs of the bulk-boundary correspondence in non-Hermitian systems \cite{Gong18,Okuma20,Zhang20,Borgnia20} cannot simply be applied to quantum channels. The bulk-boundary correspondence in topological feedback control should be formulated so as to incorporate complete positivity and trace preservation.

\subsection{Topological distinction between steady states and transient dynamics}
The complete positivity also impacts the topological classification. As discussed in Sec.~\ref{sec_mod_conj_sym}, the complete positivity (or the Hermiticity preservation that follows) imposes the modular conjugation symmetry on a quantum channel. While some symmetry sectors of a quantum channel do not necessarily satisfy the modular conjugation symmetry (as shown in Sec.~\ref{sec_QND}), a sector that includes a steady state must satisfy it since the steady-state density matrix must be Hermitian \cite{Albert14}. Thus, allowed symmetry classes for a steady-state sector of a quantum channel are restricted to those three classes (AI, AI $+$ psH$_+$, and AI $+$ psH$_-$), which include modular conjugation symmetry. Thus, the topological classification for boundary-localized steady states in Sec.~\ref{sec_boundary_SS} is more restrictive than the general topological classification of non-Hermitian systems due to the complete positivity.

If transient dynamics are considered, topological phenomena that are prohibited in steady states due to complete positivity become possible. An example is given by the $\mathbb{Z}_2$ demon in Sec.~\ref{sec_Z2}. As shown in Fig.~\ref{fig_Z2_spec}, a spin-diagonal sector $\mathcal{E}_{\uparrow\uparrow}$ that includes a steady state is topologically trivial, whereas an off-diagonal sector $\mathcal{E}_{\uparrow\downarrow}$ displays nontrivial point-gap topology owing to the absence of modular conjugation symmetry. The nontrivial topology in the off-diagonal sector manifests itself in the transient dynamics of the transverse components of the spin density, which vanish in the steady state (see Fig.~\ref{fig_Z2_dynamics}). 
The topological distinction between the steady state and transient dynamics thus enriches topological phenomena in quantum feedback control.

\section{Dissipative feedback control\label{sec_diss_fb}}
\subsection{Formalism}
In reality, there exist many sources of noise and dissipation during feedback control. 
The formalism presented so far can be generalized to such cases in which feedback operations are not unitary. Suppose that a feedback operation depending on a measurement outcome $m$ is described by a CPTP map
\begin{equation}
    \mathcal{F}_m(\hat{\rho})=\sum_\alpha \hat{F}_{m,\alpha}\hat{\rho} \hat{F}_{m,\alpha}^\dag,
    \label{eq_diss_fbCPTP}
\end{equation}
where the Kraus operators $\hat{F}_{m,\alpha}$ satisfy
\begin{equation}
\sum_\alpha \hat{F}_{m,\alpha}^\dag \hat{F}_{m,\alpha}=\hat{I}.
\label{eq_fbKraus_TP}
\end{equation}
Then, given an initial density matrix $\hat{\rho}_{\mathrm{i}}$, the density matrix after feedback control is
\begin{align}
    \hat{\rho}_{\mathrm{f}}=&\sum_m p_m \mathcal{F}_m\Bigl(\frac{1}{p_m}\hat{M}_m\hat{\rho}_{\mathrm{i}} \hat{M}_m^\dag\Bigr)\notag\\
    =&\sum_m\sum_\alpha \hat{F}_{m,\alpha}\hat{M}_m\hat{\rho}_{\mathrm{i}} \hat{M}_m^\dag \hat{F}_{m,\alpha}^\dag,
\end{align}
where $p_m=\mathrm{Tr}[\hat{M}_m^\dag \hat{M}_m \hat{\rho}_{\mathrm{i}}]$ is the probability of outcome $m$ for the measurement operator $\hat{M}_m$. Thus, feedback control is described by a quantum channel
\begin{equation}
    \mathcal{E}(\hat{\rho})=\sum_{m,\alpha}\hat{K}_{m,\alpha}\hat{\rho} \hat{K}_{m,\alpha}^\dag
    \label{eq_diss_channel}
\end{equation}
with Kraus operators
\begin{equation}
    \hat{K}_{m,\alpha}:=\hat{F}_{m,\alpha}\hat{M}_m.
\end{equation}
Note that
\begin{align}
    \sum_{m,\alpha}\hat{K}_{m,\alpha}^\dag \hat{K}_{m,\alpha}=&\sum_{m,\alpha}\hat{M}_m^\dag \hat{F}_{m,\alpha}^\dag \hat{F}_{m,\alpha} \hat{M}_m\notag\\
    =&\sum_{m}\hat{M}_m^\dag \hat{M}_m=\hat{I}
\end{align}
holds due to Eq.~\eqref{eq_fbKraus_TP} and $\sum_m \hat{M}_m^\dag \hat{M}_m=\hat{I}$.

By repeating the analysis in Sec.~\ref{sec_top_setup} for the quantum channel \eqref{eq_diss_channel}, the eigenspectrum of the quantum channel is given by that of the Bloch matrix $X(\vec{k})$, whose matrix elements are given by
\begin{align}
    X_{a,c,\vec{\mu};b,d,\vec{\nu}}(\vec{k}):=&
    \frac{1}{N_{\mathrm{cell}}}\sum_{\vec{j},\vec{j}'}\sum_{m,\alpha}(\hat{K}_{m,\alpha})_{\vec{j},a;\vec{j}',b}\notag\\
    &\times (\hat{K}_{m,\alpha})^*_{\vec{j}+\vec{\mu},c;\vec{j}'+\vec{\nu},d}e^{-i\vec{k}\cdot(\vec{R}_{\vec{j}}-\vec{R}_{\vec{j}'})}.
    \label{eq_diss_Bloch}
    \end{align}
The difference from the case with unitary feedback is the existence of the additional index $\alpha$, which can be regarded as information that cannot be specified in the feedback process.

\subsection{Dissipative Maxwell demon}

As an example of dissipative feedback control described by a CPTP map, Eq.~\eqref{eq_diss_fbCPTP}, we consider a chiral Maxwell demon (see Sec.~\ref{sec_model}) subject to dissipation. We consider a spinless particle on a 1D lattice with $L$ sites and perform a projective position measurement for this particle as in Sec.~\ref{sec_model}. We then perform a feedback operation governed by the Lindblad equation
\begin{align}
    \frac{d\hat{\rho}}{dt}=&\mathcal{L}_m(\hat{\rho})\notag\\
    =&-i[\hat{H}_m,\hat{\rho}]+\sum_j\Bigl(\hat{L}_{m,j}\hat{\rho} \hat{L}_{m,j}^\dag -\frac{1}{2}\{ \hat{L}_{m,j}^\dag \hat{L}_{m,j},\hat{\rho}\}\Bigr),
    \label{eq_fb_Lindblad}
\end{align}
where $\mathcal{L}_m$ is the generator (Lindbladian) conditioned on the measurement outcome, $H_m$ is the feedback Hamiltonian given in Sec.~\ref{sec_chiral_demon}, and $\hat{L}_{m,j}$ is the Lindblad operator acting on a local region around site $j$. Then, a CPTP map for the feedback operation is given by
\begin{equation}
    \mathcal{F}_m=e^{\mathcal{L}_m\tau},
\end{equation}
where $\tau$ is the duration of feedback. Using the Lindblad equation \eqref{eq_fb_Lindblad}, we can consider various dissipative effects on feedback control such as noise during feedback and a coupling to a reservoir. In particular, if the system is in contact with a heat bath, the topological Maxwell demon can rectify thermal fluctuations as well as quantum fluctuations to achieve unidirectional transport of a particle. However, we note that the Hamiltonian and the Lindblad operators in Eq.~\eqref{eq_fb_Lindblad} should be chosen to be local to ensure the locality of the Kraus operators and to allow truncation of the Bloch matrix into a finite-dimensional matrix (see Sec.~\ref{sec_loc_Kraus}). The locality-preserving Lindblad equation can be derived if the correlation time of the environment is sufficiently short \cite{Shiraishi24}.

Since the matrix elements of the Kraus operator $\hat{K}_{m,\alpha}=\hat{F}_{m,\alpha}\hat{M}_m$ are written as
\begin{equation}
(\hat{K}_{m,\alpha})_{i,j}=(\hat{F}_{m,\alpha}\hat{M}_m)_{i,j}=(\hat{F}_{m,\alpha})_{i,m}\delta_{j,m},
\end{equation}
the Bloch matrix of the quantum channel is given by
\begin{align}
    X_{\mu,\nu}(k)=&\frac{1}{L}\sum_{j,j'}\sum_{m,\alpha}(\hat{K}_{m,\alpha})_{j,j'}(\hat{K}_{m,\alpha})^*_{j+\mu,j'+\nu}e^{-ik(j-j')}\notag\\
    =&\delta_{\nu,0}\frac{1}{L}\sum_j\sum_{m,\alpha}(\hat{F}_{m,\alpha})_{j,m}(\hat{F}_{m,\alpha})^*_{j+\mu,m}e^{-ik(j-m)}.
\end{align}
Thus, similarly to the case with unitary feedback control, eigenvalues of the Bloch matrix are given by
\begin{align}
    \xi_0(k)=&X_{0,0}(k)\notag\\
    =&\frac{1}{L}\sum_{j}\sum_{m,\alpha}|(\hat{F}_{m,\alpha})_{j,m}|^2e^{-ik(j-m)},\\
    \xi_1(k)=&\xi_2(k)=\cdots=\xi_{L-1}(k)=0.
\end{align}
Here, we note that
\begin{align}
    \sum_\alpha|(\hat{F}_{m,\alpha})_{j,m}|^2=&\sum_\alpha \bra{j}\hat{F}_{m,\alpha}\ket{m}\bra{m}\hat{F}_{m,\alpha}^\dag\ket{j}\notag\\
    =&\bra{j}\mathcal{F}_m(\ket{m}\bra{m})\ket{j}\notag\\
    =&\bra{j}\mathcal{F}_m\Bigl(\frac{1}{p_m}\hat{M}_m\hat{\rho} \hat{M}_m^\dag\Bigr)\ket{j}\notag\\
    =&p(j|m),
\end{align}
where $p(j|m)$ is the probability of a particle being found at site $j$ after feedback control with outcome $m$. Then, the eigenvalue $\xi_0(k)$ is written as
\begin{align}
    \xi_0(k)=&\frac{1}{L}\sum_{j,m}p(j|m)e^{-ik(j-m)},
\end{align}
which takes the same form as in the case with unitary feedback control [see Eq.~\eqref{eq_chiral_xi0}]. Thus, regardless of whether feedback control is unitary or not, the eigenvalues of the quantum channel and its topology are completely determined by the probability distribution $p(j|m)$ of the position after the feedback.

\begin{figure}[t]
    \includegraphics[width=8.0cm]{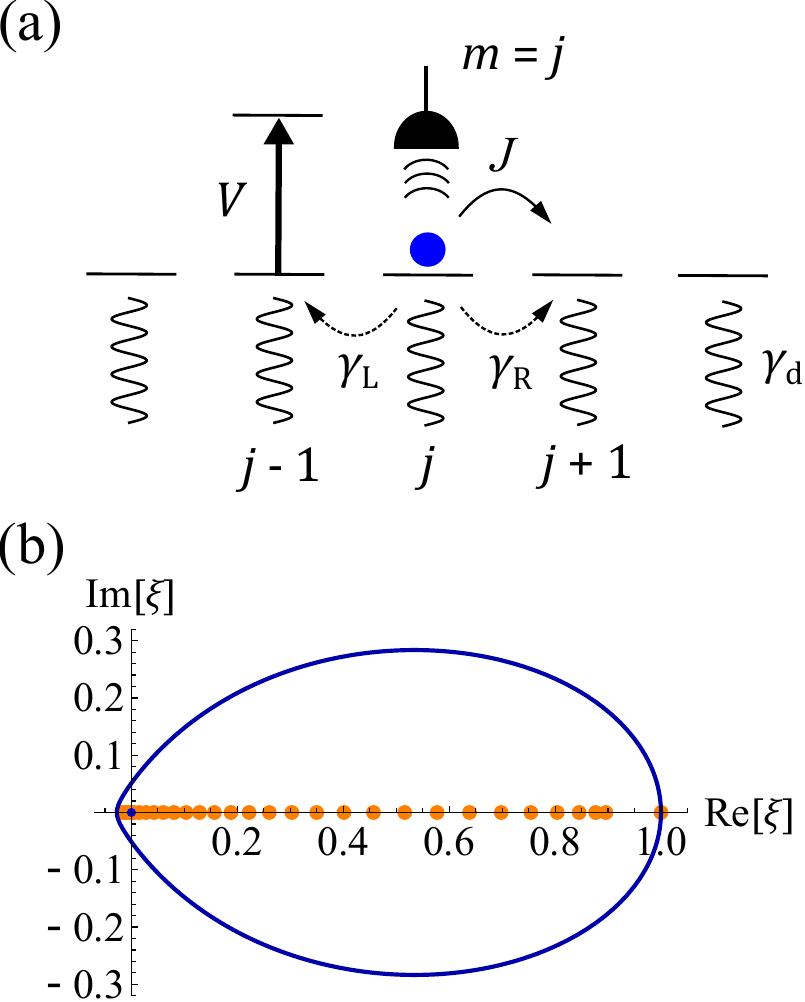}
    \caption{(a) Schematic illustration of a chiral Maxwell demon subject to dissipation. (b) Eigenspectrum of a quantum channel for a chiral Maxwell demon subject to dissipation. The parameters are set to $J=1,V=10,\gamma_{\mathrm{d}}=0.5,\gamma_{\mathrm{L}}=\gamma_{\mathrm{R}}=0.25$, and $\tau=1$. A blue curve and blue dots at the origin constitute the eigenspectrum under the PBC in the limit of infinite system size, and orange dots show the eigenspectrum under the OBC with system size $L=30$.}
    \label{fig_diss_demon}
\end{figure}

To demonstrate the robustness of the chiral Maxwell demon against dissipation during feedback control, we consider the Lindblad operators
\begin{align}
    \hat{L}_{m,j}^{(\mathrm{d})}=&\sqrt{\gamma_{\mathrm{d}}}\ket{j}\bra{j},\\
    \hat{L}_{m,j}^{(\mathrm{L})}=&\sqrt{\gamma_{\mathrm{L}}}\ket{j-1}\bra{j},\\
    \hat{L}_{m,j}^{(\mathrm{R})}=&\sqrt{\gamma_{\mathrm{R}}}\ket{j+1}\bra{j},
\end{align}
which, respectively, describe dephasing, stochastic hopping to the left, and stochastic hopping to the right [see Fig.~\ref{fig_diss_demon}(a) for a schematic illustration]. Figure \ref{fig_diss_demon}(b) shows the eigenspectrum of the quantum channel for the chiral Maxwell demon subject to dissipation. As can be seen from the figure, the loop structure of the eigenspectrum under the PBC is robust against dissipation, leading to the non-Hermitian skin effect under the OBC. This result indicates that the chiral Maxwell demon can rectify the noise induced by dissipation as well as quantum fluctuations induced by unitary evolution and achieve unidirectional transport that is robust against noise and decoherence.

\begin{figure}[t]
    \includegraphics[width=8.0cm]{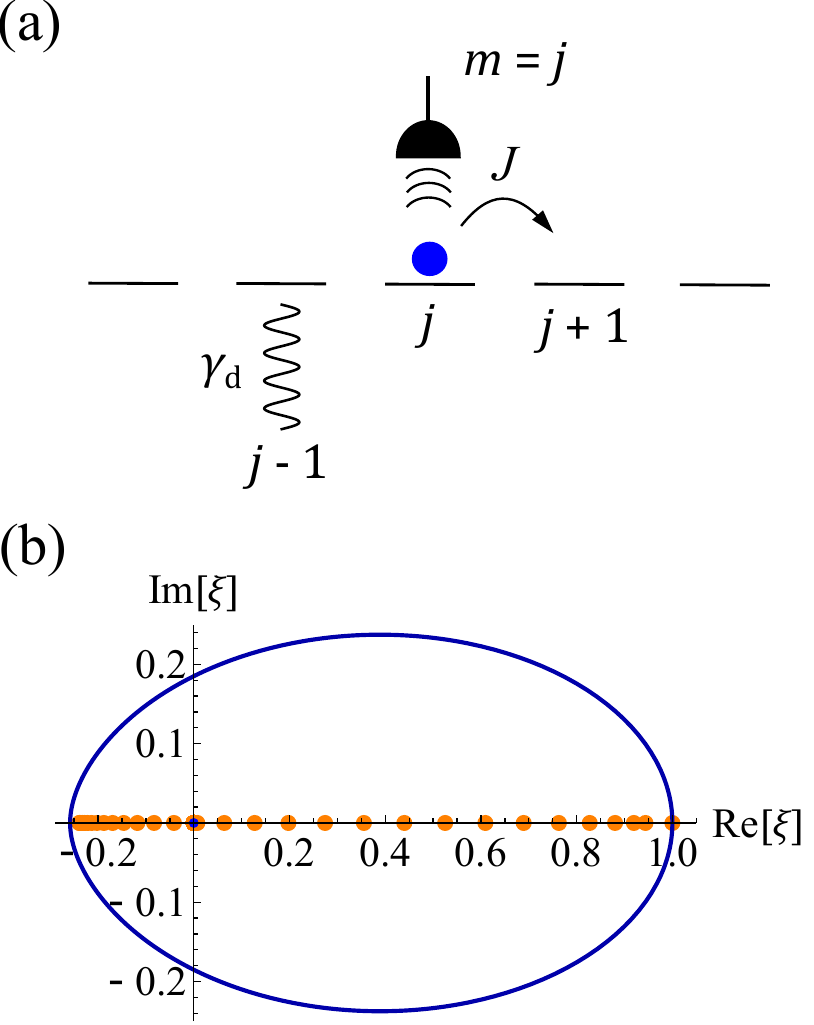}
    \caption{(a) Schematic illustration of topological feedback control with engineered dissipation. (b) Eigenspectrum of a quantum channel for topological feedback control with engineered dissipation. The parameters are set to $J=1,\gamma_{\mathrm{d}}=5$, and $\tau=1$. A blue curve and blue dots at the origin constitute the eigenspectrum under the PBC in the limit of infinite system size, and orange dots show the eigenspectrum under the OBC with system size $L=30$.}
    \label{fig_diss_eng}
\end{figure}

\subsection{Feedback control with engineered dissipation}

Dissipative feedback control not only shows the stability of topological feedback control against noise, but it can also be exploited to utilize dissipation as an essential tool for feedback control. Such a possibility can be envisaged since recent developments in quantum technologies with atoms, ions, and photons have enabled one to realize engineered dissipation \cite{Mueller12}. For example, localized dissipation can be introduced in ultracold atoms, where the strength of dissipation is controlled by laser fields \cite{Barontini13,Corman19,Huang23,Tonielli19,Dolgirev20}. Motivated by the possibility of dissipation engineering, we propose a topological feedback control that is enabled by dissipation. 

The setup of feedback control with engineered dissipation is as follows. We consider a spinless particle on a 1D lattice, perform a projective position measurement on it, and obtain a measurement outcome $m$. Next, we let the particle undergo dissipative time evolution under the Lindblad equation \eqref{eq_fb_Lindblad}. Here, we assume that the Hamiltonian does not depend on the measurement outcome $m$ and that it is given by a simple tight-binding model
\begin{align}
    \hat{H}_m=\hat{H}:=-J\sum_j(\ket{j}\bra{j+1}+\mathrm{H.c.})
    \label{eq_eng_diss_Hamil}
\end{align}
with $J>0$ being the hopping amplitude. 
The dependence of the conditioned Lindblad equation \eqref{eq_fb_Lindblad} on the measurement outcome is encoded in the Lindblad operator, where engineered dissipation is described by
\begin{equation}
    \hat{L}_{m,j-1}=\delta_{m,j}\sqrt{\gamma_{\mathrm{d}}}\ket{j-1}\bra{j-1}.
\end{equation}
In other words, we introduce localized dissipation to the left side of the particle [see Fig.~\ref{fig_diss_eng}(a) for a schematic illustration]. Here, we exploit the continuous quantum Zeno effect \cite{Daley14} in the strong-dissipation regime $\gamma_{\mathrm{d}}\gg J$ to suppress the hopping of the particle to the site that is subject to strong dissipation. Because of this feedback control of dissipation, hopping of a particle to the right is facilitated, and chiral transport is eventually achieved.

The eigenspectrum of a quantum channel for this dissipative feedback control is shown in Fig.~\ref{fig_diss_eng}(b). The eigenspectrum under the PBC forms a loop, and that under the OBC displays the non-Hermitian skin effect. Here, the loop structure of the eigenspectrum under the PBC is realized due to the dissipation since the Hamiltonian \eqref{eq_eng_diss_Hamil} does not contain a wall potential. Thus, the topological feedback control and the associated chiral transport are achieved by dissipation.

\section{Conclusion and outlook\label{sec_conclusion}}

In this paper, we have presented a class of dynamical topological phases realized with feedback control, which are characterized by the topology of quantum channels. The nontrivial point-gap topology of a quantum channel characterized by a winding number ensures robust chiral transport, which is analogous to the chiral edge current in the quantum Hall effect and leads to the emergence of the non-Hermitian skin effect under the open boundary condition. We have also provided a symmetry classification of quantum feedback control, identified ten classes compatible with projective measurements, and constructed examples of feedback control in each symmetry class. This general framework of symmetry and topology of quantum feedback control provides a guiding principle for designing stable feedback control of quantum systems since topology is insensitive to small disturbances. Furthermore, feedback control offers a unique way to control dynamical topological phases of matter distinct from other well-known classes such as periodically driven systems \cite{Thouless83,Oka09,Kitagawa10,Kitagawa10_2,Kitagawa11,Lindner11,Nathan15,RoyHarper17,Higashikawa19,Oka19,Rudner20,Harper20}, non-Hermitian systems \cite{Esaki11,Gong18,Kawabata19,Zhou19,Bergholtz21,Liu22,Okuma23}, and open quantum systems subject to engineered dissipation \cite{Bardyn13,Lieu20}.

Topological feedback control is expected to be realized in various experimental platforms. For example, the past decade has witnessed remarkable progress in site-resolved measurement techniques and control of ultracold atoms in optical lattices \cite{Bakr09,Weitenberg11,Gross21}. More recently, atoms in optical tweezer arrays have emerged as highly controllable quantum simulators in which individual atoms can be measured and addressed in a programmable manner \cite{Barredo16,Bernien17,Barredo18,Ebadi21,Bluvstein22,Bluvstein23,Browaeys20}. The high-precision position measurement and subsequent local control of a Hamiltonian for topological feedback control can be achieved by these state-of-the-art techniques. In addition to these promising platforms, quantum feedback control has widely been used in photonic systems \cite{Sayrin11,Vidrighin16}, levitated nanoparticles \cite{Kamba21,Tebbenjohanns21,Magrini21}, and superconducting qubits \cite{Vijay12,Cottet17,Nagihloo18,Masuyama18}. Given the rapid development in quantum control, topological feedback control in various experimental platforms will expand the scope of topological phases of matter.

We conclude this paper by discussing future perspectives and open problems. 
First, while we have mostly focused on feedback control with projective measurements, the present framework for topology of quantum channels is also applicable to general quantum measurements; therefore, it is natural to ask whether nonprojective measurements can be harnessed to construct topological feedback control. In particular, since the symmetry classification in Sec.~\ref{sec_BL_sym} is based on the spectral property of projective measurement channels, BL symmetry classes other than those in Table \ref{table_AZdag} are not prohibited for general quantum measurements. Since a quantum channel in a symmetry class beyond the tenfold way necessarily has a pair of eigenvalues $\pm 1$ (if one assumes the existence of a steady state), its eigenvalue spectrum is reminiscent of that of dissipative time crystals \cite{Gong18,Gong18_3}, where the eigenvalue $-1$ implies oscillating components under repeated application of a quantum channel.

Second, as discussed in Sec.~\ref{sec_top_CPTP}, a crucial distinction between quantum channels and general non-Hermitian matrices is complete positivity and trace preservation, which should be satisfied by any time evolution of the density matrix of a quantum system \cite{NielsenChuang_book}. Although the topological classification of non-Hermitian matrices without the CPTP condition has been obtained \cite{Kawabata19,Zhou19}, it is unclear whether the CPTP condition alters the topological classification. For example, in Sec.~\ref{sec_sym_proj_example}, we have shown that the CPTP condition places a nontrivial constraint on the Bloch matrix with transposition symmetry, which turns out to be characterized by an integer topological invariant rather than a $\mathbb{Z}_2$ invariant. It merits further study to investigate the generality of this phenomenon. From a broader perspective, this problem may be related to a classification of time evolution of open quantum systems described by CPTP maps. 

Last but not least, topological feedback control suggests a novel connection between topology and Maxwell's demon. For example, an experimental realization of Maxwell's demon in Ref.~\cite{Toyabe10} demonstrated chiral transport of a particle due to feedback control, in close analogy with the chiral Maxwell demon in Sec.~\ref{sec_model} (see also Ref.~\cite{LiuNakagawaUeda23}). Maxwell's demon has attracted century-long interest in thermodynamics and statistical physics, and an information-theoretic understanding of the role of Maxwell's demon has led to the integration of thermodynamics and information theory \cite{Parrondo15}. Thus, investigating the topology of feedback control may shed light on a hitherto unnoticed role of topology in thermodynamics.

\begin{acknowledgments}
M.N. acknowledges helpful discussions with Yuto Ashida, Kohei Kawabata, Masatoshi Sato, Kenji Shimomura, and Ken Shiozaki during the workshop YITP-T-24-03 on ``Recent Developments and Challenges in Topological Phases'' held at the Yukawa Institute for Theoretical Physics.
This work was supported by KAKENHI Grants No.~JP20K14383, No.~JP22H01152, and No.~JP24K16989 from the Japan Society for the Promotion of Science.
We gratefully acknowledge support from the CREST program ``Quantum Frontiers" (Grant No.~JPMJCR23I1) by the Japan Science and Technology Agency.
\end{acknowledgments}

\appendix

\section{Bloch theory of quantum channels\label{sec_Bloch}}

The Bloch theory lies at the heart of topological band theory \cite{MoessnerMoore_book}. However, how to generalize this theory to feedback-controlled quantum systems is nontrivial since feedback control involves not only unitary time evolution but also measurement processes. Here, we develop a Bloch theory of quantum channels described by CPTP maps. The following formalism is not limited to discrete quantum feedback control but can generally be applied to any translationally invariant quantum channels in single-particle systems.

\subsection{CPTP map as a non-Hermitian operator\label{sec_CJ}}

We begin by representing a CPTP map in terms of a non-Hermitian matrix. We consider a single-particle quantum system on a lattice (see Sec.~\ref{sec_top_setup} for details). We first vectorize the density matrix
    \begin{equation}
        \hat{\rho}=\sum_{\vec{i},\vec{j}}\sum_{a,b}\rho_{\vec{i},a;\vec{j},b}\ket{\vec{i},a}\bra{\vec{j},b}
        \label{eq_rho_CJ}
    \end{equation}
    by using the Choi-Jamio\l kowski isomorphism \cite{Jamiolkowski72,Choi75,Jiang13,Wilde_textbook} as
    \begin{align}
    \ket{\hat{\rho}}:=&(\hat{\rho}\otimes \hat{I})\sqrt{D}\ket{\Omega}\notag\\
    =&\sum_{\vec{i},\vec{j}}\sum_{a,b}\rho_{\vec{i},a;\vec{j},b}\ket{\vec{i},a}\otimes\ket{\vec{j},b},
    \label{eq_ketrho}
    \end{align}
    where $\{\ket{\vec{i},a}\}$ is an orthonormal basis set of the $D$-dimensional Hilbert space $\mathcal{H}$ of a single-particle system on a lattice ($D=N_{\mathrm{cell}}D_{\mathrm{loc}}$; see Sec.~\ref{sec_top_setup}), and
    \begin{equation}
    \ket{\Omega}=\frac{1}{\sqrt{D}}\sum_{\vec{i}}\sum_{a}\ket{\vec{i},a}\otimes\ket{\vec{i},a}
    \end{equation}
    is a maximally entangled state on a doubled Hilbert space $\mathcal{H}\otimes\mathcal{H}$. Consequently, we obtain a mapping from a CPTP map
    \begin{equation}
    \mathcal{E}(\hat{\rho})=\sum_m \hat{K}_m\hat{\rho} \hat{K}_m^\dag
    \end{equation}
    to a non-Hermitian operator $\tilde{\mathcal{E}}$ acting on $\mathcal{H}\otimes\mathcal{H}$ as
    \begin{align}
    \tilde{\mathcal{E}}\ket{\hat{\rho}}:=&\ket{\mathcal{E}(\hat{\rho})}\notag\\
    =&(\mathcal{E}(\hat{\rho})\otimes \hat{I})\sqrt{D}\ket{\Omega}\notag\\
    =&\sum_m\sum_{\vec{i},\vec{j},\vec{l}}\sum_{a,b,c}\rho_{\vec{i},a;\vec{j},b}\hat{K}_m\ket{\vec{i},a}\bra{\vec{j},b}\hat{K}_m^\dag\ket{\vec{l},c}\otimes\ket{\vec{l},c}\notag\\
    =&\sum_m\sum_{\vec{i},\vec{j}}\sum_{a,b}\rho_{\vec{i},a;\vec{j},b}\hat{K}_m\ket{\vec{i},a}\otimes \hat{K}_m^*\ket{\vec{j},b}\notag\\
    =&\Bigl(\sum_m\hat{K}_m\otimes \hat{K}_m^*\Bigr)\ket{\hat{\rho}}.
    \label{eq_mat_rep_derivation}
    \end{align}
    Thus, the matrix representation of $\mathcal{E}$ is
    \begin{equation}
    \tilde{\mathcal{E}}=\sum_m\hat{K}_m\otimes \hat{K}_m^*.
    \end{equation}
    The eigenvalues of $\tilde{\mathcal{E}}$ are the same as those of $\mathcal{E}$ since
    \begin{align}
    \tilde{\mathcal{E}}\ket{\hat{\rho}_n^R}=&\ket{\mathcal{E}(\hat{\rho}_n^R)}\notag\\
    =&\ket{\xi_n\hat{\rho}_n^R}\notag\\
    =&\xi_n\ket{\hat{\rho}_n^R},
    \label{eq_eigen_duality}
    \end{align}
    where we use Eq.~\eqref{eq_ketrho} in deriving the last equality.

\subsection{Momentum-space representation\label{sec_mom}}

Here, we consider a translationally invariant quantum channel satisfying
    \begin{equation}
    \sum_m(\hat{T}_\lambda \hat{K}_m\hat{T}_\lambda^{-1})\hat{\rho}(\hat{T}_\lambda \hat{K}_m\hat{T}_\lambda^{-1})^\dag=\sum_m\hat{K}_m\hat{\rho} \hat{K}_m^\dag
    \label{eq_app_trans_sym1}
    \end{equation}
    for arbitrary $\hat{\rho}$, with $\hat{T}_\lambda\ (\lambda=1,\cdots,d)$ being a translation operator:
    \begin{equation}
    \hat{T}_\lambda\ket{\vec{i},a}=\ket{\vec{i}+\vec{e}_\lambda,a}.
    \end{equation}
In the matrix representation $\tilde{\mathcal{E}}$ of the CPTP map, the translational symmetry in Eq.~\eqref{eq_app_trans_sym1} reads
    \begin{align}
    (\hat{T}_\lambda\otimes \hat{T}_\lambda^*)\tilde{\mathcal{E}}(\hat{T}_\lambda\otimes \hat{T}_\lambda^*)^{-1}=\tilde{\mathcal{E}}.
    \label{eq_app_trans_sym2}
    \end{align}
In the context of open quantum systems, this symmetry can be regarded as weak symmetry of the CPTP map (see Sec.~\ref{sec_unitary_sym}) \cite{Buca12, Albert14}. In fact, since the operator $\tilde{\mathcal{E}}$ commutes with the tensor-product translation operator $\hat{T}_\lambda\otimes \hat{T}_\lambda^*$ appearing in Eq.~\eqref{eq_app_trans_sym2}, the operator $\tilde{\mathcal{E}}$ can be block diagonalized with respect to eigenvalues of $\hat{T}_\lambda\otimes \hat{T}_\lambda^*$. Eigenvectors of $\hat{T}_\lambda\otimes \hat{T}_\lambda^*$ are constructed from the eigenstates 
\begin{equation}
    \ket{\vec{k},a}=\frac{1}{\sqrt{N_{\mathrm{cell}}}}\sum_{\vec{j}}e^{i\vec{k}\cdot\vec{R}_{\vec{j}}}\ket{\vec{j},a}   
\end{equation}
of the translation operator $\hat{T}_\lambda$ with eigenvalue $e^{-i\vec{k}\cdot\vec{a}_\lambda}$ as
\begin{align}
(\hat{T}_\lambda\otimes \hat{T}_\lambda^*)\ket{\vec{k},a}\otimes\ket{-\vec{k}',b}=e^{-i(\vec{k}-\vec{k}')\cdot\vec{a}_\lambda}\ket{\vec{k},a}\otimes\ket{-\vec{k}',b}.
\label{eq_trans_eigen}
\end{align}
Thus, eigenvectors of $\tilde{\mathcal{E}}$ are classified according to the eigenvalues of $\hat{T}_\lambda\otimes \hat{T}_\lambda^*$. Since the vector $\ket{\vec{k},a}\otimes\ket{-\vec{k}',b}$ corresponds to an operator $\ket{\vec{k},a}\bra{\vec{k}',b}$ through the Choi-Jamio\l kowski isomorphism [see Eqs.~\eqref{eq_rho_CJ} and \eqref{eq_ketrho}], $\vec{k}-\vec{k}'$ in Eq.~\eqref{eq_trans_eigen} is physically interpreted as the difference between crystal momenta of the ket and bra degrees of freedom of the density matrix. 

By using the expression of the Kraus operator
\begin{equation}
\hat{K}_m=\sum_{\vec{i},\vec{j}}\sum_{a,b}(\hat{K}_m)_{\vec{i},a;\vec{j},b}\ket{\vec{i},a}\bra{\vec{j},b},
\end{equation}
the matrix representation $\tilde{\mathcal{E}}$ of the CPTP map is written as
\begin{align}
\tilde{\mathcal{E}}=&\sum_m\hat{K}_m\otimes \hat{K}_m^*\notag\\
=&\sum_m\sum_{\vec{i},\vec{j},\vec{i}',\vec{j}'}\sum_{a,b,c,d}(\hat{K}_m)_{\vec{i},a;\vec{j},b}(\hat{K}_m)^*_{\vec{i}',c;\vec{j}',d}\notag\\
&\times\ket{\vec{i},a}\bra{\vec{j},b}\otimes\ket{\vec{i}',c}\bra{\vec{j}',d}\notag\\
=&\sum_{\vec{i},\vec{j}}\sum_{a,b,c,d}\sum_{\vec{\mu},\vec{\nu}}t_{\vec{i},\vec{j}}^{a,c,\vec{\mu};b,d,\vec{\nu}}\notag\\
&\times\ket{\vec{i},a}\bra{\vec{j},b}\otimes\ket{\vec{i}+\vec{\mu},c}\bra{\vec{j}+\vec{\nu},d},
\label{eq_Etilde_app}
\end{align}
where
\begin{equation}
t_{\vec{i},\vec{j}}^{a,c,\vec{\mu};b,d,\vec{\nu}}:=\sum_m(\hat{K}_m)_{\vec{i},a;\vec{j},b}(\hat{K}_m)^*_{\vec{i}+\vec{\mu},c;\vec{j}+\vec{\nu},d}.
\label{eq_hopping_def}
\end{equation}
For later convenience, we assume that components $\mu_\lambda$ and $\nu_\lambda$ in $\vec{\mu}=(\mu_1,\cdots,\mu_d)$ and $\vec{\nu}=(\nu_1,\cdots,\nu_d)$ take a value from $-L/2, -L/2+1, \cdots, L/2-1$ [$-(L-1)/2, -(L-1)/2+1, \cdots, (L-1)/2$] for even (odd) $L$ without loss of generality by using the periodic boundary condition \eqref{eq_PBC}.
The translational symmetry \eqref{eq_app_trans_sym2} exists if
\begin{equation}
t_{\vec{i}+\vec{e}_\lambda,\vec{j}+\vec{e}_\lambda}^{a,c,\vec{\mu};b,d,\vec{\nu}}=t_{\vec{i},\vec{j}}^{a,c,\vec{\mu};b,d,\vec{\nu}},
\label{eq_trans_sym_t}
\end{equation}
where $\vec{e}_\lambda$ is a unit vector defined by $(\vec{e}_\lambda)_{\lambda'}=\delta_{\lambda,\lambda'}$. 
For notational simplicity, we introduce an auxiliary creation operator $\tilde{c}_{\vec{i},a,c,\vec{\mu}}^\dag$ as
\begin{equation}
\tilde{c}_{\vec{i},a,c,\vec{\mu}}^\dag(\ket{\mathrm{vac}}\otimes\ket{\mathrm{vac}})=\ket{\vec{i},a}\otimes\ket{\vec{i}+\vec{\mu},c},
\label{eq_c_i_app}
\end{equation}
where $\ket{\mathrm{vac}}$ is the vacuum state of the system. 
Then, the matrix representation of a translationally invariant CPTP map takes the form of a translationally invariant non-Hermitian tight-binding model as
    \begin{equation}
    \tilde{\mathcal{E}}=\sum_{\vec{i},\vec{j}}\sum_{a,b,c,d}\sum_{\vec{\mu},\vec{\nu}}t_{\vec{i},\vec{j}}^{a,c,\vec{\mu};b,d,\vec{\nu}}\tilde{c}_{\vec{i},a,c,\vec{\mu}}^\dag \tilde{c}_{\vec{j},b,d,\vec{\nu}}.
    \label{eq_tightbinding}
    \end{equation}
Here, we consider a single-particle system, although Eq.~\eqref{eq_tightbinding} is written like the second-quantized form.
We also note that the operator $\tilde{c}_{\vec{i},a,c,\vec{\mu}}^\dag$ has two additional indices $c$ and $\vec{\mu}$ compared with the original basis $\ket{\vec{i},a}$ of the Hilbert space since $\tilde{\mathcal{E}}$ acts on the doubled Hilbert space $\mathcal{H}\otimes\mathcal{H}$. Physically, this point corresponds to the fact that the density matrix has bra and ket degrees of freedom. 

Using the translational invariance, we obtain the momentum-space representation of $\tilde{\mathcal{E}}$ as
    \begin{align}
    \tilde{\mathcal{E}}=&\sum_{\vec{k}}\tilde{\bm{c}}_{\vec{k}}^\dag X(\vec{k})\tilde{\bm{c}}_{\vec{k}}\notag\\
    =&\sum_{\vec{k}}\sum_{a,b,c,d}\sum_{\vec{\mu},\vec{\nu}}X_{a,c,\vec{\mu};b,d,\vec{\nu}}(\vec{k})\tilde{c}_{\vec{k},a,c,\vec{\mu}}^\dag \tilde{c}_{\vec{k},b,d,\vec{\nu}},
    \label{eq_Etilde_k_app}
    \end{align}
where $\tilde{\bm{c}}_{\vec{k}}=(\tilde{c}_{\vec{k},a,c,\vec{\mu}})_{a,c,\vec{\mu}}$ is a column vector with
    \begin{equation}
    \tilde{c}_{\vec{k},a,c,\vec{\mu}}:=\frac{1}{\sqrt{N_{\mathrm{cell}}}}\sum_{\vec{j}}\tilde{c}_{\vec{j},a,c,\vec{\mu}}e^{-i\vec{k}\cdot\vec{R}_{\vec{j}}}
    \label{eq_c_k_app}
    \end{equation}
and
    \begin{align}
    X_{a,c,\vec{\mu};b,d,\vec{\nu}}(\vec{k}):=&\frac{1}{N_{\mathrm{cell}}}\sum_{\vec{j},\vec{j}'}t_{\vec{j},\vec{j}'}^{a,c,\vec{\mu};b,d,\vec{\nu}}e^{-i\vec{k}(\vec{R}_{\vec{j}}-\vec{R}_{\vec{j}'})}\notag\\
    =&\frac{1}{N_{\mathrm{cell}}}\sum_{\vec{j},\vec{j}'}\sum_m(\hat{K}_m)_{\vec{j},a;\vec{j}',b}(\hat{K}_m)^*_{\vec{j}+\vec{\mu},c;\vec{j}'+\vec{\nu},d}\notag\\
    &\times e^{-i\vec{k}\cdot(\vec{R}_{\vec{j}}-\vec{R}_{\vec{j}'})}.
    \label{eq_Bloch_app}
    \end{align}
The $N_{\mathrm{cell}}D_{\mathrm{loc}}^2\times N_{\mathrm{cell}}D_{\mathrm{loc}}^2$ matrix $X(\vec{k})$ is a counterpart of the Bloch Hamiltonian in band theory; therefore, we call it the Bloch matrix.

\subsection{Mode expansion\label{sec_mode_expansion}}

Let us express the density matrix $\hat{\rho}_{\mathrm{f}}=\mathcal{E}(\hat{\rho}_{\mathrm{i}})$ after feedback control in terms of eigenvalues and eigenvectors of the Bloch matrix $X(\vec{k})$. Let $v_n^L(\vec{k})$ [$v_n^R(\vec{k})$] be a left (right) eigenvector of $X(\vec{k})$ with eigenvalue $\xi_n(\vec{k})$. We assume that the Bloch matrix $X(\vec{k})$ is diagonalizable and that the eigenvectors satisfy the biorthogonal condition $(v_m^L(\vec{k}),v_n^R(\vec{k}))=\delta_{m,n}$ [here, $(u,v)$ denotes the inner product of vectors $u$ and $v$] 
and the completeness relation $\sum_n v_n^R(\vec{k})(v_n^{L}(\vec{k}))^\dag=1$. Then, the Bloch matrix is written as
    \begin{gather}
    X(\vec{k})=\sum_n\xi_n(\vec{k})v_n^R(\vec{k})(v_n^{L}(\vec{k}))^\dag,
    \end{gather}
    the matrix element of which is given by
    \begin{gather}
    X_{a,c,\vec{\mu};b,d,\vec{\nu}}(\vec{k})=\sum_n\xi_n(\vec{k})(v_n^R(\vec{k}))_{a,c,\vec{\mu}}(v_n^L(\vec{k}))^*_{b,d,\vec{\nu}}.
    \end{gather}
    Substituting this result into Eq.~\eqref{eq_Etilde_k_app}, we have
    \begin{align}
    \tilde{\mathcal{E}}=&\sum_{\vec{k}}\sum_n\sum_{a,b,c,d}\sum_{\vec{\mu},\vec{\nu}}\xi_n(\vec{k})(v_n^R(\vec{k}))_{a,c,\vec{\mu}}(v_n^L(\vec{k}))^*_{b,d,\vec{\nu}}\notag\\
    &\times\tilde{c}_{\vec{k},a,c,\vec{\mu}}^\dag \tilde{c}_{\vec{k},b,d,\vec{\nu}}\notag\\
    =&\sum_{\vec{k}}\sum_n\xi_n(\vec{k})\bar{a}_n(\vec{k})a_n(\vec{k}),
    \end{align}
where
    \begin{align}
    \bar{a}_n(\vec{k}):=&\sum_{a,c}\sum_{\vec{\mu}} (v_n^R(\vec{k}))_{a,c,\vec{\mu}}\tilde{c}_{\vec{k},a,c,\vec{\mu}}^\dag,\\
    a_n(\vec{k}):=&\sum_{b,d}\sum_{\vec{\nu}} (v_n^L(\vec{k}))^*_{b,d,\vec{\nu}}\tilde{c}_{\vec{k},b,d,\vec{\nu}}.
    \end{align}
From Eqs.~\eqref{eq_c_i_app} and \eqref{eq_c_k_app}, we find that the operator $\tilde{c}_{\vec{k},a,c,\vec{\mu}}^\dag$ creates a state
    \begin{equation}
    \tilde{c}_{\vec{k},a,c,\vec{\mu}}^\dag(\ket{\mathrm{vac}}\otimes\ket{\mathrm{vac}})=\frac{1}{\sqrt{N_{\mathrm{cell}}}}\sum_{\vec{j}}e^{i\vec{k}\cdot\vec{R}_{\vec{j}}}\ket{\vec{j},a}\otimes\ket{\vec{j}+\vec{\mu},c},
    \end{equation}
which is an eigenstate of $\hat{T}_\lambda\otimes \hat{T}_\lambda^*$ with eigenvalue $e^{-i\vec{k}\cdot\vec{a}_\lambda}$. Thus, the right eigenvector of $\tilde{\mathcal{E}}$ with eigenvalue $\xi_n(\vec{k})$ is given by
    \begin{align}
    \ket{\hat{\rho}_n^R(\vec{k})}=&\frac{1}{\sqrt{N_{\mathrm{cell}}}}\sum_{\vec{j}}\sum_{a,c}\sum_{\vec{\mu}} (v_n^R(\vec{k}))_{a,c,\vec{\mu}} e^{i\vec{k}\cdot\vec{R}_{\vec{j}}}\notag\\
    &\times\ket{\vec{j},a}\otimes\ket{\vec{j}+\vec{\mu},c},
    \end{align}
and the corresponding left eigenvector is given by
    \begin{align}
    \ket{\hat{\rho}_n^L(\vec{k})}=&\frac{1}{\sqrt{N_{\mathrm{cell}}}}\sum_{\vec{j}}\sum_{b,d}\sum_{\vec{\nu}} (v_n^L(\vec{k}))_{b,d,\vec{\nu}} e^{i\vec{k}\cdot\vec{R}_{\vec{j}}}\notag\\
    &\times\ket{\vec{j},b}\otimes\ket{\vec{j}+\vec{\nu},d}.
    \end{align}
    Using the Choi-Jamio\l kowski isomorphism [see Eq.~\eqref{eq_eigen_duality}], we obtain the right and left eigenoperators of the original CPTP map $\mathcal{E}$ as
    \begin{align}
    \hat{\rho}_n^R(\vec{k})=&\frac{1}{\sqrt{N_{\mathrm{cell}}}}\sum_{\vec{j}}\sum_{a,c}\sum_{\vec{\mu}} (v_n^R(\vec{k}))_{a,c,\vec{\mu}} e^{i\vec{k}\cdot\vec{R}_{\vec{j}}}\notag\\
    &\times\ket{\vec{j},a}\bra{\vec{j}+\vec{\mu},c},
    \label{eq_r_eigenop_k}
    \end{align}
    and
    \begin{align}
    \hat{\rho}_n^L(\vec{k})=&\frac{1}{\sqrt{N_{\mathrm{cell}}}}\sum_{\vec{j}}\sum_{b,d}\sum_{\vec{\nu}} (v_n^L(\vec{k}))_{b,d,\vec{\nu}} e^{i\vec{k}\cdot\vec{R}_{\vec{j}}}\notag\\
    &\times\ket{\vec{j},b}\bra{\vec{j}+\vec{\nu},d},
    \label{eq_l_eigenop_k}
    \end{align}
    respectively. In terms of the eigensystem of the CPTP map, the density matrix $\hat{\rho_{\mathrm{f}}}$ after feedback control is expanded as [see Eq.~\eqref{eq_mode_expansion}]
    \begin{align}
    \hat{\rho_{\mathrm{f}}}=\mathcal{E}(\hat{\rho}_{\mathrm{i}})=&\sum_{\vec{k}}\sum_n\xi_n(\vec{k})\mathrm{Tr}[(\hat{\rho}_n^{L}(\vec{k}))^\dag\hat{\rho}_{\mathrm{i}}]\hat{\rho}_n^R(\vec{k}).
    \label{eq_expansion_k}
    \end{align}
    Note that $\langle\hat{\rho}_m^L(\vec{k}),\hat{\rho}_n^R(\vec{k})\rangle=\delta_{mn}$.

\subsection{Fundamental constraints on the Bloch matrix\label{sec_Bloch_constraint}}

    Since a quantum channel must satisfy the complete positivity and the trace preservation, the Bloch matrix $X(\vec{k})$ cannot be a general non-Hermitian matrix. We note that a superoperator satisfies the complete positivity if it has the Kraus representation \eqref{eq_CPTP} (see e.g., Ref.~\cite{Wilde_textbook}). 
    The Kraus representation implies Hermiticity preservation, which leads to an antiunitary symmetry \eqref{eq_mod_conj_sym} of a quantum channel (see Sec.~\ref{sec_mod_conj_sym}). Using Eq.~\eqref{eq_tightbinding}, the right-hand side of the symmetry condition \eqref{eq_mod_conj_sym} is written as
    \begin{align}
        \tilde{\mathcal{E}}
            =&\sum_{\vec{i},\vec{j}}\sum_{a,b,c,d}\sum_{\vec{\mu},\vec{\nu}}t_{\vec{i},\vec{j}}^{a,c,\vec{\mu};b,d,\vec{\nu}}\notag\\
        &\times\ket{\vec{i},a}\bra{\vec{j},b}\otimes\ket{\vec{i}+\vec{\mu},c}\bra{\vec{j}+\vec{\nu},d},
        \end{align}
    and the left-hand side of Eq.~\eqref{eq_mod_conj_sym} is given by
    \begin{align}
    \tilde{J}\tilde{\mathcal{E}}\tilde{J}^{-1}=&\sum_{\vec{i},\vec{j}}\sum_{a,b,c,d}\sum_{\vec{\mu},\vec{\nu}}(t_{\vec{i},\vec{j}}^{a,c,\vec{\mu};b,d,\vec{\nu}})^*\notag\\
    &\times\ket{\vec{i}+\vec{\mu},c}\bra{\vec{j}+\vec{\nu},d}\otimes\ket{\vec{i},a}\bra{\vec{j},b}\notag\\
    =&\sum_{\vec{i},\vec{j}}\sum_{a,b,c,d}\sum_{\vec{\mu},\vec{\nu}}(t_{\vec{i}+\vec{\mu},\vec{j}+\vec{\nu}}^{c,a,-\vec{\mu};d,b,-\vec{\nu}})^*\notag\\
    &\times\ket{\vec{i},a}\bra{\vec{j},b}\otimes\ket{\vec{i}+\vec{\mu},c}\bra{\vec{j}+\vec{\nu},d}.
    \end{align}
    Thus, the symmetry \eqref{eq_mod_conj_sym} holds if
    \begin{equation}
        (t_{\vec{i}+\vec{\mu},\vec{j}+\vec{\nu}}^{c,a,-\vec{\mu};d,b,-\vec{\nu}})^*=t_{\vec{i},\vec{j}}^{a,c,\vec{\mu};b,d,\vec{\nu}}.
        \label{eq_mod_conj_cond_hopping}
    \end{equation}
    We can check Eq.~\eqref{eq_mod_conj_cond_hopping} directly from the definition \eqref{eq_hopping_def} as
    \begin{align}
    (t_{\vec{i}+\vec{\mu},\vec{j}+\vec{\nu}}^{c,a,-\vec{\mu},d,b,-\vec{\nu}})^*=&\sum_m(\hat{K}_m)^*_{\vec{i}+\vec{\mu},c;\vec{j}+\vec{\nu},d}(\hat{K}_m)_{\vec{i},a;\vec{j},b}\notag\\
    =&t_{\vec{i},\vec{j}}^{a,c,\vec{\mu};b,d,\vec{\nu}}.
    \end{align}
    In terms of the Bloch matrix, the relation \eqref{eq_mod_conj_cond_hopping} leads to
    \begin{align}
    X_{a,c,\vec{\mu};b,d,\vec{\nu}}(\vec{k})=&\frac{1}{N_{\mathrm{cell}}}\sum_{\vec{j},\vec{j}'}t_{\vec{j},\vec{j}'}^{a,c,\vec{\mu};b,d,\vec{\nu}}e^{-i\vec{k}\cdot(\vec{R}_{\vec{j}}-\vec{R}_{\vec{j}'})}\notag\\
    =&\frac{1}{N_{\mathrm{cell}}}\sum_{\vec{j},\vec{j}'}(t_{\vec{j}+\vec{\mu},\vec{j}'+\vec{\nu}}^{c,a,-\vec{\mu};d,b,-\vec{\nu}})^*e^{-i\vec{k}\cdot(\vec{R}_{\vec{j}}-\vec{R}_{\vec{j}'})}\notag\\
    =&\frac{1}{N_{\mathrm{cell}}}\sum_{\vec{l},\vec{l}'}(t_{\vec{l},\vec{l}'}^{c,a,-\vec{\mu};d,b,-\vec{\nu}})^*e^{-i\vec{k}\cdot(\vec{R}_{\vec{l}-\vec{\mu}}-\vec{R}_{\vec{l}'-\vec{\nu}})}\notag\\
    =&e^{i\vec{k}\cdot(\vec{R}_{\vec{\mu}}-\vec{R}_{\vec{\nu}})}[X_{c,a,-\vec{\mu};d,b,-\vec{\nu}}(-\vec{k})]^*,
    \label{eq_mod_conj_cond_Bloch}
    \end{align}
    where we use $\vec{R}_{\vec{l}-\vec{\mu}}-\vec{R}_{\vec{l}'-\vec{\nu}}=\vec{R}_{\vec{l}}-\vec{R}_{\vec{l}'}-\vec{R}_{\vec{\mu}}+\vec{R}_{\vec{\nu}}$ in deriving the last equality.
    
    Next, we consider the condition for trace preservation. The superoperator preserves the trace of the density matrix if
    \begin{equation}
    \sum_m \hat{K}_m^\dag \hat{K}_m=\hat{I}.
    \label{eq_TP_cond}
    \end{equation}
    The condition \eqref{eq_TP_cond} for the trace preservation is rewritten as
    \begin{align}
    \sum_{\vec{i}}\sum_a\sum_m (\hat{K}_m)^*_{\vec{i},a;\vec{j},b}(\hat{K}_m)_{\vec{i},a;\vec{l},d}=\delta_{\vec{j},\vec{l}}\delta_{b,d},
    \end{align}
    and thus the hopping amplitudes and the Bloch matrix satisfy
    \begin{equation}
    \sum_{\vec{i}}\sum_a t_{\vec{i},\vec{j}}^{a,a,\vec{0};b,d,\vec{\nu}}=\delta_{\vec{\nu},\vec{0}}\delta_{b,d}
    \end{equation}
    and
    \begin{equation}
    \sum_a X_{a,a,\vec{0};b,d,\vec{\nu}}(\vec{k}=0)=\delta_{\vec{\nu},\vec{0}}\delta_{b,d},
    \label{eq_TP_cond_k}
    \end{equation}
    respectively. 
    Equations \eqref{eq_mod_conj_cond_Bloch} and \eqref{eq_TP_cond_k} place fundamental constraints on the Bloch matrix.

\section{Feedback control with projective measurement\label{sec_proj}}

Here we consider quantum feedback control with a projective measurement, where the measurement operators are given by $\hat{M}_m=\ket{\vec{j},a}\bra{\vec{j},a}$ and a measurement outcome is denoted by $m=(\vec{j},a)$. In this case, the matrix elements of the Kraus operator $\hat{K}_m=\hat{U}_m\hat{M}_m$ are written as
    \begin{equation}
    (\hat{K}_m)_{\vec{j}_1,a_1;\vec{j}_2,a_2}=
    (\hat{U}_m)_{\vec{j}_1,a_1;\vec{j},a}\delta_{\vec{j}_2,\vec{j}}\delta_{a_2,a}.
    \end{equation}
    Thus, the momentum-space representation of the CPTP map is given by
    \begin{align}
    X_{a,c,\vec{\mu};b,d,\vec{\nu}}(\vec{k})=&\frac{\delta_{b,d}\delta_{\vec{\nu},\vec{0}}}{N_{\mathrm{cell}}}\sum_{\vec{j},\vec{j}'}(\hat{U}_{\vec{j}',b})_{\vec{j},a;\vec{j}',b}(\hat{U}_{\vec{j}',b})^*_{\vec{j}+\vec{\mu},c;\vec{j}',b}\notag\\
    &\times e^{-i\vec{k}\cdot(\vec{R}_{\vec{j}}-\vec{R}_{\vec{j}'})},
    \end{align}
which indicates that the Bloch matrix can have nonzero matrix elements only in $D_{\mathrm{loc}}$ columns. Therefore, nonzero eigenvalues of the Bloch matrix coincide with those of a $D_{\mathrm{loc}}\times D_{\mathrm{loc}}$ matrix $X_{\mathrm{trunc}}(\vec{k}):=(X_{a,a,\vec{0};b,b,\vec{0}}(\vec{k}))$, which simplifies the spectral analysis of the CPTP map.

In particular, if a system has no internal degrees of freedom (i.e., $D_{\mathrm{loc}}=1$), the analysis is further simplified. In this case, we can omit the indices for internal degrees of freedom, and the matrix elements of the Bloch matrix are written as
\begin{align}
    X_{\vec{\mu},\vec{\nu}}(\vec{k}):=&\frac{\delta_{\vec{\nu},\vec{0}}}{N_{\mathrm{cell}}}\sum_{\vec{j},\vec{j}'}(\hat{U}_{\vec{j}'})_{\vec{j};\vec{j}'}(\hat{U}_{\vec{j}'})^*_{\vec{j}+\vec{\mu};\vec{j}'}
     e^{-i\vec{k}\cdot(\vec{R}_{\vec{j}}-\vec{R}_{\vec{j}'})}.
\end{align}
Since the Bloch matrix has nonzero matrix elements only in one column, its eigenvalues $\xi_{\vec{\mu}}(\vec{k})$ are given by its diagonal entries as
    \begin{align}
    \xi_{\vec{0}}(\vec{k})=&X_{\vec{0},\vec{0}}(\vec{k})\notag\\
    =&\frac{1}{N_{\mathrm{cell}}}\sum_{\vec{j},\vec{j}'}|(\hat{U}_{\vec{j}'})_{\vec{j},\vec{j}'}|^2e^{-i\vec{k}\cdot(\vec{R}_{\vec{j}}-\vec{R}_{\vec{j}'})}\notag\\
    =&\frac{1}{N_{\mathrm{cell}}}\sum_{\vec{j},\vec{j}'}p(\vec{j}|\vec{j}')e^{-ik(\vec{R}_{\vec{j}}-\vec{R}_{\vec{j}'})},\label{eq_xi_0}\\
    \xi_{\vec{\mu}}(\vec{k})=&0\ (\vec{\mu}\neq \vec{0}),\label{eq_xi_n}
    \end{align}
    where
    \begin{equation}
    p(\vec{j}|\vec{j}'):=|(\hat{U}_{\vec{j}'})_{\vec{j},\vec{j}'}|^2
    \end{equation}
    is the probability of finding a particle at site $\vec{j}$ after feedback control with measurement outcome $m=\vec{j}'$. Because of the translational invariance of the CPTP map, this probability should be a function of $\vec{r}=\vec{R}_{\vec{j}}-\vec{R}_{\vec{j}'}$. Hence, we have
    \begin{align}
    \xi_{\vec{0}}(\vec{k})=\sum_{\vec{r}}p(\vec{r})e^{-i\vec{k}\cdot\vec{r}},
    \label{eq_xi_0_2}
    \end{align}
    where $p(\vec{r})=p(\vec{R}_j-\vec{R}_{j'}):=p(\vec{j}|\vec{j}')$. This result indicates that the nonzero eigenvalue of the CPTP map can be obtained from the characteristic function of the probability distribution $\{p(\vec{r})\}$ of the displacement of a particle during feedback control. We note that $\xi_{\vec{0}}(\vec{k}=0)=\sum_{\vec{r}}p(\vec{r})=1$, which is consistent with the trace-preserving condition \eqref{eq_TP_cond_k}.

The Bloch matrix for the $D_{\mathrm{loc}}=1$ case is diagonalizable if $X_{\vec{0},\vec{0}}(\vec{k})\neq 0$. If $X_{\vec{0},\vec{0}}(\vec{k})=0$, it cannot be diagonalized except for the trivial case with $X(\vec{k})=0$. Thus, the point at which $X_{\vec{0},\vec{0}}(\vec{k})=\xi_{\vec{0}}(\vec{k})=0$ corresponds to an exceptional (nondiagonalizable) point \cite{Ashida20}. Specifically, if $X_{\vec{0},\vec{0}}(\vec{k})\neq 0$, the left eigenvector $v_{\vec{\nu}}^L(\vec{k})$ and the right eigenvector $v_{\vec{\nu}}^R(\vec{k})$ corresponding to the eigenvalue $\xi_{\vec{\nu}}(\vec{k})$ are given by
\begin{align}
    (v_{\vec{0}}^L(\vec{k}))_{\vec{\mu}}=&\delta_{\vec{\mu},\vec{0}},\label{eq_u0_proj}\\
    (v_{\vec{0}}^R(\vec{k}))_{\vec{\mu}}=&X_{\vec{\mu},\vec{0}}(\vec{k})/X_{\vec{0},\vec{0}}(\vec{k}),\\
    (v_{\vec{\nu}}^L(\vec{k}))_{\vec{\mu}}=&-\delta_{\vec{\mu},\vec{0}}X_{\vec{\nu},\vec{0}}^*(\vec{k})/X_{\vec{0},\vec{0}}^*(\vec{k})+\delta_{\vec{\mu},\vec{\nu}}\ (\vec{\nu}\neq\vec{0}),\\
    (v_{\vec{\nu}}^R(\vec{k}))_{\vec{\mu}}=&\delta_{\vec{\mu},\vec{\nu}}\ (\vec{\nu}\neq\vec{0}).\label{eq_vn_proj}
\end{align}
These eigenvectors satisfy the biorthogonal condition $(v_{\vec{\mu}}^L(\vec{k}),v_{\vec{\nu}}^R(\vec{k}))=\delta_{\vec{\mu},\vec{\nu}}$. 
    Substituting Eqs.~\eqref{eq_u0_proj}-\eqref{eq_vn_proj} into the expansions \eqref{eq_r_eigenop_k}-\eqref{eq_expansion_k}, we obtain the density matrix after feedback control as
    \begin{align}
    \mathcal{E}(\hat{\rho}_{\mathrm{i}})=&\frac{1}{\sqrt{N_{\mathrm{cell}}}}\sum_{\vec{k}}\sum_{\vec{j}}\sum_{\vec{\mu}} \xi_0(\vec{k})c(\vec{k})e^{i\vec{k}\cdot\vec{R}_{\vec{j}}}(v_{\vec{0}}^R(\vec{k}))_{\vec{\mu}}\notag\\
    &\times\ket{\vec{j}}\bra{\vec{j}+\vec{\mu}}\notag\\
    =&\frac{1}{\sqrt{N_{\mathrm{cell}}}}\sum_{\vec{k}}\sum_{\vec{j}}\xi_0(\vec{k})c(\vec{k})e^{i\vec{k}\cdot\vec{R}_{\vec{j}}}\ket{\vec{j}}\bra{\vec{j}}\notag\\
    &+\frac{1}{\sqrt{N_{\mathrm{cell}}}}\sum_{\vec{k}}\sum_{\vec{j}}\sum_{\vec{\mu}\neq 0}X_{\vec{\mu},0}(\vec{k})c(\vec{k})e^{i\vec{k}\cdot\vec{R}_{\vec{j}}}\notag\\
    &\times\ket{\vec{j}}\bra{\vec{j}+\vec{\mu}},
    \end{align}
    where the expansion coefficients
    \begin{align}
    c(\vec{k})=&\frac{1}{\sqrt{N_{\mathrm{cell}}}}\sum_{\vec{j}}\bra{\vec{j}}\hat{\rho}_{\mathrm{i}}\ket{\vec{j}}e^{-i\vec{k}\cdot\vec{R}_{\vec{j}}}\notag\\
    =&\frac{1}{\sqrt{N_{\mathrm{cell}}}}\sum_{\vec{j}}p_{\vec{j}}e^{-i\vec{k}\cdot\vec{R}_{\vec{j}}}
    \end{align}
    are determined by the probability distribution $p_{\vec{j}}=\mathrm{Tr}[\hat{M}_{\vec{j}}^\dag \hat{M}_{\vec{j}}\hat{\rho}_{\mathrm{i}}]$ of measurement outcomes. 
    The probability distribution of the position after feedback control is given by
    \begin{align}
    \bra{\vec{j}}\mathcal{E}(\hat{\rho}_{\mathrm{i}})\ket{\vec{j}}=&\frac{1}{\sqrt{N_{\mathrm{cell}}}}\sum_{\vec{k}}\xi_0(\vec{k})c(\vec{k})e^{i\vec{k}\cdot\vec{R}_{\vec{j}}}\notag\\
    =&\frac{1}{N_{\mathrm{cell}}}\sum_{\vec{k}}\sum_{\vec{j}'}\xi_0(\vec{k})p_{\vec{j}'}e^{i\vec{k}\cdot(\vec{R}_{\vec{j}}-\vec{R}_{\vec{j}'})}\notag\\
    =&\sum_{\vec{j}'}p(\vec{j}|\vec{j}')p_{\vec{j}'}.
    \end{align}

\section{Eigenvalues of a quantum channel for projective measurement\label{sec_proj_channel}}
Here, we show that eigenvalues of a quantum channel for a projective measurement are either zero or one. We consider a quantum channel
\begin{equation}
    \mathcal{E}_{\mathrm{proj}}(\hat{\rho})=\sum_m \hat{P}_m\hat{\rho} \hat{P}_m,
    \label{eq_CPTP_meas_app}
\end{equation}
in which the Kraus operators are given by projective operators, i.e., $\hat{P}_m^\dag=\hat{P}_m$ and $\hat{P}_m\hat{P}_n=\delta_{m,n}\hat{P}_m$.
The projective measurement channel \eqref{eq_CPTP_meas_app} is positive semidefinite since
\begin{align}
    \langle \hat{A},\mathcal{E}_{\mathrm{proj}}(\hat{A})\rangle=&\mathrm{Tr}[\hat{A}^\dag \sum_m \hat{P}_m \hat{A} \hat{P}_m]\notag\\
    =&\sum_m\mathrm{Tr}[\hat{A}^\dag \hat{P}_m^2 \hat{A} \hat{P}_m^2]\notag\\
    =&\sum_m\mathrm{Tr}[\hat{P}_m\hat{A}^\dag \hat{P}_m \hat{P}_m \hat{A} \hat{P}_m]\notag\\
    =&\sum_m\langle \hat{P}_m \hat{A} \hat{P}_m, \hat{P}_m \hat{A} \hat{P}_m\rangle\notag\\
    \geq& 0
\end{align}
for an arbitrary operator $\hat{A}$. Thus, eigenvalues of $\mathcal{E}_{\mathrm{proj}}$ are real and non-negative. These eigenvalues must be either zero or one since the measurement channel satisfies the idempotency condition:
\begin{align}
    \mathcal{E}^2_{\mathrm{proj}}(\hat{\rho})=&\sum_m\sum_{m'}\hat{P}_m\hat{P}_{m'}\hat{\rho} \hat{P}_{m'} \hat{P}_m\notag\\
    =&\sum_m\sum_{m'}\delta_{m,m'}\hat{P}_{m}\hat{\rho} \delta_{m,m'}\hat{P}_m\notag\\
    =&\sum_m \hat{P}_m\hat{\rho} \hat{P}_m\notag\\
    =&\mathcal{E}_{\mathrm{proj}}(\hat{\rho}).
\end{align}
This completes the proof.

\section{Correspondence between symmetry of a time-evolution operator and that of its generator\label{sec_sym_corresp}}
Suppose that a time-evolution operator has a generator. To investigate the symmetry of a time-evolution operator, it is helpful to consider the symmetry of its generator. Here, we summarize the correspondence between symmetry of a time-evolution operator and that of its generator to compare the former with the symmetry of quantum channels discussed in Sec.~\ref{sec_BL_sym}. 

In closed quantum systems, a time-evolution operator $\hat{U}=e^{-i\hat{H}t}$ is generated by a Hermitian Hamiltonian $\hat{H}$. The Altland-Zirnbauer symmetries of a Hermitian Hamiltonian are written as
\begin{subequations}
\begin{align}
    \hat{V}_C\hat{H}^*\hat{V}_C^{-1}=&\hat{H},\ \hat{V}_C\hat{V}_C^*=\pm \hat{I},\\
    -\hat{V}_K\hat{H}^*\hat{V}_K^{-1}=&\hat{H},\ \hat{V}_K\hat{V}_K^*=\pm \hat{I},\\
    -\hat{\Gamma} \hat{H}\hat{\Gamma}^{-1}=&\hat{H},\ \hat{\Gamma}^2=\hat{I},
\end{align}
\end{subequations}
where $\hat{V}_C,\hat{V}_K,\hat{\Gamma}$ are unitary operators \cite{Altland97}. The corresponding symmetries of a unitary time-evolution operator derived from these symmetries of a Hamiltonian are written as
\begin{subequations}
\begin{align}
    \hat{V}_C\hat{U}^T\hat{V}_C^{-1}=&\hat{U},\ \hat{V}_C\hat{V}_C^*=\pm \hat{I},\label{eq_TRS_U}\\
    \hat{V}_K\hat{U}^*\hat{V}_K^{-1}=&\hat{U},\ \hat{V}_K\hat{V}_K^*=\pm \hat{I},\label{eq_PHS_U}\\
    \hat{\Gamma} \hat{U}^\dag\hat{\Gamma}^{-1}=&\hat{U},\ \hat{\Gamma}^2=\hat{I}.\label{eq_CS_U}
\end{align}
\end{subequations}
A CPTP map of unitary time evolution is given by $\mathcal{E}(\hat{\rho})=\hat{U}\hat{\rho} \hat{U}^\dag$, which is mapped to a unitary operator $\tilde{\mathcal{E}}=\hat{U}\otimes \hat{U}^*$ acting on the doubled Hilbert space. It follows from Eqs.~\eqref{eq_TRS_U}-\eqref{eq_CS_U} that the corresponding symmetries of the CPTP map are given by
\begin{subequations}
\begin{align}
    (\hat{V}_C\otimes \hat{V}_C^*)\tilde{\mathcal{E}}^T(\hat{V}_C\otimes \hat{V}_C^*)^{-1}=\tilde{\mathcal{E}}
\end{align}
with $(\hat{V}_C\otimes \hat{V}_C^*)(\hat{V}_C\otimes \hat{V}_C^*)^*=+\hat{I}\otimes\hat{I}$,
\begin{align}
    (\hat{V}_K\otimes \hat{V}_K^*)\tilde{\mathcal{E}}^*(\hat{V}_K\otimes \hat{V}_K^*)^{-1}=\tilde{\mathcal{E}}
\end{align}
with $(\hat{V}_K\otimes \hat{V}_K^*)(\hat{V}_K\otimes \hat{V}_K^*)^*=+\hat{I}\otimes\hat{I}$, and
\begin{align}
    (\hat{\Gamma}\otimes\hat{\Gamma}^*)\tilde{\mathcal{E}}^\dag (\hat{\Gamma}\otimes\hat{\Gamma}^*)^{-1}=\tilde{\mathcal{E}}
\end{align}
with $(\hat{\Gamma}\otimes\hat{\Gamma}^*)^2=\hat{I}\otimes\hat{I}$.
\end{subequations} 
Thus, these symmetries of unitary channels constitute a subset of symmetry classes with $\eta_C=\eta_K=+1$ in Eqs.~\eqref{eq_TRSdag}-\eqref{eq_PH}.

In Markovian open quantum systems, the generator (Lindbladian) $\mathcal{L}$ of a CPTP map $\mathcal{E}=e^{\mathcal{L}t}$ for time evolution features the Gorini-Kossakowski-Sudarshan-Lindblad form \cite{BreuerPetruccione}. The matrix representation of a CPTP map through the Choi-Jamio\l kowski isomorphism (see Appendix~\ref{sec_CJ}) is denoted by $\tilde{\mathcal{E}}=e^{\tilde{\mathcal{L}}t}$. Here, we note that a Lindbladian cannot be a general non-Hermitian matrix since the real part of its eigenvalues must be nonpositive. Because of this constraint, only the following set of Bernard-LeClair symmetries are possible for a Lindbladian \cite{Lieu20} (see also Refs.~\cite{Kawabata22,Sa22}):
\begin{subequations}
\begin{align}
    \tilde{\mathcal{V}}_C\tilde{\mathcal{L}}^T\tilde{\mathcal{V}}_C^{-1}=&\tilde{\mathcal{L}},\ \tilde{\mathcal{V}}_C\tilde{\mathcal{V}}_C^*=\pm \hat{I}\otimes\hat{I},\label{eq_TRS_L}\\
    \tilde{\mathcal{V}}_K\tilde{\mathcal{L}}^*\tilde{\mathcal{V}}_K^{-1}=&\tilde{\mathcal{L}},\ \tilde{\mathcal{V}}_K\tilde{\mathcal{V}}_K^*=\pm \hat{I}\otimes\hat{I},\label{eq_PHS_L}\\
    \tilde{\mathcal{V}}_Q \tilde{\mathcal{L}}^\dag\tilde{\mathcal{V}}_Q^{-1}=&\tilde{\mathcal{L}},\ \tilde{\mathcal{V}}_Q^2=\hat{I}\otimes\hat{I},\label{eq_CS_L}
\end{align}
\end{subequations}
where $\tilde{\mathcal{V}}_C,\tilde{\mathcal{V}}_K,\tilde{\mathcal{V}}_Q$ are unitary operators acting on the doubled Hilbert space. 
The corresponding symmetries of a CPTP map are given by
\begin{subequations}
\begin{align}
    \tilde{\mathcal{V}}_C\tilde{\mathcal{E}}^T\tilde{\mathcal{V}}_C^{-1}=&\tilde{\mathcal{E}},\ \tilde{\mathcal{V}}_C\tilde{\mathcal{V}}_C^*=\pm \hat{I}\otimes\hat{I},\label{eq_TRS_CPTPL}\\
    \tilde{\mathcal{V}}_K\tilde{\mathcal{E}}^*\tilde{\mathcal{V}}_K^{-1}=&\tilde{\mathcal{E}},\ \tilde{\mathcal{V}}_K\tilde{\mathcal{V}}_K^*=\pm \hat{I}\otimes\hat{I},\label{eq_PHS_CPTPL}\\
    \tilde{\mathcal{V}}_Q \tilde{\mathcal{E}}^\dag\tilde{\mathcal{V}}_Q^{-1}=&\tilde{\mathcal{E}},\ \tilde{\mathcal{V}}_Q^2=\hat{I}\otimes\hat{I}.\label{eq_CS_CPTPL}
\end{align}
\end{subequations}
The symmetries \eqref{eq_TRS_CPTPL}-\eqref{eq_CS_CPTPL} coincide with those of quantum channels in Eqs.~\eqref{eq_TRSdag}-\eqref{eq_PH}.

\section{Symmetry-preserving boundary condition for a $\mathbb{Z}_2$ demon\label{sec_symOBC}}

In this appendix, we describe how to implement a symmetry-preserving boundary condition for the $\mathbb{Z}_2$ demon in Sec.~\ref{sec_Z2}. The transition probability amplitudes for a feedback control in class AII$^\dag$ must satisfy Eqs.~\eqref{eq_QND_qcond1}-\eqref{eq_QND_qcond3}. Here we perform a feedback control near the boundary of the system to preserve these symmetry conditions. We consider a 1D lattice with $L$ unit cells under the OBC, and the site $j=1$ ($j=L$) denotes the left (right) boundary. Feedback control under the PBC satisfies the conditions \eqref{eq_QND_qcond1}-\eqref{eq_QND_qcond3}. However, if we perform the same feedback under the OBC, the reflection of a particle at the boundary violates these conditions. A feedback protocol that preserves the symmetry at the boundary is given as follows. Let $m_1=(j_1,a_1)$ be a measurement outcome of the first measurement. If $j_1=1$ or $j_1=L$, we do nothing (i.e., the feedback Hamiltonian is set to zero: $\hat{H}_{1,a_1}=\hat{H}_{L,a_1}=0$). The transition amplitudes from the boundary sites therefore vanish. 
If $j_1\neq 1,L$, we perform feedback control with the same Hamiltonian $\hat{H}_{j_1,a_1}$ as that for the PBC case described in Sec.~\ref{sec_Z2}. Next, we perform the second projective measurement and obtain an outcome $m_2=(j_2,a_2)$. If $j_2=2,\cdots,L-1$, we perform the same additional feedback control as the PBC case shown in Eq.~\eqref{eq_Z2_additionalFB}. If $j_2=1$ or $j_2=L$, we perform a feedback operation with the unitary operator
\begin{align}
    \hat{U}_{m_1,m_2}^{\prime\prime}=\hat{S}_{m_1,m_2}\exp[-i\theta_{m_1,m_2}\hat{S}^z],
\end{align}
where
\begin{align}
    \hat{S}_{m_1,m_2}=\sum_\sigma(\ket{j_1,a_1,\sigma}\bra{j_2,a_2,\sigma}+\ket{j_2,a_2,\sigma}\bra{j_1,a_1,\sigma})    
\end{align}
is a swap operator. This unitary operation returns a particle back to the site before the first feedback control, thereby making the conditions \eqref{eq_QND_qcond1}-\eqref{eq_QND_qcond3} satisfied for the boundary sites. The phase factor $\theta_{m_1,m_2}$ is chosen so that the condition \eqref{eq_QND_qcond1} is satisfied for $j_2=j_1$.

Specifically, for the model in Sec.~\ref{sec_Z2}, the symmetry is preserved if the phase factors satisfy
\begin{align}
    &p(1,A|2,A)e^{i\theta_{(2,A),(1,A)}}+p(1,B|2,A)e^{i\theta_{(2,A),(1,B)}}\notag\\
    =&p(1,A|2,B)e^{i\theta_{(2,B),(1,A)}}+p(1,B|2,B)e^{i\theta_{(2,B),(1,B)}}
    \label{eq_Z2app_pcondL}
\end{align}
and
\begin{align}
    &p(L,A|L-1,A)e^{i\theta_{(L-1,A),(L,A)}}\notag\\
    &+p(L,B|L-1,A)e^{i\theta_{(L-1,A),(L,B)}}\notag\\
    =&p(L,A|L-1,B)e^{i\theta_{(L-1,B),(L,A)}}\notag\\
    &+p(L,B|L-1,B)e^{i\theta_{(L-1,B),(L,B)}},
    \label{eq_Z2app_pcondR}
\end{align}
where
\begin{equation}
    p(j_2,a_2|j_1,a_1)=|(\hat{U}_{j_1,a_1})_{j_2,a_2,\sigma;j_1,a_1,\sigma}|^2.
    \label{eq_Z2app_p}
\end{equation}
The existence of a solution of Eqs.~\eqref{eq_Z2app_pcondL} and \eqref{eq_Z2app_pcondR} depends on the details of the transition probability \eqref{eq_Z2app_p}. For example, if
\begin{align}
    p(1,A|2,B)=p(1,B|2,A)
    \label{eq_Z2app_solcond1}
\end{align}
and
\begin{align}
    2p(1,A|2,B)\geq |p(1,B|2,B)-p(1,A|2,A)|
    \label{eq_Z2app_solcond2}
\end{align}
are satisfied, a solution of Eq.~\eqref{eq_Z2app_pcondL} is given by
\begin{align}
    \theta_{(2,A),(1,A)}=&\theta_{(2,B),(1,B)}=0,\label{eq_Z2app_thetasol1}\\
    \theta_{(2,A),(1,B)}=&\arccos\frac{p(1,B|2,B)-p(1,A|2,A)}{2p(1,A|2,B)},\label{eq_Z2app_thetasol2}\\
    \theta_{(2,B),(1,A)}=&\pi-\theta_{(2,A),(1,B)}.\label{eq_Z2app_thetasol3}
\end{align}
A solution of Eq.~\eqref{eq_Z2app_pcondR} can be obtained similarly. For the parameters employed in the numerical simulation in Fig.~\ref{fig_Z2_spec}, the conditions \eqref{eq_Z2app_solcond1} and \eqref{eq_Z2app_solcond2} are met, and therefore, we set the phase factors as Eqs.~\eqref{eq_Z2app_thetasol1}-\eqref{eq_Z2app_thetasol3} in Fig.~\ref{fig_Z2_spec}.

It is worthwhile to note that conditions \eqref{eq_Z2app_pcondL} and \eqref{eq_Z2app_pcondR} may not have a solution. In that case, the symmetry of a quantum channel should be broken under the OBC, and the symmetry-protected non-Hermitian skin effect is not ensured. However, if the symmetry is broken only near the boundary, the transport of a particle towards the boundary due to the skin effect is partially maintained for a certain time before a particle approaches the boundary since the dynamics in the bulk is unaltered \cite{Gong18}.

\begin{figure}[t]
    \includegraphics[width=8.0cm]{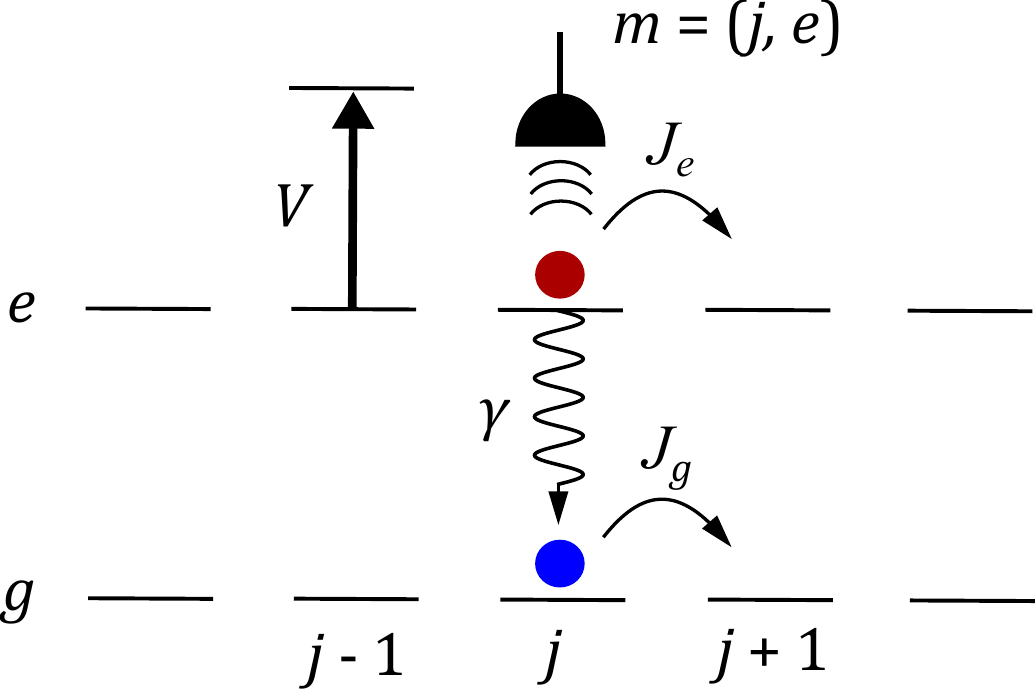}
    \caption{Schematic illustration of topological feedback control that shows the non-Hermitian skin effect but with no boundary-localized steady state. A two-level atom with internal states $g$ and $e$ on a 1D lattice is considered. The excited state $e$ spontaneously decays into the ground state $g$ with rate $\gamma$. If a projective measurement finds the atom in the excited state, a potential wall with height $V$ is introduced on the left side of the particle.}
    \label{fig_spont_demon}
\end{figure}

\begin{figure}[t]
    \includegraphics[width=8.0cm]{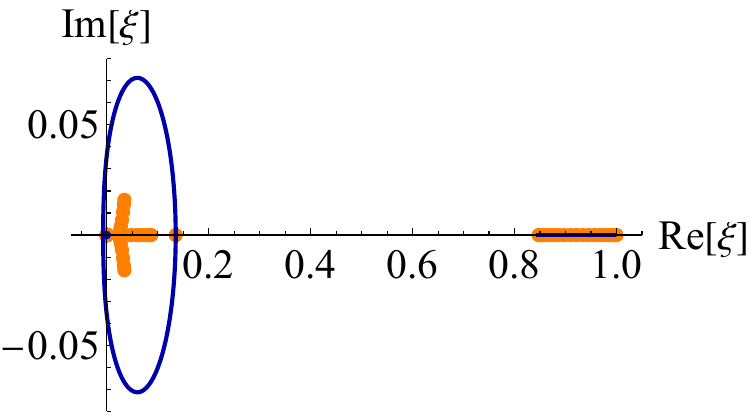}
    \caption{Eigenspectrum of a quantum channel in Eq.~\eqref{eq_E_spont}. The parameters are set to $J_e=1,J_g=0.2,V=10,\gamma=2$, and $\tau=1$. Blue curves and blue dots at the origin constitute the eigenspectrum under the PBC for infinite system size, and orange dots show the eigenspectrum under the OBC for the system size $L=20$.}
    \label{fig_spont_spec}
\end{figure}

\section{Topological feedback control that shows the non-Hermitian skin effect but with no boundary-localized steady state\label{sec_spont_demon}}
The models of topological feedback control presented in Secs.~\ref{sec_model}, \ref{sec_SSH}-\ref{sec_Z2}, and \ref{sec_diss_fb} show the non-Hermitian skin effects and boundary-localized steady states due to nontrivial point-gap topology of quantum channels. While the former is widely observed in non-Hermitian topological phases \cite{Gong18,Okuma20,Zhang20,Borgnia20,Okuma23,Lin23}, the latter does not exist in conventional non-Hermitian systems and thus is unique to topological quantum channels that preserve the positivity and trace of a density matrix (see Sec.~\ref{sec_boundary_SS}). Here, to distinguish these two topological phenomena, we present a model that shows the non-Hermitian skin effect but with no boundary-localized steady states. We consider a two-level atom on a 1D lattice with $L$ sites. The internal states of the atom consist of a ground state ($g$) and an excited state ($e$). We perform a projective measurement of the position and the internal state of the atom described by the measurement operator $\hat{M}_{j,a}=\ket{j,a}\bra{j,a}$, where $\ket{j,a}$ denotes a quantum state of the atom at site $j=1,\cdots,L$ and internal state $a=g,e$. Next, depending on the measurement outcome $m=(j,a)$, we let the atom undergo time evolution under the Lindblad equation
\begin{align}
    \frac{d\hat{\rho}}{dt}=&\mathcal{L}_{m}(\hat{\rho})\notag\\
    =&-i[\hat{H}_{j,a},\hat{\rho}]+\sum_i\Bigl(\hat{L}_{i}\hat{\rho} \hat{L}_{i}^\dag -\frac{1}{2}\{ \hat{L}_{i}^\dag \hat{L}_{i},\hat{\rho}\}\Bigr),
    \label{eq_fb_Lindblad_spont}
\end{align}
where
\begin{align}
\hat{H}_{j,a}=&-\sum_{i,a'}J_{a'}(\ket{i,a'}\bra{i+1,a'}+\ket{i+1,a'}\bra{i,a'})\notag\\
&+\delta_{a,e}V\ket{j-1,e}\bra{j-1,e}
\end{align}
is the feedback Hamiltonian with hopping amplitudes $J_e,J_g\in\mathbb{R}$ and potential height $V>0$, and $\hat{L}_i=\sqrt{\gamma}\ket{i,g}\bra{i,e}$ is a Lindblad operator that describes spontaneous decay with rate $\gamma>0$ from the excited state to the ground state. In other words, we impose a potential wall only if the internal state of the atom is detected to be the excited state. See Fig.~\ref{fig_spont_demon} for a schematic illustration. The quantum channel for feedback control is thus given by (see also Sec.~\ref{sec_diss_fb})
\begin{align}
\mathcal{E}(\hat{\rho})=\sum_{m}e^{\mathcal{L}_m\tau}(\hat{M}_{m}\hat{\rho}\hat{M}_m^\dag),
\label{eq_E_spont}
\end{align}
where $\tau$ is the duration of feedback control.

Figure \ref{fig_spont_spec} shows the eigenspectrum of the quantum channel \eqref{eq_E_spont}. Here, a salient feature is that the band in the PBC spectrum that involves the steady state (the unit eigenvalue) does not form a loop and thus has a vanishing winding number. Physically, this feature occurs because the atom is in the internal ground state in the steady state and thus is not localized near the edges of the system. Meanwhile, the band located away from the steady state exhibits a loop structure. This topological structure leads to the non-Hermitian skin effect of an atom in the internal excited state, which has a finite lifetime due to spontaneous decay. Thus, the non-Hermitian skin effect in topological feedback control is not always accompanied by a boundary-localized steady state.

\bibliography{feedback_ref.bib}

\end{document}